\documentclass[longauth]{aa}  

\usepackage{graphicx}
\usepackage{txfonts}
\usepackage{booktabs,threeparttable}
\usepackage{sistyle}
\usepackage{amsmath}
\usepackage{array,multirow,makecell}
\usepackage{subfig}
\usepackage[export]{adjustbox}
\usepackage{color}
\usepackage{textcomp}
\usepackage{gensymb}
\usepackage{multirow}
\usepackage[utf8]{inputenc}
\usepackage{placeins} 
\bibpunct{(}{)}{;}{a}{}{,} 

\makeatletter
\renewcommand*\aa@pageof{, page \thepage{} of \pageref*{LastPage}}
\makeatother

\usepackage{hyperref}
\hypersetup{
    colorlinks=true,
    linkcolor=blue,filecolor=magenta,
    citecolor=blue,urlcolor=blue,
}

\usepackage[capitalise]{cleveref} 

\newcommand{\kms}{\ensuremath{\rm km\,s^{-1}}}
\newcommand{\Msol}{\ensuremath{M_{\odot}}}

\newcommand{\sub}[1]{_{\mathrm{#1}}}

\newcommand{\bsens}{\texttt{bsens}\xspace}
\newcommand{\cleanest}{\texttt{cleanest}\xspace}

\begin{document} 

	\title{ALMA-IMF. V. Prestellar and protostellar core populations in the W43 cloud complex 
 }

  \author{T. Nony\inst{1}
          \and R. Galván-Madrid\inst{1}
          \and F. Motte\inst{2}
          \and Y. Pouteau\inst{2}
          \and N. Cunningham\inst{2}
          \and F. Louvet\inst{2}
          \and A. M. Stutz\inst{3}
          \and B. Lefloch\inst{2}
          \and S. Bontemps\inst{4}
          \and N. Brouillet\inst{4}    
          \and A. Ginsburg\inst{5}
          \and I. Joncour\inst{2}
          \and F. Herpin\inst{4}
          \and P. Sanhueza\inst{6,7}
          \and T. Csengeri\inst{4}
          \and A. P. M. Towner\inst{5}
          \and M. Bonfand\inst{4}
          \and M. Fernández-López\inst{8}
          \and T. Baug\inst{9}
          \and L. Bronfman\inst{10}
          \and G. Busquet\inst{11,12,13}
          \and J. Di Francesco\inst{14}
          \and A. Gusdorf\inst{15,16}
          \and X. Lu\inst{17}
          \and F. Olguin\inst{18}
          \and M. Valeille-Manet\inst{4}
          \and A. P. Whitworth\inst{19}
          }

   \institute{Instituto de Radioastronomía y Astrofísica, Universidad Nacional Autónoma de México, Morelia, Michoacán 58089, México
    \email{t.nony@irya.unam.mx}
   \and Univ. Grenoble Alpes, CNRS, IPAG, 38000 Grenoble, France
   \and Departamento de Astronom\'{i}a, Universidad de Concepci\'{o}n,Casilla 160-C, Concepci\'{o}n, Chile
   \and OASU/LAB, Univ. de Bordeaux - CNRS/INSU, 36615 Pessac, France
   \and Department of Astronomy, University of Florida, PO Box 112055, USA
   \and National Astronomical Observatory of Japan, National Institutes of Natural Sciences, 2-21-1 Osawa, Mitaka, Tokyo 181-8588, Japan 
   \and Department of Astronomical Science, SOKENDAI (The Graduate University for Advanced Studies), 2-21-1 Osawa, Mitaka, Tokyo 181-8588, Japan
   \and Instituto Argentino de Radioastronom\'\i a (CCT-La Plata, CONICET; CICPBA), C.C. No. 5, 1894, Villa Elisa, Buenos Aires, Argentina 
   \and S. N. Bose National Centre for Basic Sciences, Block JD, Sector III, Salt Lake, Kolkata 700106, India
   \and Departamento de Astronomía, Universidad de Chile, Casilla 36-D, Santiago, Chile 
   \and Departament de Física Quàntica i Astrofísica (FQA), Universitat de Barcelona (UB), c. Martí i Franquès, 1, 08028 Barcelona, Spain 
   \and Institut de Ciències del Cosmos (ICCUB), Universitat de Barcelona (UB), c. Martí i Franquès, 1, 08028 Barcelona, Spain
   \and Institut d'Estudis Espacials de Catalunya (IEEC), c. Gran Capità, 2-4, 08034 Barcelona, Spain 
   \and Herzberg Astronomy and Astrophysics Research Centre, National Research Council of Canada, 5071 West Saanich Road, Victoria, BC CANADA V9E 2E7 
   \and Laboratoire de Physique de l’Ecole Normale Supérieure, ENS, Université PSL, CNRS, Sorbonne Université, Université Paris Cité, F-75005, Paris, France
   \and Observatoire de Paris, PSL University, Sorbonne Université, LERMA, 75014, Paris, France
   \and Shanghai Astronomical Observatory, Chinese Academy of Sciences, 80 Nandan Road, Shanghai 200030, People’s Republic of China 
   \and Institute of Astronomy, National Tsing Hua University, Hsinchu 30013, Taiwan
   \and School of Physics and Astronomy, Cardiff University, Cardiff, UK
   			}

   \date{}

  \abstract
  {The origin of the stellar initial mass function (IMF) and its relation with the core mass function (CMF) are actively debated issues with important implications in astrophysics. Recent observations in the W43 molecular complex of top-heavy CMFs, with an excess of high-mass cores compared to the canonical mass distribution, raise questions about our understanding of the star formation processes and their evolution in space and time. }
  {We aim to compare populations of protostellar and prestellar cores in three regions imaged in the ALMA-IMF Large Program.}
  {We created an homogeneous core catalogue in W43, combining a new core extraction in W43-MM1 with the catalogue of W43-MM2\&MM3 presented in a previous work. Our detailed search for protostellar outflows enabled us to identify between 23 and 30 protostellar cores out of 127 cores in W43-MM1 and between 42 and 51 protostellar cores out of 205 cores in W43-MM2\&MM3. Cores with neither outflows nor hot core emission are classified as prestellar candidates.}
  {We found a similar fraction of cores which are protostellar in the two regions, about 35\%. This fraction strongly varies in mass, from $f_{\rm pro} \simeq 15-20 $\% at low mass, between 0.8 and 3\,$\Msol$ up to $f_{\rm pro} \simeq 80 $\% above 16\,$\Msol$. Protostellar cores are found to be, on average, more massive and smaller in size than prestellar cores. Our analysis also revealed that the high-mass slope of the prestellar CMF in W43, $\alpha=-1.46_{-0.19}^{+0.12}$, is consistent with the Salpeter slope, and thus the top-heavy form measured for the global CMF, $\alpha=-0.96 \pm 0.09$, is due to the protostellar core population.}
  {Our results could be explained by 'clump-fed' models in which cores grow in mass, especially during the protostellar phase, through inflow from their environment. The difference between the slopes of the prestellar and protostellar CMFs moreover implies that high-mass cores grow more in mass than low-mass cores.}

               
   \maketitle
%

\section{Introduction}  
\label{s:intro}

In the field of high-mass star and cluster formation, many important topics remain open, from the way stars grow in mass during their early stages of evolution to the origin and universality of their mass distributions, the initial mass function (IMF).
One of the first models for the formation of high-mass stars, referred to as 'core collapse', 
was proposed by \cite{Mckee2003}. 
It could also be labelled as a 'core-fed' model, in the sense that the mass reservoir available for the formation of the star is limited to the mass of an isolated prestellar core accumulated before its collapse. This model 
requires the existence of a massive, turbulent prestellar core that has formed quasi-statically. This pre-assembled core is in virial equilibrium, as observed in low-mass prestellar cores, and supported by some magnetic and/or supersonic turbulent pressure that keep effective Jeans masses large enough to prevent sub-fragmentation.
The core collapse model implies a one-to-one relation between the mass distribution of prestellar cores and the stellar IMF \citep{Tan14}. 

Alternatively, other models propose the rapid growth of cores via competitive accretion from a common cloud mass reservoir \citep[e.g.][]{Bonnel06} or via accretion streams associated with the global hierarchical collapse of clouds \citep[e.g.][]{smith09,Vazquez19}.
These models are often referred to as 'clump-fed', since they involve a gas mass reservoir on larger scales than core-fed models, with dynamical processes between the individual cores and their parental clump to cloud environment. 

Over the past decades, results from numerical simulations and observations have challenged the core-fed models, while bringing more support for clump-fed models.
Observationally, evidence showing the importance of dynamical processes have accumulated \citep[e.g.][]{schneider10, Galvan10,csengeri11b, peretto13, henshaw14, louvet16,Contreras18}, while only a few high-mass prestellar core candidates have been reported \citep{Bontemps2010,duarte2013co,Wang2014,Nony18}.
In the empirical evolutionary sequence proposed for the formation of high-mass stars by \cite{motte18a},
high-mass protostellar cores form from low-mass protostellar cores, which accrete further material from their parental clump. This model thus does not rely on the existence of high-mass prestellar cores.

In this context, measuring core mass functions (CMFs) is of major importance to understand core mass growth and constrain models. Historically, the first observations were conducted towards nearby, low- to intermediate-mass star-forming regions using ground-based telescopes \citep{Motte98,Enoch08} and the \textit{Herschel} Space Observatory \citep{Polychroni13,Konyves15,Konyves20}.
These studies found CMFs similar to the stellar IMF, with high-mass end slopes consistent with the canonical, so-called 'Salpeter slope', of the IMF (dN/dlogM $\propto$ M$^{\alpha}$ with $\alpha=-1.35$). This led to the idea that stellar masses are directly related to core masses through fragmentation processes, as suggested by the core collapse models, although the inferred direct link from the CMF to the IMF relies on questionable hypotheses on core-to-star mass efficiency, fragmentation, and timescales \citep[see][]{Clark07,Offner14,Pelkonen21}. 
More recently, the deployment of the Atacama Large Millimeter/submillimeter Array (ALMA) interferometer has enabled observations of more distant high-mass star-forming regions, resulting in new measures of CMFs that depart from the canonical form of the IMF \citep[][]{Motte18b,Liu18,Kong19,Sanhueza19,Lu20,ONeil21}. This illustrates that extending observations to different environments than
the solar neighbourhood is necessary to get a complete view of star-formation processes, especially in the high-mass regime. \\

The pioneering study of \cite{Motte18b} towards W43-MM1 was one of the first to measure a CMF significantly different from the canonical IMF.
The molecular cloud complex W43, located at 5.5\,kpc from the Sun \citep{zhang14} at the junction point of the Scutum-Centaurus spiral arm and the galactic bar, is known as a particularly active star-forming region.
Among the parsec-scale clumps identified in W43 by \cite{motte03}, MM1 stands out as one of the most extreme protoclusters of the Milky Way. With $2\times10^4\,M_\odot$ within $6\,$pc$^2$ and a star-formation rate reminiscent of that of starburst galaxies, SFR\,$\sim\,6000\,\Msol\,$Myr$^{-1}$ \citep{nguyen11b, nguyen13,louvet14}, the W43-MM1 ridge qualifies as a `mini-starburst'. W43-MM1 has been observed at 1.3~mm with ALMA, revealing a cluster of 131 cores with masses ranging from 1 to 100 $\Msol$ \citep{Motte18b}. The combined CMF of prestellar and protostellar cores was found to be significantly flatter than the IMF (slope of $\alpha = -0.96 \pm 0.12$), that is with an excess of high-mass cores. 
The observation of a few selected emission lines, simultaneous to the continuum, also enabled the detection of a high-mass prestellar core candidate of about $60\,\Msol$ by \cite{Nony18}, which was further characterised through a detailed chemical analysis by \cite{Molet19}. Later on, \cite{Nony20} carried out a survey of molecular outflows in W43-MM1 using CO (2--1) and SiO (5--4) lines and revealed a rich cluster of 46 outflow lobes from 27 cores. 

The top-heavy CMF measured in W43-MM1 has raised questions concerning the relation between the CMF and the IMF, which motivated the ALMA-IMF Large Program
\citep{Motte22}. 
Among the targets are the neighbouring regions W43-MM2 and W43-MM3, the second and third most massive clumps identified in W43 by \cite{motte03} ($1.6\times10^3\,M_\odot$ and $1\times10^3\,M_\odot$, respectively). Whereas W43-MM2 is likely as young as W43-MM1 given their similar 1.3 mm to 3 mm flux ratios (see \citealt{Motte22}), the presence of an ultra-compact H$_{\rm II}$ 
region in W43-MM3 indicates a more evolved region. Using ALMA-IMF continuum maps of W43-MM2\&MM3, \cite{Pouteau22a} detected 205 cores with masses up to about 70$\,\Msol$ and a slope similar to that of W43-MM1, $\alpha = -0.95 \pm 0.04$. \\

In this work, we aim to further discriminate protostellar and prestellar cores using outflow detections, in order to address the question of the origin of the top-heavy CMFs measured by \cite{Motte18b} and \cite{Pouteau22a}. 
So far, protostellar cores have received less attention than prestellar cores in CMF studies. Yet, being at an intermediate evolutionary stage between prestellar cores and young stars, they are key to understanding the evolution of the CMF to the IMF.
In \cref{s:obs-cat}, we present ALMA observations of continuum and line data for the three regions of W43: MM1, MM2 and MM3, and the detection of cores within. 
We start the analysis in \cref{s:analy} by presenting the detection of outflows, then we compare core properties between regions and between protostellar and prestellar core populations, focusing on their mass. A discussion on the high-mass prestellar core candidates and on core mass growth processes is proposed in \cref{s:disc}, and \cref{s:conclu} summarises our main results and conclusions.

\section{Observations and core catalogues}
\label{s:obs-cat}

\subsection{Continuum and line observations}
\label{su:obs}

\begin{table*}[htb]
\begin{center}
   \begin{threeparttable}
          \caption[]{\label{tab:data}Parameters of continuum images and CO datacubes.}
        \begin{tabular}{ c c c c c c c c c }
    \hline
    \hline
    Region &  Image     &   $\nu_{\rm obs}$\tnote{a}  & Bandwidth\tnote{b}  & pixel & \multicolumn{2}{c}{Resolution}  & $\sigma_{\rm rms}\tnote{c}$ \\
       &    &  [GHz]        &   [GHz]    & [$\arcsec$] & [$\arcsec \times \arcsec$] & [km~s$^{-1}$]  & [mJy\,beam$^{-1}$] \\
     \hline
    W43-MM1 & Cont. $\bsens^*$  & 229.494  &  2.82 & 0.07 & $0.51 \times 0.36$  & -  & 0.08 \\
    W43-MM2 & Cont. \bsens   & 228.901  &  3.45 & 0.1 & $0.52 \times 0.41$  & -  & 0.13 \\
    W43-MM3 & Cont. \bsens   & 228.902  &  3.45 & 0.1 & $0.51 \times 0.43$  & -  & 0.09 \\
    W43-MM1 & CO (2--1)  &  230.462   & 0.169  & 0.1 & $0.55 \times 0.39$ & 1.3 & 2.3 \\
    W43-MM2 & CO (2--1)  &  230.466   & 0.183  & 0.09 & $0.61 \times 0.50$ & 1.3 & 2.0 \\
    W43-MM3 & CO (2--1)  &  230.466   & 0.148  & 0.14 & $0.62 \times 0.53$ & 1.3 & 2.0 \\
    \hline
        \end{tabular} 
        \begin{tablenotes}
\item [a] Reference observed frequency: central frequency calculated for \bsens continuum images using a spectral index $\alpha=3.5$ \citep[see Table D.1 of][]{Ginsburg22} and central frequency of the CO (2--1) line. 
\item [b] Bandwidth used to build the continuum images and bandwidth of the CO cube.
\item [c] 1 mJy\,beam$^{-1}$ corresponds to 92~mK at 230 GHz for a 0.5$\arcsec$ beam.
    \end{tablenotes}
\end{threeparttable}
\end{center} 
\end{table*}
%

Observations of W43-MM1 at 1.3~mm (Band 6) were carried out in Cycle 2 between July 2014 and June 2015 (project \#2013.1.01365.S), with the ALMA 12-m array covering baselines ranging from 7.6~m to 1045~m. W43-MM1 was imaged with a $78\arcsec\times$ 53$\arcsec$ (2.1 pc $\times$ 1.4 pc) mosaic composed of 33 fields. The primary beam FWHM is  26.7$\arcsec$ and the maximum detectable scale is $\sim 12\arcsec$. 
Observations of W43-MM2 and W43-MM3 were carried out between December 2017 and December 2018 as part of the ALMA-IMF Large Program \citep[project \#2017.1.01355.L, PIs: Motte, Ginsburg, Louvet, Sanhueza, see][]{Motte22} with similar spatial and spectral setup. Mosaics are composed of 27 fields and the maximum detectable scale is $\sim 11\arcsec$.
Our observations with the 7-m array configuration are not used in this work. The combined 12-m\,+\,7-m data indeed have higher noise levels than the 12-m only data \citep[see][]{Pouteau22a} without bringing information significant for this work, whether on the detection of compact cores or the identification of collimated outflows.

Continuum data for the three regions were processed with CASA 5.4 using the data reduction pipeline developed by the ALMA-IMF consortium \citep[described in detail in][]{Ginsburg22}.
In short, the pipeline performs several iterations of cleaning and phase self-calibration using masks of increasing size and decreasing thresholds. We used the multi-scale multi-frequency synthesis (MS-MFS) method of tclean with two Taylor terms and scales of [0,3,9,27] pixels, corresponding to point sources and several larger scales. For the final clean of W43-MM1, an additional scale of 54 pixels was used.

CO (2--1) cubes of W43-MM2\&MM3 were also processed using the ALMA-IMF data pipeline, although without self-calibration (see \citealt{Cunningham23} for more detail on line data reduction). We used the multiscale method of tclean with scales of [0,6,18,54] and [0,4,12,24] pixels for W43-MM2 and W43-MM3, respectively, corresponding to point sources and multiples of the beam size. The resulting cubes have a similar beam of about 0.61$\arcsec \times$ 0.52$\arcsec$. The CO cube of W43-MM1 was presented and published by \cite{Nony20}. 
Parameters of the various continuum images and CO cubes are summarised in \cref{tab:data}. 
Continuum has been subtracted from the cubes in the image plane using the imcontsub task of CASA.
CO cubes were created after ALMA-IMF data were reprocessed in QA3 to correct the known spectral normalisation issue\footnote{See ALMA ticket \url{https://help.almascience.org/kb/articles/what-errors-could-originate-from-the-correlator-spectral-normalization-and-tsys-calibration}}.

\subsection{Continuum maps and core catalogues} 
\label{su:cont}

The ALMA-IMF pipeline provides two different continuum images \citep[see][]{Ginsburg22}. The first one, called \cleanest, is produced from a selection of channels free of line contamination using the \texttt{findcont} routine of CASA. The second one, called \bsens, covers the whole frequency setup to favour the best sensitivity possible but can be contaminated by strong line emission, especially that from the CO (2--1) line.
The following analysis of the three W43 regions is based on the deep core extraction method on these maps, developed by \cite{Pouteau22a} for W43-MM2\&MM3 and applied to W43-MM1.

\subsubsection{W43-MM2 and MM3}
\label{sub:contMM23}
The neighbouring W43-MM2 and W43-MM3 regions were first imaged separately using the ALMA-IMF data reduction pipeline (see \cref{su:obs}). The resulting continuum maps have the same size of $92\arcsec \times 97\arcsec$ at 1.3~mm, similar angular resolution of about $0.51\arcsec \times 0.42\arcsec$ and share a common area of $10\arcsec \times 90\arcsec$. The two maps have then been combined in a single image of the W43-MM2\&MM3 region, used in the following analyses.
\cite{Pouteau22a} showed that applying a "denoising" process on the continuum map before running the core extraction increases the number of detected cores thanks to an increased sensitivity, without introducing spurious sources or degrading the quality of flux measurements. In short, the denoising process relies on MnGSeg \citep{Robitaille19}, a wavelet-based method which decomposes an image into a Gaussian component associated with the cloud structure and the noise and a coherent component containing the hierarchical star-forming structures. By removing part of the Gaussian component, the noise level decreased by $\sim30\%$ in the "denoised" image of W43-MM2\&MM3.

\begin{figure*}
    \centering
    \includegraphics[width=0.93\hsize]{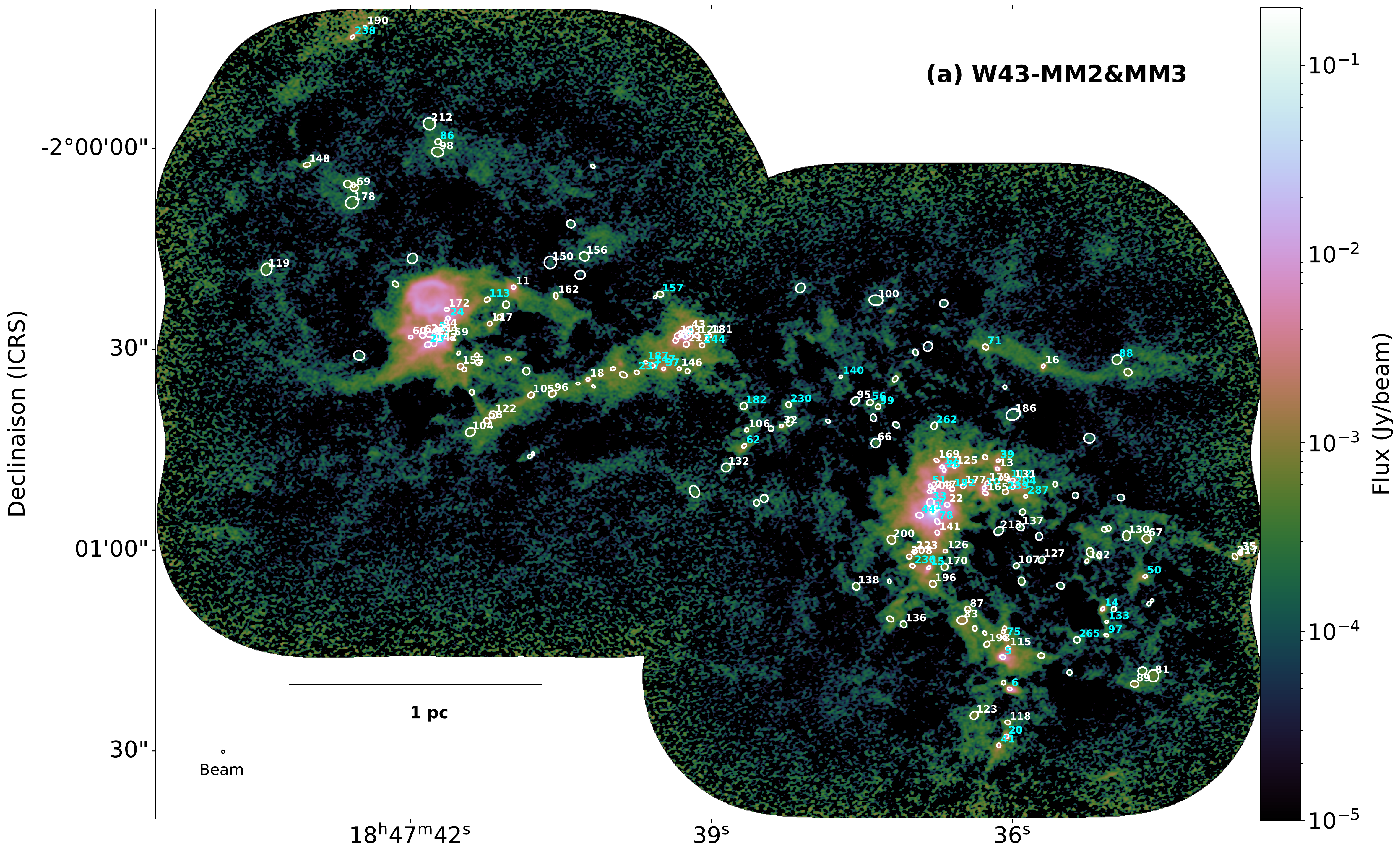}
    \includegraphics[width=0.93\hsize]{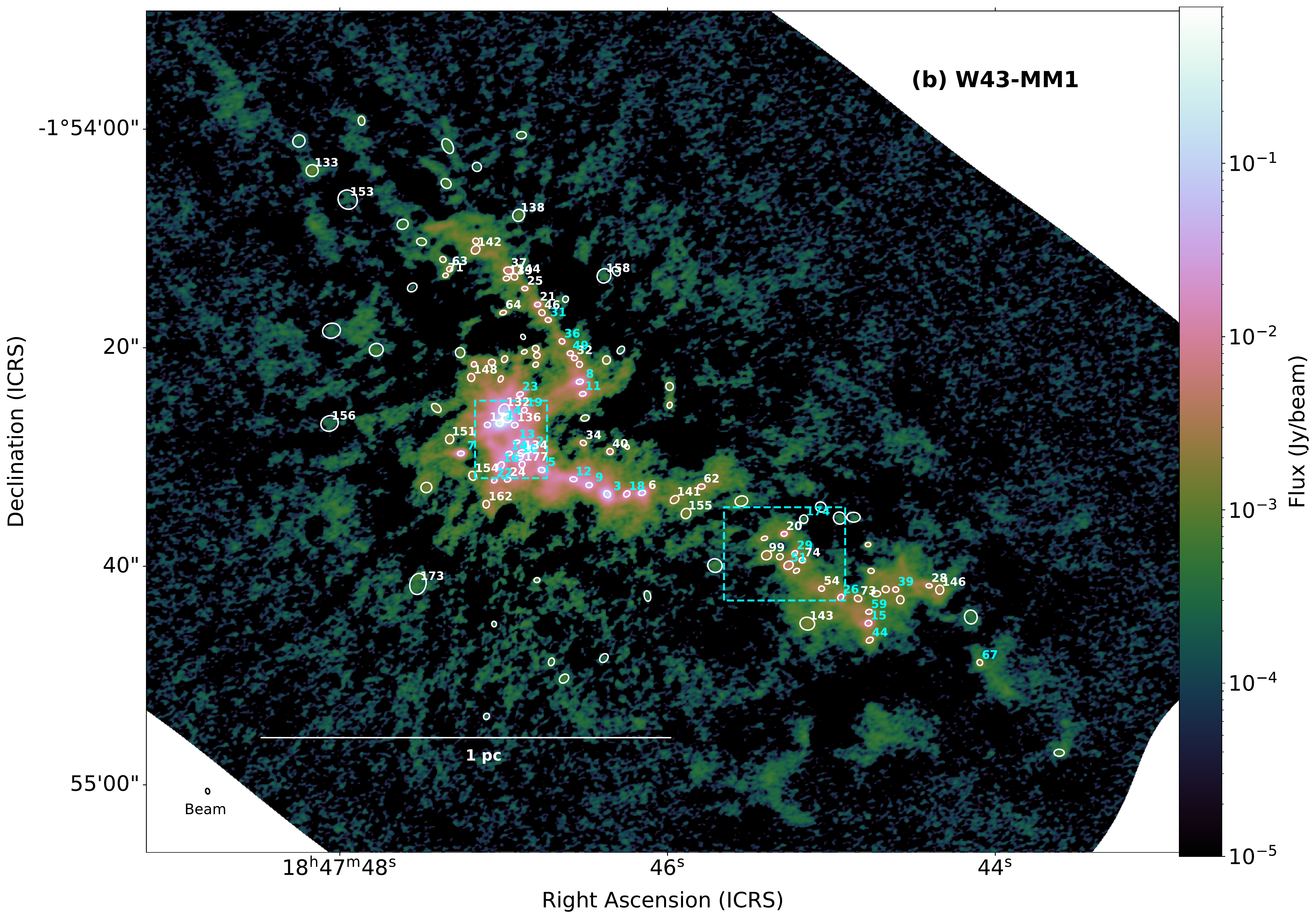}
    \caption{W43-MM2\&MM3 (in \textit{a}) and W43-MM1 (in \textit{b}) protoclusters, as imaged at 1.3 mm by the ALMA 12-m array. The sources extracted with \textit{getsf} are represented by ellipses showing their size at FWHM. 
    Cores listed in \cref{tab:nature-coreMM23,tab:nature-coreMM1} are numbered in cyan for protostellar cores and in white for prestellar cores above the completeness limit (1.6$\,\Msol$ and 0.8$\,\Msol$ in \textit{a} and \textit{b}, resp.). \bsens and $\bsens^*$ continuum images are shown in \textit{a} and \textit{b}, respectively. 
    Rectangles in cyan in \textit{b} indicate the position of the zooms provided in \cref{fig:app-flow-MM1-zoom} to highlight a new high-mass prestellar core candidate and new core-outflow associations. Ellipses in the lower left corners represent the angular resolution of the image and scale bars indicate the size in physical units. 
    }
    \label{fig:cont-MM123}
\end{figure*}

The \textit{getsf} algorithm \citep{Mensh21} has been chosen for the core extraction. \textit{getsf} is the successor of the sources and filaments extraction methods \textit{getsources} and \textit{getfilaments} \citep{Mensh12,Mensh13} 
and is well suited to extract compact cores in complex environments. 
\cite{Pouteau22a} used the denoised \bsens image of W43-MM2\&MM3 for the detection step and the denoised \bsens and \cleanest images for measurements. 
The catalogue of the combined regions contains 208 sources passing the recommended post-selection filters (i.e. size lower than four times the beam, ellipticity lower than 2, peak and integrated flux signal-to-noise ratios above 2). Three sources with high free-free contamination were further discarded, leading to a final sample of 205 cores. Cores are shown on the 1.3\,mm continuum map in \cref{fig:cont-MM123}a.
For 14 cores with significant 1.3~mm flux contamination, \cleanest measurements were used instead of \bsens measurements. These cores were identified from their \bsens/\cleanest flux ratios significantly above unity and consist in four cores with emission of complex organic molecules, associated with hot cores, and ten cores with contamination from other bright lines such as CO (2--1).
The full catalogue of cores including positions, sizes, peak and integrated fluxes, and masses has been published by \cite{Pouteau22a}. A subset of these parameters is included in \cref{tab:nature-coreMM23}.

\subsubsection{W43-MM1 }
\label{sub:contMM1}
We reprocessed the previously obtained W43-MM1 observations at 1.3~mm using the pipeline developed for ALMA-IMF 
\citep[see also Appendix F of][]{Ginsburg22}. 
The resulting continuum maps show significant reduction of sidelobes around the central region compared to the maps presented by \cite{Motte18b} (see \cref{fig:app-cont-MM1}).
We used a modified \bsens image, $\bsens^*$, excluding from the \bsens frequency selection the brightest lines such as CO isotopologues, SiO and SO. The detailed frequency selection for this $\bsens^*$ map is provided in \cref{s:app-freq-MM1}.

Following the approach developed by \cite{Pouteau22a} towards W43-MM2\&MM3 (see \cref{sub:contMM23}), we performed the source extraction with \textit{getsf} using the denoised $\bsens^*$ image for the detection step and the denoised $\bsens^*$ and \cleanest images for measurements. 
127 cores were found after post-selection filtering\footnote{The choice of taking post-selection filters on the original continuum image instead of the primary-beam corrected image (as done in this work) affects marginally the final catalogue, with 12 additional cores with masses below the completeness limit being detected.}. 
We took \cleanest measurements instead of $\bsens^*$ measurements for five cores with large line contamination, associated with known hot cores \citep{Brouillet22}. 
In \cref{fig:app-hc-conta}, we show the core   $\bsens^*$/\cleanest flux ratios as a function of signal-to-noise ratio. The five corrected cores shown in red stand out with flux ratios above 3$\sigma$. 

Cores from this catalogue are outlined with ellipses on the 1.3 mm continuum map of Fig~\ref{fig:cont-MM123}b. A short comparison with the previous \textit{getsources} catalogue of 131 cores from \cite{Motte18b} is provided in \cref{s:app-compar-cat}. We used temperatures from the map presented by \cite{Motte18b} and calculated masses similarly as in W43-MM2\&MM3, using an equation correcting for the dust optical thickness \citep[see][]{Motte18b,Pouteau22a}.
Cores have diameters ranging from 1200 to 9300~au once deconvolved from the 0.44$\arcsec$-beam and masses ranging from 0.2$\,\Msol$ to 109$\,\Msol$.  
\cref{tab:nature-coreMM1} lists the parameters (Gaussian FWHM sizes, fluxes, temperatures and masses) of the complete core sample (see \cref{su:complete}).

\subsection{Completeness}
\label{su:complete}

In order to determine the completeness level of the source extraction, we performed synthetic source extractions using 1.3~mm continuum maps of each region as background images. 
In W43-MM2\&MM3, \cite{Pouteau22a} evaluated a 90\% completeness level of 0.8$\,\Msol$ (see their Appendix C), with 113 cores out of 205 above this level. 

We performed similar analyses in W43-MM1 and found a 90\% completeness level of 1.6$\,\Msol$ (see \cref{s:app-completeness}), consistent with the completeness estimation of \cite{Motte18b}. This higher completeness mass could be due to the brighter and more centrally concentrated continuum emission in W43-MM1 compared to W43-MM2\&MM3, which produces higher structural noise (see also \cref{su:disc-cpl-proto}).
Among the 127 cores of the new core catalogue, 68 fall above the 1.6$\,\Msol$ completeness level. 59 of these (87\%) are also detected in the \cleanest catalogue of Louvet et al. (subm.) and 48 (71\%) are in the core catalogue of \cite{Motte18b} (see also \cref{s:app-compar-cat}).
The complete core sample of W43, with masses above $1.6\,\Msol$, has 126 cores (69 cores in W43-MM1 and 57 cores in W43-MM2\&MM3). In addition, 56 cores in W43-MM2\&MM3 with masses in the [0.8-1.6] $\Msol$ interval are included in the following analysis.

\begin{figure*}[ht]
    \centering
    \includegraphics[width=0.9\hsize]{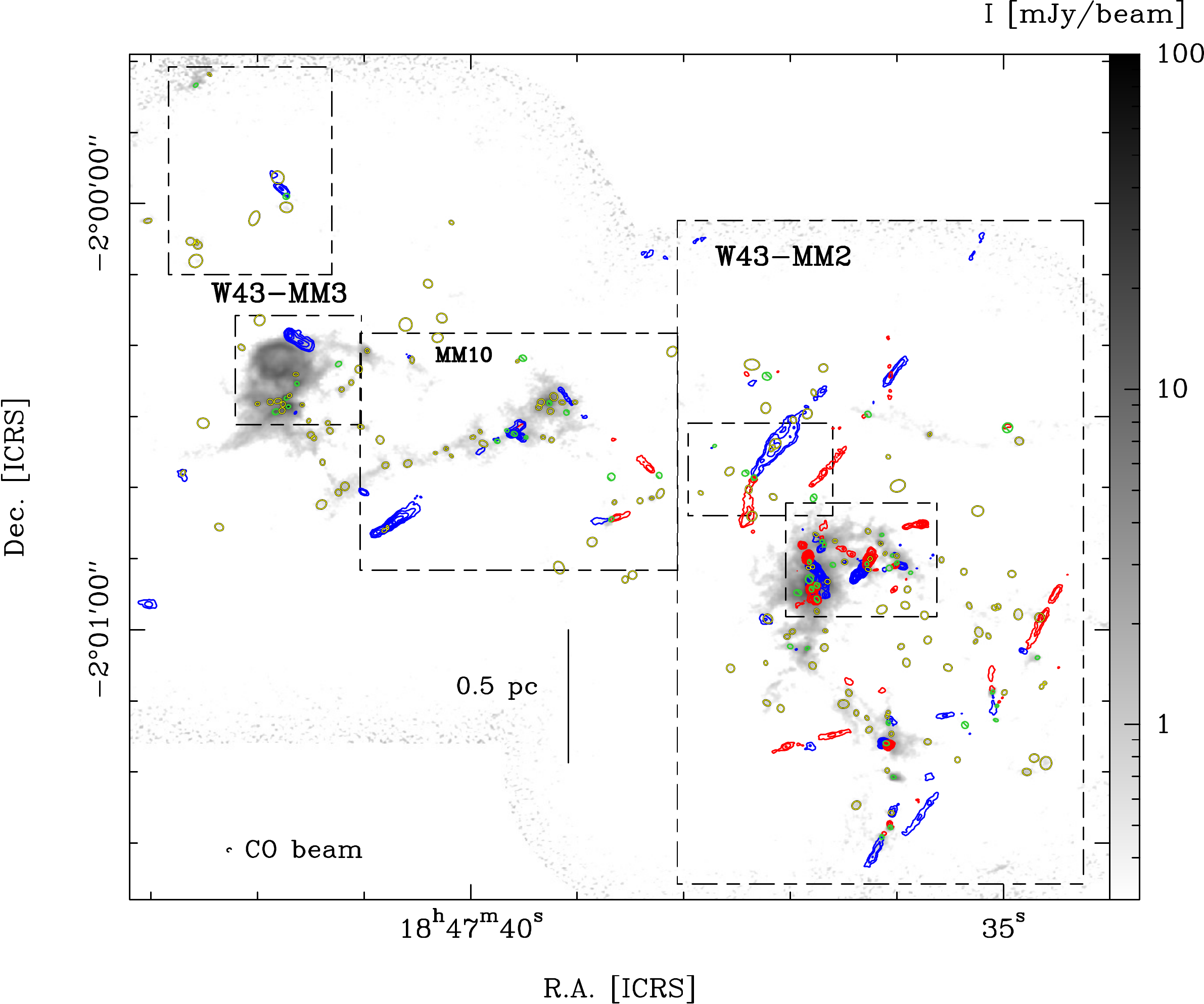}
    \caption{Molecular outflows in the W43-MM2\&MM3 region overlaid on the 1.3-mm continuum emission (grey scale). The CO (2--1) blue- and red-shifted line wings, integrated over 53.6-61.2 $\kms$ and 125.9-133.6 $\kms$ respectively, are overlaid as blue and red contours on the 1.3 mm $\bsens^*$ continuum image. Contours are 5, 10 to Max by steps of 10 in units of $\sigma$, with $\sigma=10$ mJy\,beam$^{-1}\,$km\,s$^{-1}$ and 13 mJy\,beam$^{-1}\,$km\,s$^{-1}$ in MM2 and MM3, respectively. Cores driving outflows are represented by green ellipses showing their FWHM size, while other cores are represented by yellow ellipses. 
    The beam of the CO cubes is represented in the lower left corner and a scale bar is shown. The position of the zooms shown in \cref{fig:flow-MM2,fig:flow-MM3} are delimited with dashed lines.}
    \label{fig:flow-overview}
\end{figure*}

\section{Analysis}
\label{s:analy}

\subsection{Identification of protostellar cores using outflows}
\label{su:flow}

Molecular outflows are the most common signature of the protostellar phase of cores in the millimetre wavelength range \citep[e.g.][]{Bontemps96}. Therefore, we used the presence or absence of outflows as the main criterion to identify protostellar cores.
In addition to this, most of massive protostellar cores are also associated with compact emission of complex organic molecules, referred to as \textit{hot core} emission. The presence of hot core emission has been used as a secondary criterion to identify protostellar cores, based on the systematic survey carried out in W43-MM2\&MM3 by Bonfand et al. (in prep) and in W43-MM1 by \cite{Brouillet22}.

We assume the protostellar core sample has the same mass completeness as the total core sample (see \cref{su:complete}), because outflows are detected for cores from the lowest to the highest masses (see \cref{tab:nature-coreMM1,tab:nature-coreMM23}). In agreement, \cite{Nony20} suggested that a dense and dynamic environment is the main limiting factor for outflow detection, more than the mass of the driving core. The mass completeness of protostellar core detection is discussed in further detail in \cref{su:disc-cpl-proto}. Uncertainties on the characterisation of protostellar cores are taken into account through the uncertainties in outflow detection or attribution (see the distinction between robust and tentative in the following).

\subsubsection{Protostellar cores in W43-MM2\&MM3}
\label{su:flowMM23}

\begin{figure}[htp]
    \centering
    \includegraphics[width=\hsize]{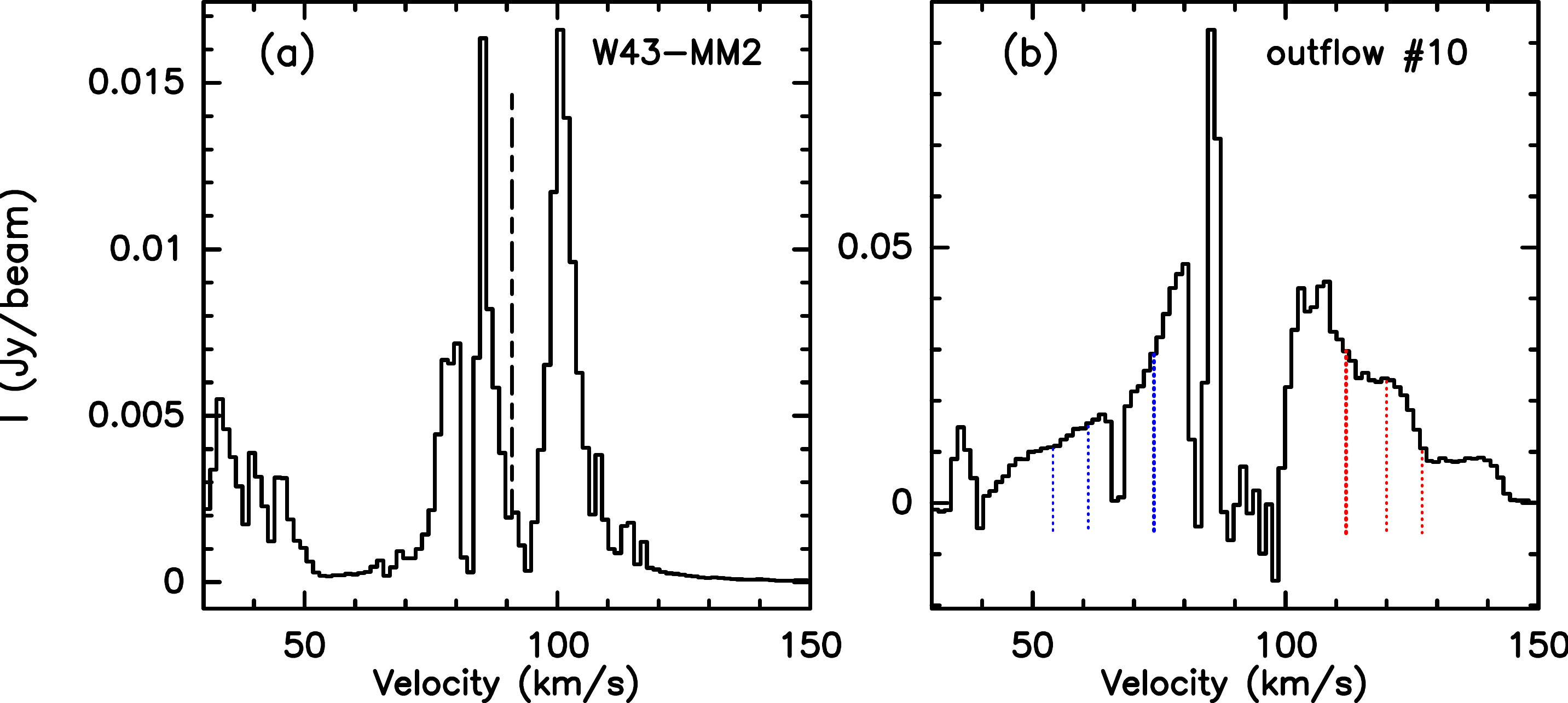}
    \includegraphics[width=\hsize]{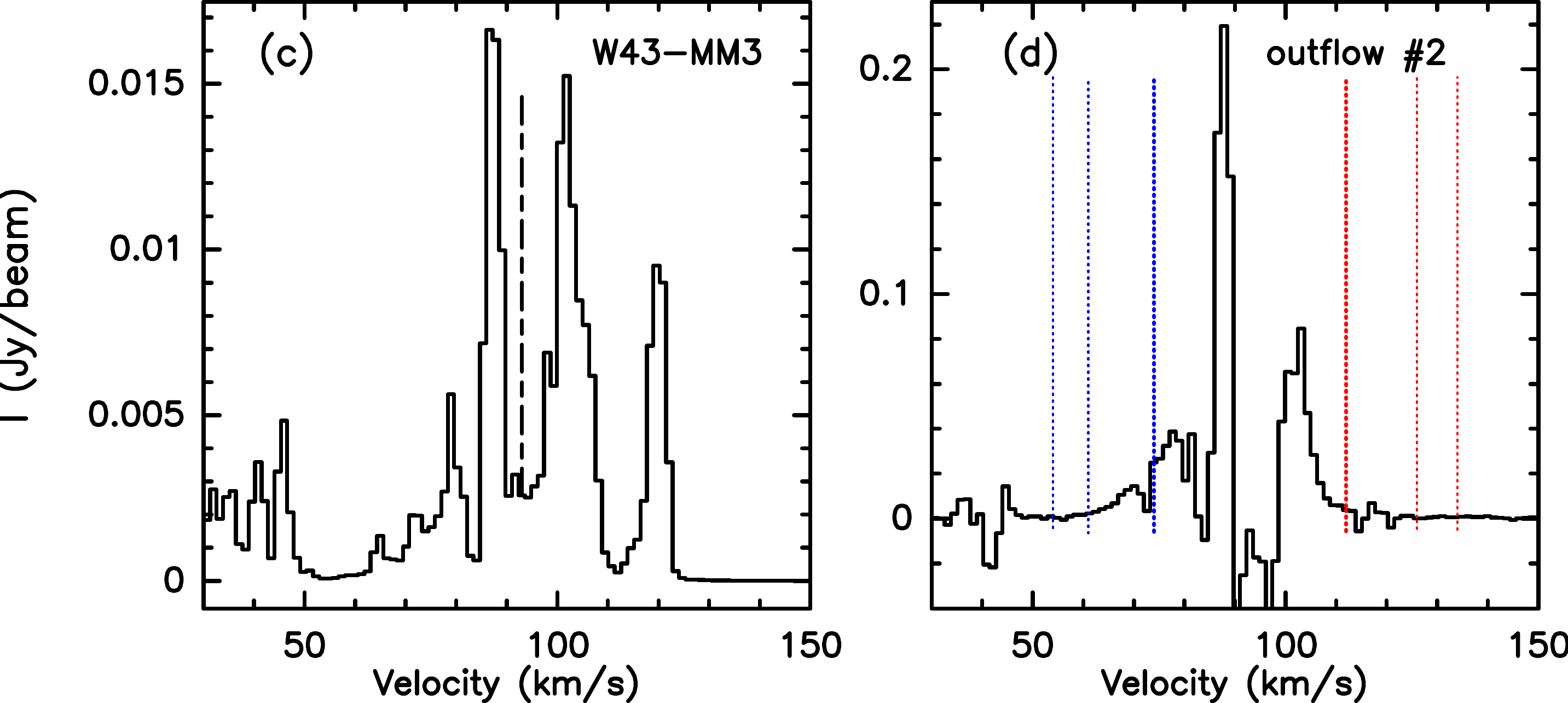}
    \caption{Spectra of the CO (2--1) line towards the entire W43-MM2 and W43-MM3 regions (\textit{a} and \textit{c}) and towards bright outflows (\textit{b} and \textit{d}). The systemic velocities, V$_{\rm LSR}$ = 91$\,\kms$ and V$_{\rm LSR}$ = 93$\,\kms$ for W43-MM2 and W43-MM3, respectively, 
    are measured with the DCN(3-2) line \citep{Cunningham23} and are indicated by black dashed lines in (\textit{a}) and (\textit{c}). The limits of the high- and low-velocity intervals are indicated by blue and red dotted lines in (\textit{b}) and (\textit{d}).}
    \label{fig:spec}
\end{figure}

\begin{figure*}
    \centering
    \begin{tabular}{cc}
    \adjustbox{valign=b}{\subfloat{\includegraphics[width=0.55\hsize,valign=c]{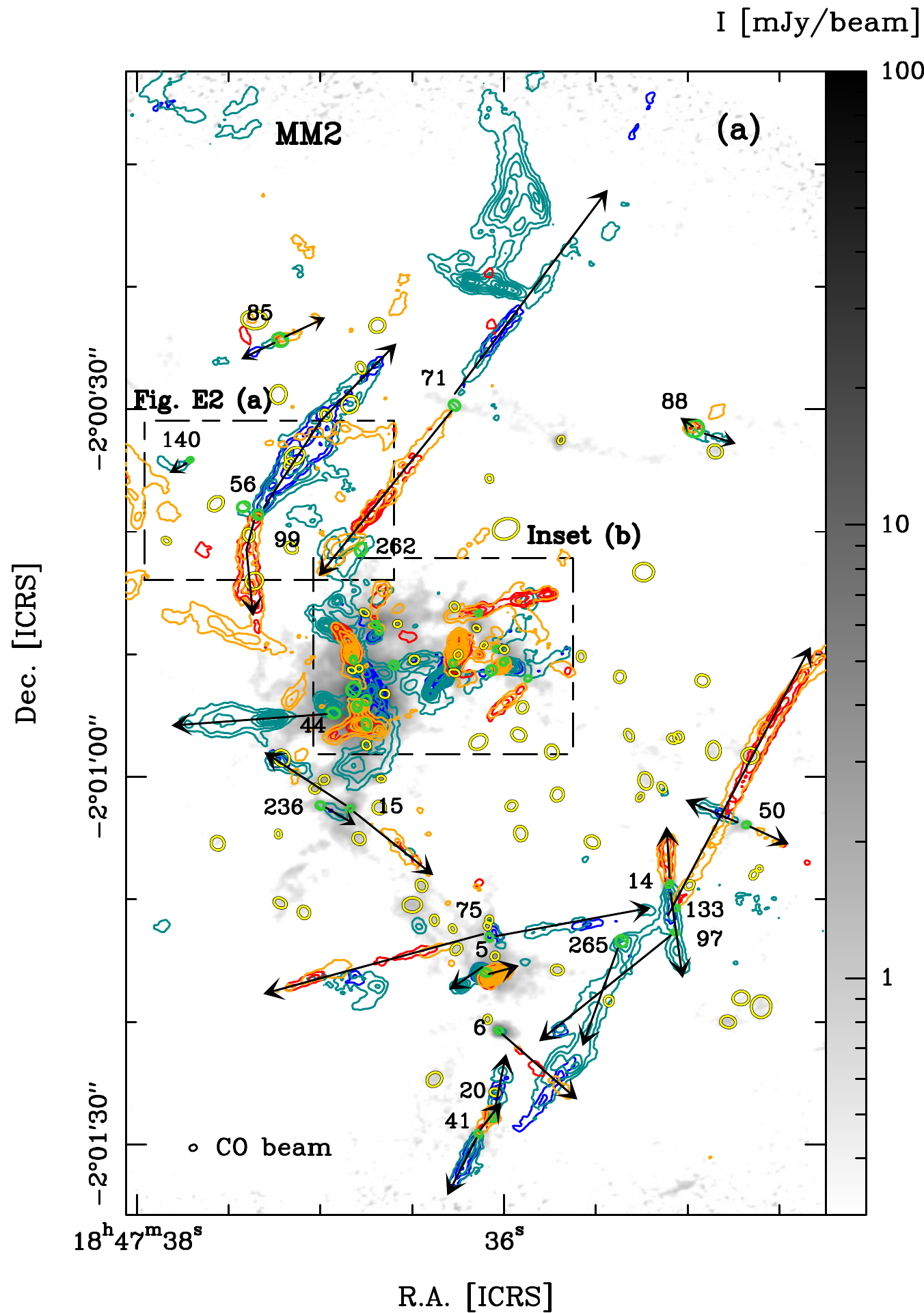}}}
    \adjustbox{valign=b}{\begin{tabular}{@{}c@{}}
    \subfloat{\includegraphics[width=0.47\hsize]{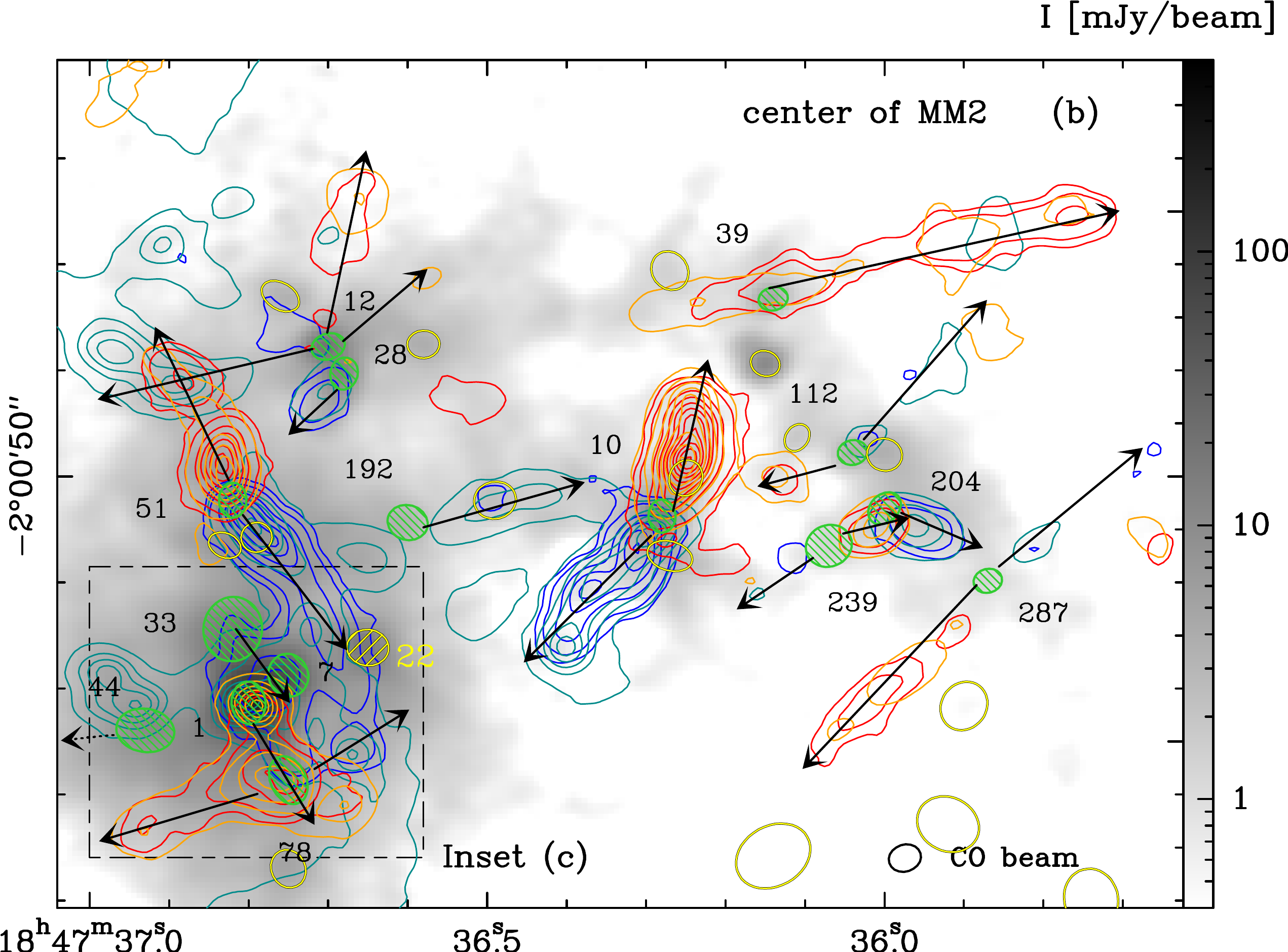}} \\ \subfloat{\includegraphics[width=0.43\hsize]{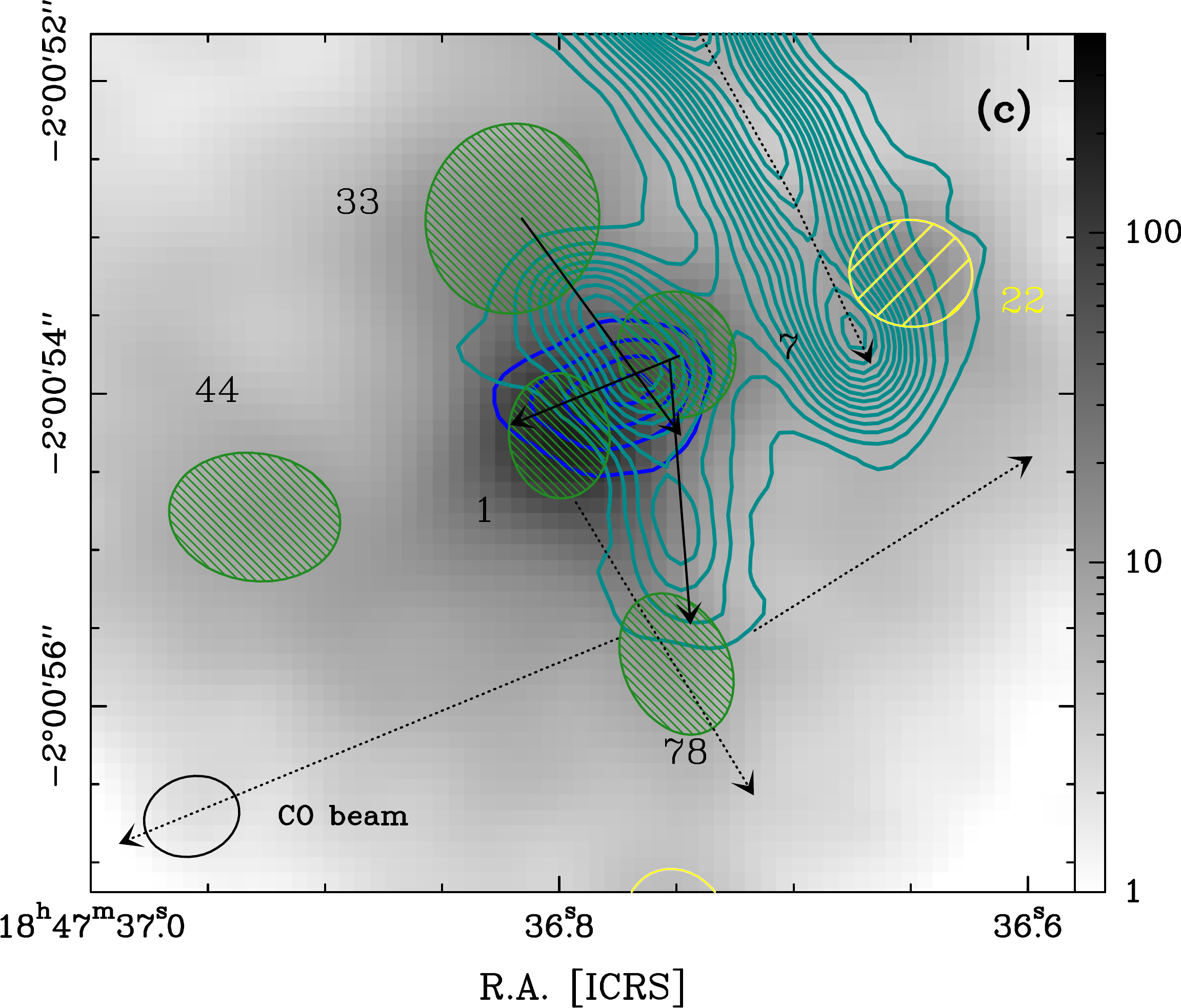}}\end{tabular}}
    \end{tabular}
    \caption{Zoom-in from \cref{fig:flow-overview} towards W43-MM2. \textit{(a-b)} The CO (2--1) blue-shifted line wing is integrated over 53.6-61.2 $\kms$ (blue contours) and 73.9-75.1 $\kms$ (cyan contours), the red-shifted wing over 110.7-112.0 $\kms$ (orange contours) and 119.6-127.2 $\kms$ (red contours).
    \textit{(a)} Contours are 5, 10 to Max by steps of 10 in units of $\sigma_{\rm HV2}=$ 10 mJy\,beam$^{-1}\,$km\,s$^{-1}$ at high velocity and 5, 15 to Max by 15 in units of $\sigma_{\rm LV2}=$ 4 mJy\,beam$^{-1}\,$km\,s$^{-1}$  at low velocity.
   \textit{(b)} Inset: contours are 5, 11 to Max by steps of 20 in units of $\sigma_{\rm HV2}$ at high velocity and 10 to Max by steps of 20 in units of $\sigma_{\rm LV2}$ at low velocity.
   \textit{(c)} Inset: The CO (2--1) blue-shifted line wing is integrated over 19-22 $\kms$ (blue contours) and 42-49 $\kms$ (cyan contours). Contours are 5 to Max by steps of 10 in units of $\sigma=$ 5 and 8 mJy\,beam$^{-1}\,$km\,s$^{-1}$, respectively.
   \textit{(a-c)} Cores driving outflows are represented by green ellipses showing their FWHM size, hatched for robust detections and empty otherwise, and numbered. Arrows indicate the direction of their outflows. Two blue lobes could be attributed to core\#7 in \textit{c}. Prestellar core candidates are represented by yellow ellipses, hatched for the massive core \#22.
   An ellipse representing the angular resolution of the CO cube is shown in the lower right.
   An additional zoomed-in figure highlighting outflows from cores \#56 and \#262 is provided in Appendix (\cref{fig:app-flow-MM2-zooms}a). }
   \label{fig:flow-MM2}
\end{figure*}

\begin{figure*}[!htp]
    \centering
    \includegraphics[width=0.75\hsize]{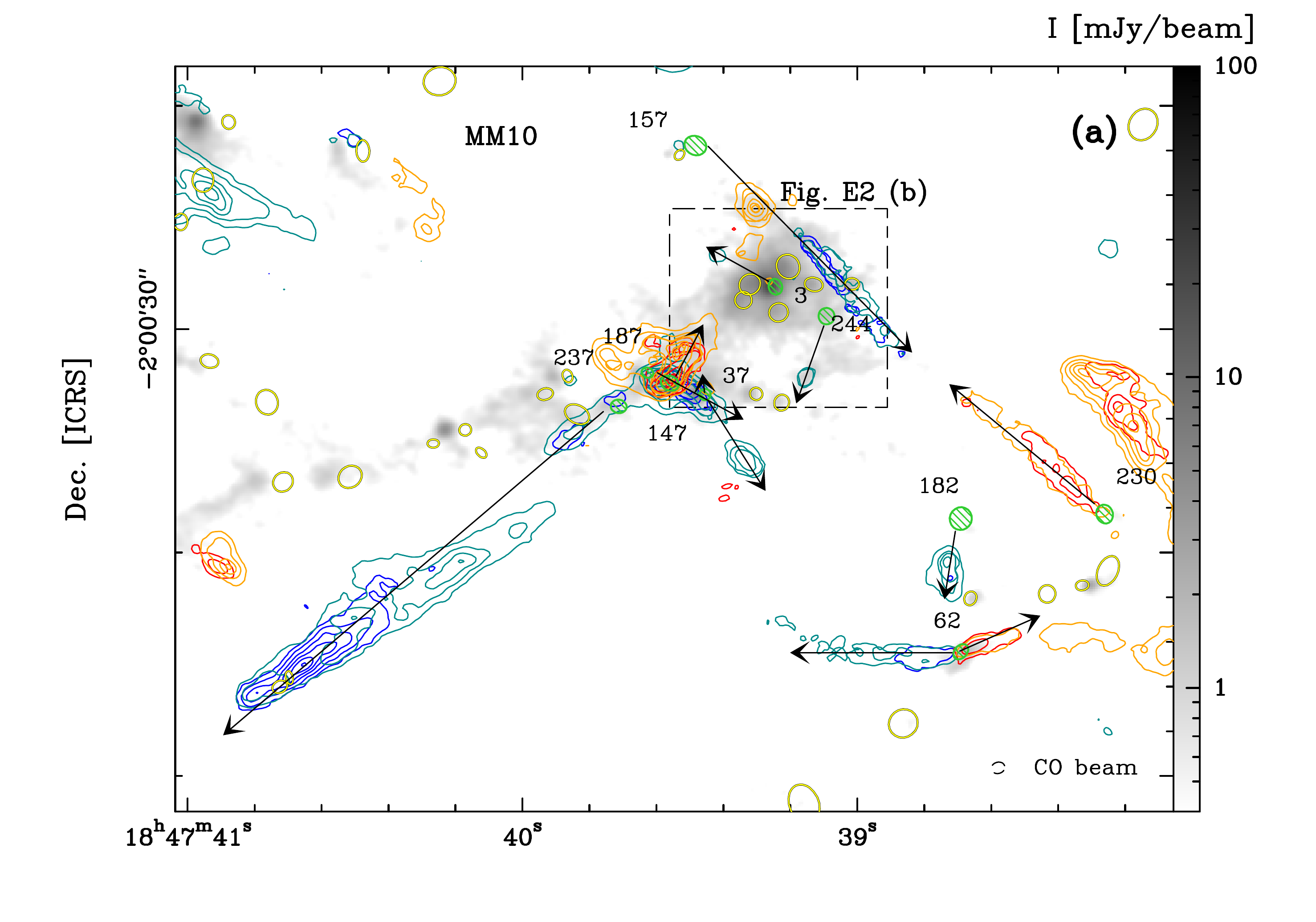}
    \subfloat{\includegraphics[width=0.53\hsize]{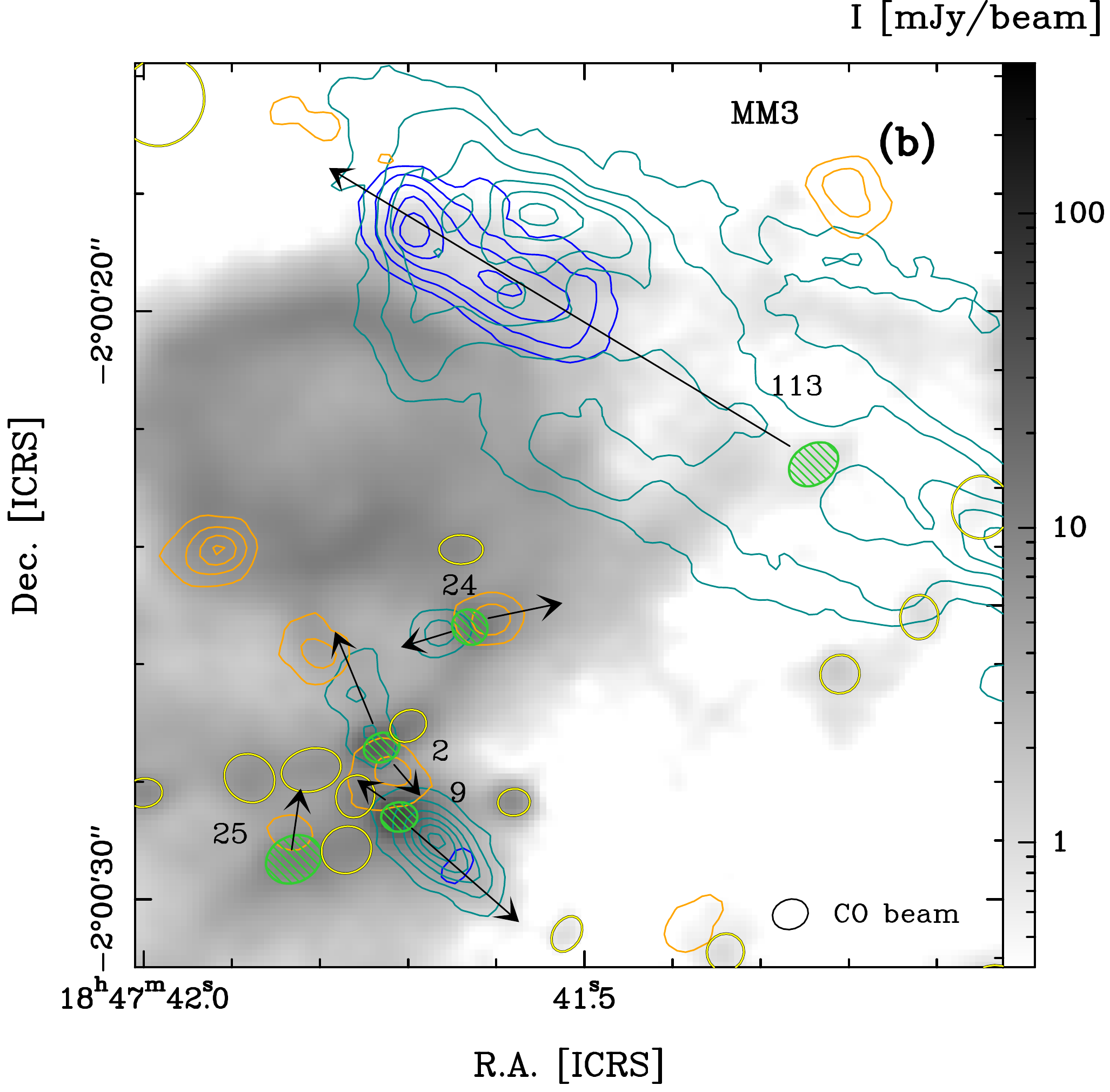}}
    \subfloat{\includegraphics[width=0.44\hsize]{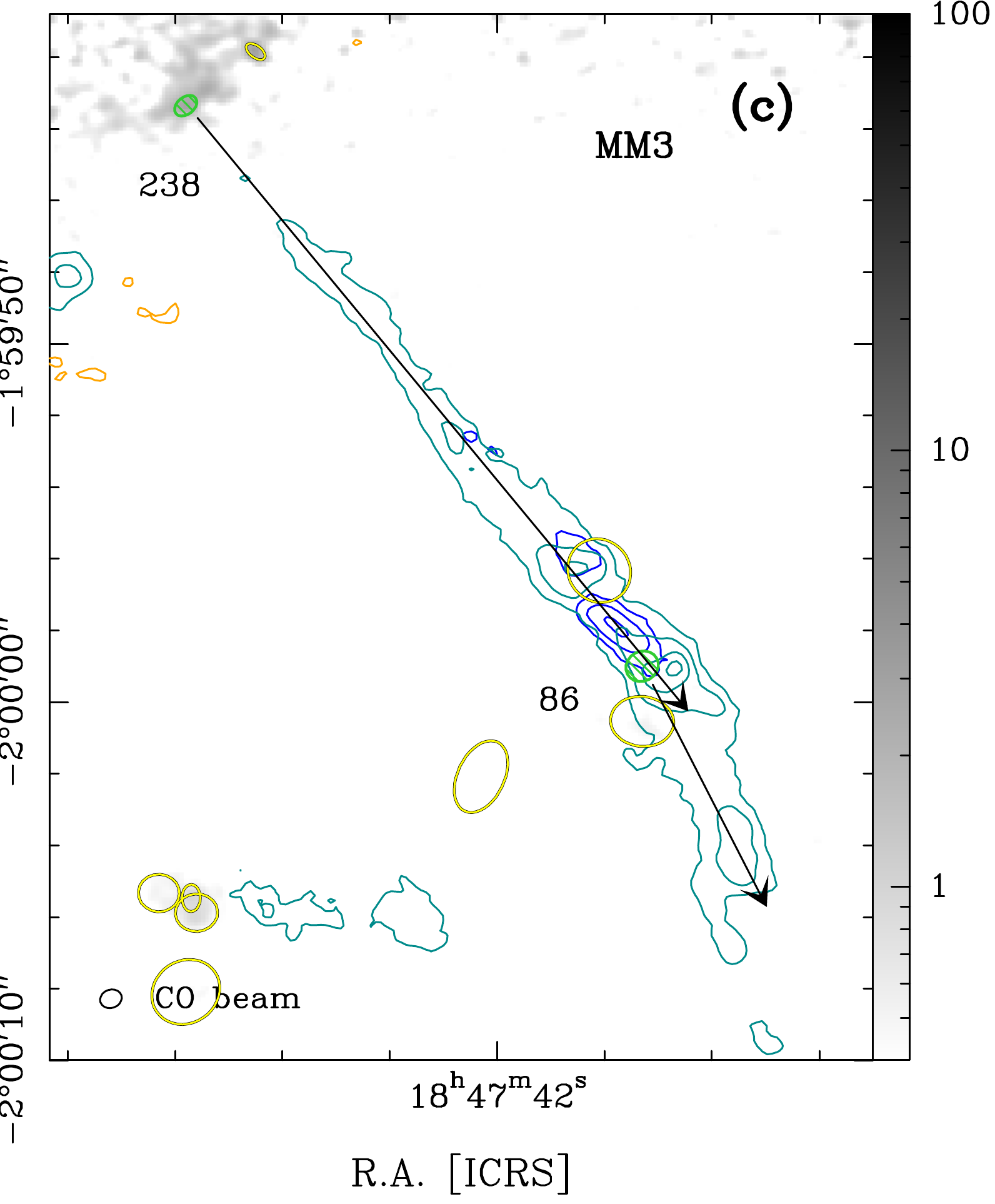}}
    \caption{Zoom from \cref{fig:flow-overview} towards W43-MM3, with the same colour code as in \cref{fig:flow-MM2} except for the high-velocity integration of CO (in red contours) which is 125.9-133.6 $\kms$ instead of 119.6-127.2 $\kms$.
    Contours are 4, 10 to Max by steps of 10 in units of $\sigma_{\rm HV3}=$ 13 mJy\,beam$^{-1}\,$km\,s$^{-1}$  at high velocity and 5, 15 to Max by steps of 15 in units of $\sigma_{\rm LV3}=$ 5 mJy\,beam$^{-1}\,$km\,s$^{-1}$  at low velocity.
    Same convention as in \cref{fig:flow-MM2} for lines and ellipses. An additional zoomed-in figure highlighting outflows from cores \#3 and \#244 is provided in Appendix (\cref{fig:app-flow-MM2-zooms}b).}
    \label{fig:flow-MM3}
\end{figure*}

\cref{fig:flow-overview} provides an overview of molecular outflows in W43-MM2\&MM3. The blue- and red- shifted lines of CO (2--1) were integrated over 53.6-61.2 $\kms$ and 125.9-133.6 $\kms$, respectively. These velocity ranges have been chosen to be centred on the average velocity at rest (V$_{\rm LSR}$) of W43-MM2 and W43-MM3, 92 $\kms$, and be free of contamination by large-scale foreground emission (present in 29-52 $\kms$ and 64-69 $\kms$ for both MM2 and MM3, 113-117 $\kms$ for MM2 and 70-73 $\kms$, 115-123 $\kms$ for MM3, see spectra shown in \cref{fig:spec}).
The identification of outflow lobes has been done through a detailed inspection of the CO cube. The minimum requirement for the detection of a lobe is to show emission above 5$\sigma$ in three consecutive channels.
We preferentially looked for the highest velocity component of outflows, which is the most collimated and the easiest to distinguish from the cloud emission. Lower velocity CO emission has been used also to detect a handful of outflows which do not appear at high velocity. 

The association between cores and outflows is highlighted with arrows in Figs.\,\ref{fig:flow-MM2} and \ref{fig:flow-MM3}. In these zoomed-in figures, both high and low velocity emission are shown in contours, with integrations over 53.6-61.2 $\kms$ and 119.6-127.2 $\kms$, 73.9-75.1 $\kms$ and 110.7-112.0 $\kms$, respectively. The high-velocity intervals correspond to -38 to -31 $\kms$ and +28 to +35 $\kms$ compared to an average V$_{\rm LSR}$ of 92 $\kms$, while the low-velocity intervals correspond to -18 to -17 $\kms$ and +19 to +20 $\kms$. The limits of these velocity intervals are shown on the CO spectra in \cref{fig:spec}. \cref{tab:nature-coreMM23} indicates whether a blue lobe, a red lobe, or both have been found. 
Over the 205 cores, 26 drive a single blue- or red-shifted lobe (20 and six cores, respectively) and 24 drive a bipolar outflow. A single core, \#5, is associated with two bipolar outflows, and two cores (\#7 and \#12) are tentatively associated with an additional monopolar lobe. In total, we detected 80 outflow lobes associated with 51 cores.
The association between cores and outflows is considered robust for 41 cores and tentative for 10 cores driving monopolar outflows.
Among those tentative associations, blue lobes of cores \#265 and \#86 could be driven by other cores in their vicinity (indicated as "conf" in \cref{tab:nature-coreMM23}, see \cref{fig:flow-MM2}a and \cref{fig:flow-MM3}c). In three other cases, for cores \#3, 25 and 97, the identification of CO emission as outflow itself is tentative (see \cref{fig:app-flow-MM2-zooms}b, \cref{fig:flow-MM3}b, \cref{fig:flow-MM2}a). 
Core \#3 is however considered as a robust protostellar core due to its hot core emission (Bonfand et al. in prep.). 
Therefore, the number of protostellar cores lies between 42 (robust outflows or hot core detections) and 51 (robust and tentative detections), which we express in the following as 46.5 $\pm$ 4.5 protostellar cores.
This represents a fraction $f_{\rm pro} = 23 \pm 2$\% of the 205 cores detected in W43-MM2\&MM3. 
When considering the complete core sample, the fraction of protostellar core is $f_{\rm pro} = 27 \pm 3$\% (31 $\pm$ 3 cores out of 113 cores above $0.8\,\Msol$). This fraction rises to 34 $\pm 4$\% for the population of cores above $1.6\,\Msol$.

\subsubsection{Protostellar cores in W43-MM1}
\label{su:flowMM1}

Using the CO (2--1) and SiO (5--4) lines, \cite{Nony20} found that 27 cores among the 131 identified by \cite{Motte18b} drive outflows (see our \cref{tab:nature-coreMM1} and Table 2 of \citealt{Nony20}). 
Of the 27 cores driving outflows, 19 were found to have robust outflow detection and eight were determined to have tentative detections.
The latter are flagged as such in \cref{tab:nature-coreMM1} and consist of outflows developing in confused environments, with several lobes overlapping, or outflows only detected at low velocity.
We verified that the new core extraction does not modify the core-outflow association of \cite{Nony20}, since the 27 cores with outflows are all detected in the new core catalogue. 
The comparison between the new core catalogue and the CO (2--1) cube only added two cores with tentative outflows, \#51 and \#174 (new), both associated with a single outflow lobe (see \cref{fig:app-flow-MM1-zoom}b). 

In W43-MM1, the numbers of cores with robust and tentative outflow detection are thus 19 and 10, respectively. In addition, four cores with tentative (\#4, \#10 and \#11) or absence of (\#5) outflow detection are confirmed as being protostellar due to the detection of significant hot core emission \citep[][]{Brouillet22}. Therefore, the number of protostellar cores lies between 23 (robust outflows or hot core detections) and 30 (robust and tentative detections), or equivalently there are 26.5 $\pm$ 3.5 protostellar cores. This represents a fraction $f_{\rm pro} = 21 \pm 3\%$ of the 127 cores detected in W43-MM1. 
The complete sample, including only cores with masses above 1.6$\,\Msol$, contains 25 $\pm$ 3 protostellar cores out of 69.
This represents a fraction of protostellar cores of $f_{\rm pro} = 36 \pm 4$\%, which is comparable within uncertainties to the value obtained in W43-MM2\&MM3 ($f_{\rm pro} = 34 \pm 4\%$ for cores above the same threshold of $1.6\,\Msol$, see \cref{su:flowMM23}). 

\subsection{Physical properties of cores in regions}
\label{su:compar-regions}  

\begin{figure*}
    \centering
    \includegraphics[width=0.85\hsize]{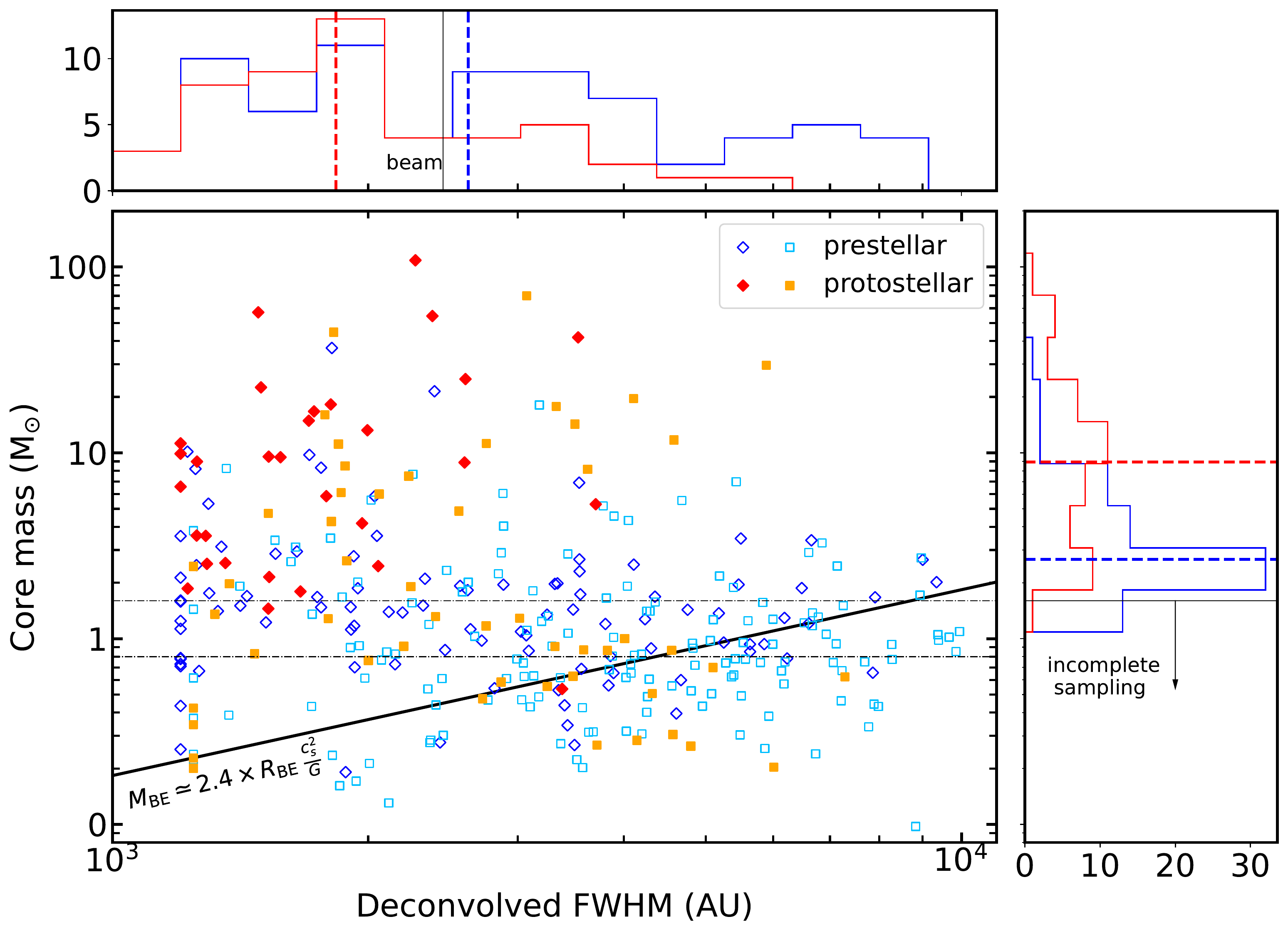}
    \caption{Mass vs size diagram for cores in W43, compared to the mass-size relation of critical Bonnor-Ebert spheres at $T_{\rm dust}=23$~K (black solid line). The core FWHMs are deconvolved from the beams, where the minimum deconvolved size corresponds to a half-beam angular diameter.
    In the main panel, the core sample is divided between prestellar core (represented by empty, blue and cyan symbols) and protostellar cores (filled, red and orange symbols), as well as between cores lying in W43-MM1 (diamonds) and in W43-MM2\&MM3 (squares). Completeness levels in W43-MM1 and W43-MM2\&MM3, 1.6$\,\Msol$ and 0.8$\,\Msol$ respectively, are represented by horizontal dashed lines.
    In the lateral panels are represented the marginal distributions of masses and FWHM sizes for the combined W43 region (W43-MM1 + W43-MM2\&MM3), excluding cores below the common 1.6$\,\Msol$ completeness limit. Distributions of prestellar and protostellar core parameters are displayed in blue and red, respectively. The dashed lines show their median values.}
    \label{fig:M-R}
\end{figure*}

Using outflows and hot cores emission, we identified 26.5 $\pm$ 3.5 protostellar cores out of 127 cores in W43-MM1 and 46.5 $\pm$ 4.5 out of 205 cores in W43-MM2\&MM3 (see \cref{su:flow}). Cores with neither outflows nor hot core, 100.5 $\pm$ 3.5 cores in W43-MM1 and 158.5 $\pm$ 4.5 in W43-MM2\&MM3, 
are good candidates to be in a preceding evolutionary stage, corresponding to prestellar cores.
In the following analysis, one source (\#132) is discarded from the sample of prestellar candidates because of its mass, size, and structure (see \cref{su:disc-hm-prestellar}). 
Figure~\ref{fig:M-R} displays the cores from both regions of W43 in a mass-to-size diagram.
Whereas prestellar cores from W43-MM1 and W43-MM2\&MM3 span a similar range of parameters, protostellar cores in W43-MM2\&MM3 are detected down to lower masses than those in W43-MM1 (median mass of 1.3$\,\Msol$ versus 8.9$\,\Msol$). 
This deficit of low-mass protostellar cores in W43-MM1 could be partly explained by a poorer completeness level (1.6$\,\Msol$ versus 0.8$\,\Msol$, see also discussion in \cref{su:disc-cpl-proto}). 
Indeed, when compared above a common completeness level of 1.6$\,\Msol$, protostellar core populations of the two regions have similar median masses of about 9$\,\Msol$ and cannot be statistically distinguished (Kolgomorov-Smirnov (KS) statistic of 0.14 with p-value of 0.94). Similar results are obtained between prestellar core populations (KS statistic of 0.16 with p-value 0.63). 

Figure~\ref{fig:M-R} also displays for comparison the mass-size relation of critical Bonnor-Ebert spheres at the median dust temperature of cores ($T_{\rm dust}=23$~K).
Following the analysis of the \cleanest core catalogue of ALMA-IMF \citep{Motte22}, we found that 109 cores from the 127 in W43-MM1 (86\%) and 119 cores from the 205 in W43-MM2\&MM3 (58\%) have $M>M_{\rm BE}$ and could be considered as gravitationally bound. 
These results are in line with the previous observation that cores in W43-MM1 have higher masses
compared to W43-MM2\&MM3. 
It has, however, little impact on our analysis, as all cores above the 1.6$\,\Msol$ completeness level in W43-MM1 and W43-MM2\&MM3 are bound, as well as 44 out of 56 cores in [0.8-1.6]$\,\Msol$ in W43-MM2\&MM3. Furthermore, this evaluation of cores gravitational boundedness should be considered with caution, as Bonnor-Ebert spheres are not the most appropriate models for the massive and dense cores in our observations. 
The gravitational boundedness of cores will be evaluated with greater precision in subsequent ALMA-IMF studies measuring turbulent support and evaluating the overall energy budget of cores.

\subsection{Mass distribution of prestellar and protostellar cores}
\label{su:cmf}

In \cref{su:compar-regions} we compared properties of cores from two regions of W43, W43-MM1 and W43-MM2\&MM3, and showed that, with regard to their mass, cores come from the same parent population.
In the following, we analyse together W43-MM1 and W43-MM2\&MM3 and compare the mass distributions
of the combined populations of protostellar and prestellar cores in W43, 73 $\pm$ 8 cores and 259 $\pm$ 8 cores, respectively. 
The populations of protostellar and prestellar cores in W43 above a completeness threshold of 1.6$\,\Msol$ are 45 $\pm$ 5 and 80 $\pm$ 5, respectively.

From the lateral panels of \cref{fig:M-R}, one observes that protostellar cores are significantly more massive and smaller in size than prestellar cores.
In detail, the median deconvolved sizes and masses of protostellar cores are 1800~au and 8.9$\,\Msol$, respectively, while those of prestellar cores are 2600~au and 2.7$\,\Msol$.  
As a consequence, the population of protostellar cores is about 10 times denser than that of prestellar cores, with median volume densities of 3.3 $\times$ 10$^{7}$ cm$^{-3}$ versus 4.7 $\times$ 10$^{6}$ cm$^{-3}$.
Our result that protostellar cores are more massive than prestellar cores is consistent with observations in other regions by \cite{Massi19} and \cite{Kong21}. However, both studies also find that protostellar cores have similar or slightly larger sizes than prestellar cores, in contradiction to what we observe in W43. This discrepancy could be due to differences in the definition taken by each algorithm to extract a source.

\begin{figure}
    \centering
    \includegraphics[width=\hsize]{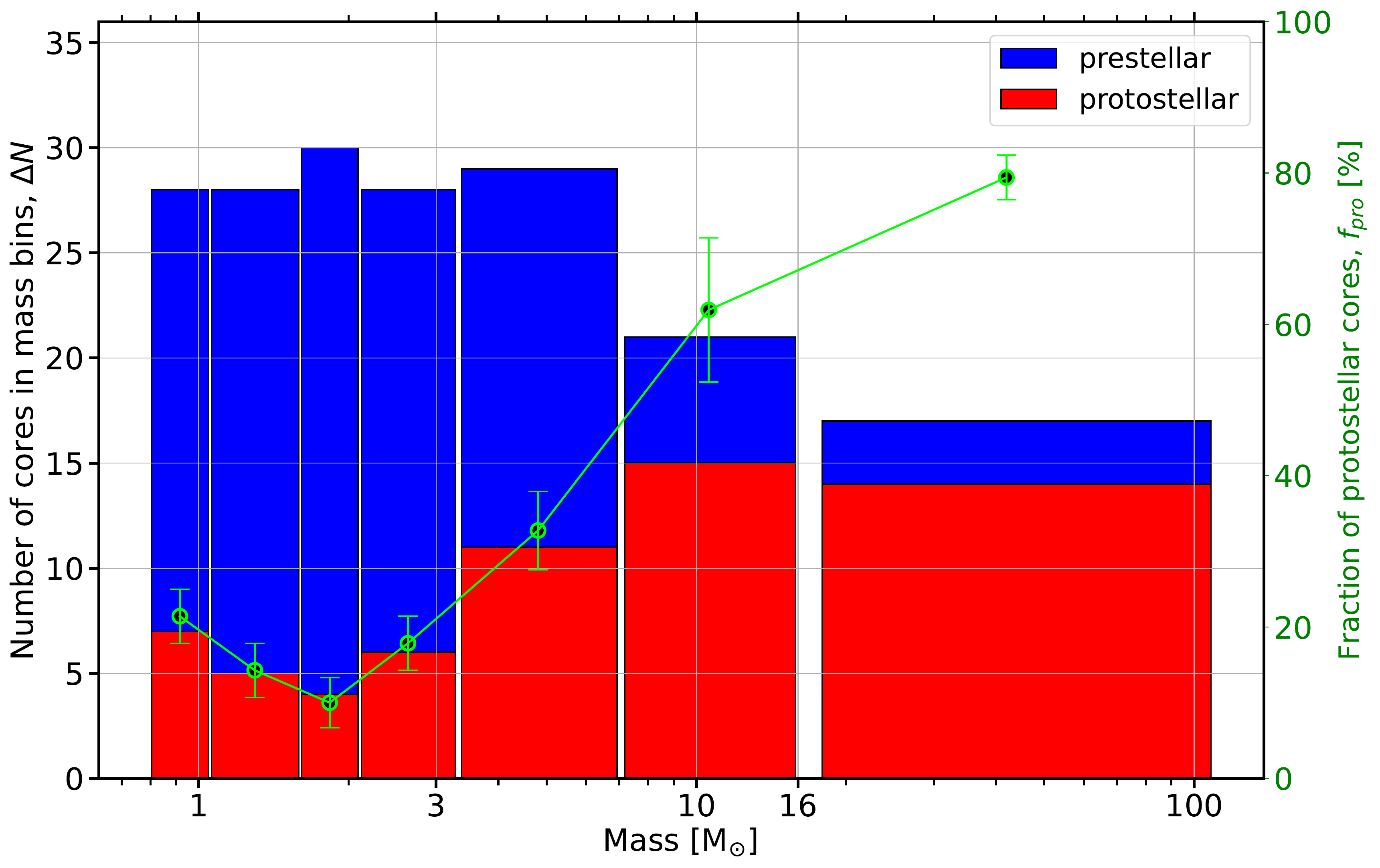}
    \caption{Mass histogram of prestellar and protostellar cores in W43. Cores are divided into seven mass bins, from 0.8 to 110$\,\Msol$. 
    Green circles indicate, in each bin, the percentage of protostellar cores among all, protostellar and prestellar, cores (see right axis).}
    \label{fig:histomass}
\end{figure}

In \cref{fig:histomass} we compare the numbers of prestellar and protostellar cores in seven mass intervals, ranging from 0.8 to 110$\,\Msol$. The bins have been constructed to preserve a roughly constant number (28 or 29) of cores in each of the five first bins and with two reference masses, 1.6 and 16$\,\Msol$. The former corresponds to the completeness level of W43-MM1, cores in the two first bins below 1.6$\,\Msol$ are exclusively located in W43-MM2\&MM3. The latter delimits the last bin and corresponds to the minimum mass of a core able to form high-mass stars, assuming a 50\% core-to-star mass efficiency.
In \cref{fig:histomass}, the fraction of cores which is protostellar, $f_{\rm pro}$ is also represented in each bin. Its uncertainty is derived from that on the count of protostellar cores (see \cref{su:flow}). When all protostellar cores in a bin are labelled as robust, a minimum of $\pm 2$ cores is used to compute the uncertainty. 
The fraction of protostellar cores stays at a roughly constant level of $f_{\rm pro} \simeq 15-20\%$ in the four first bins, between 0.8 and $\simeq 3 \,\Msol$. It increases in the subsequent intermediate and high-mass bins, from $f_{\rm pro} = 33 \pm 5 \%$ in [3.3-7] $\,\Msol$ to $f_{\rm pro} = 80 \pm 3 \%$ in [16-110] $\,\Msol$. 
The $f_{\rm pro} \simeq 15-20\%$ measured between 0.8 and 3 $\,\Msol$ is comparable to that observed in other regions \citep[e.g. 14$\%$ in a sample of Infrared Dark Clouds (IRDCs) of][]{Li20}, although the global fraction of protostellar cores, $f_{\rm pro} \simeq 35\%$, is higher.

\begin{figure}
    \centering
    \includegraphics[width=\hsize]{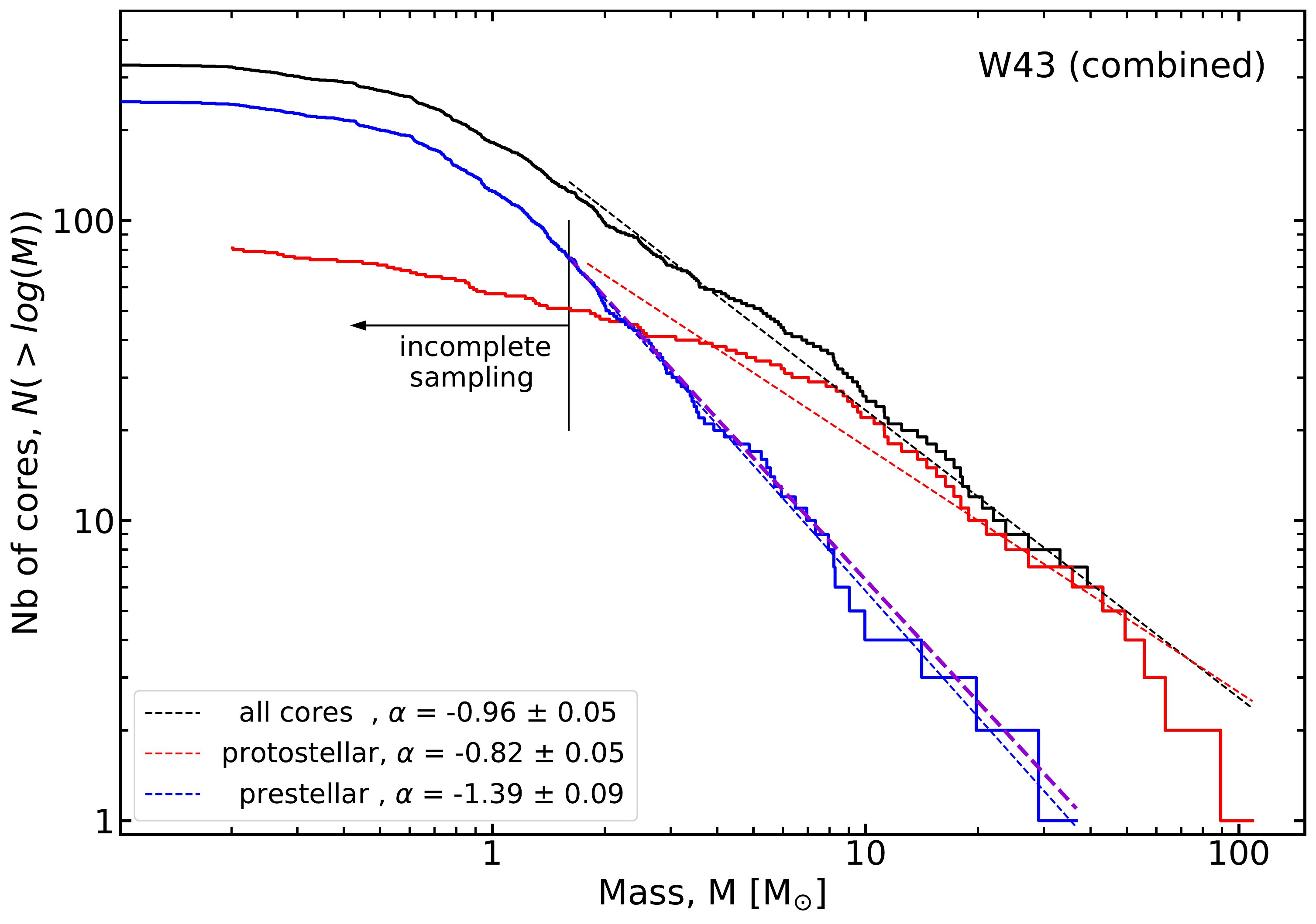}
    \caption{CMFs for the prestellar (in blue) and protostellar (in red) cores in W43. CMFs are represented in the form of a cumulative histogram and fitted with a least square method. The canonical Salpeter slope of the IMF (-1.35 in this form) is represented with a dashed purple line.}
    \label{fig:CMFs_W43_Chi2}
\end{figure}
 
\begin{figure*}

\begin{minipage}[t]{0.48\linewidth}
	\centering
    \includegraphics[width=1.05\linewidth]{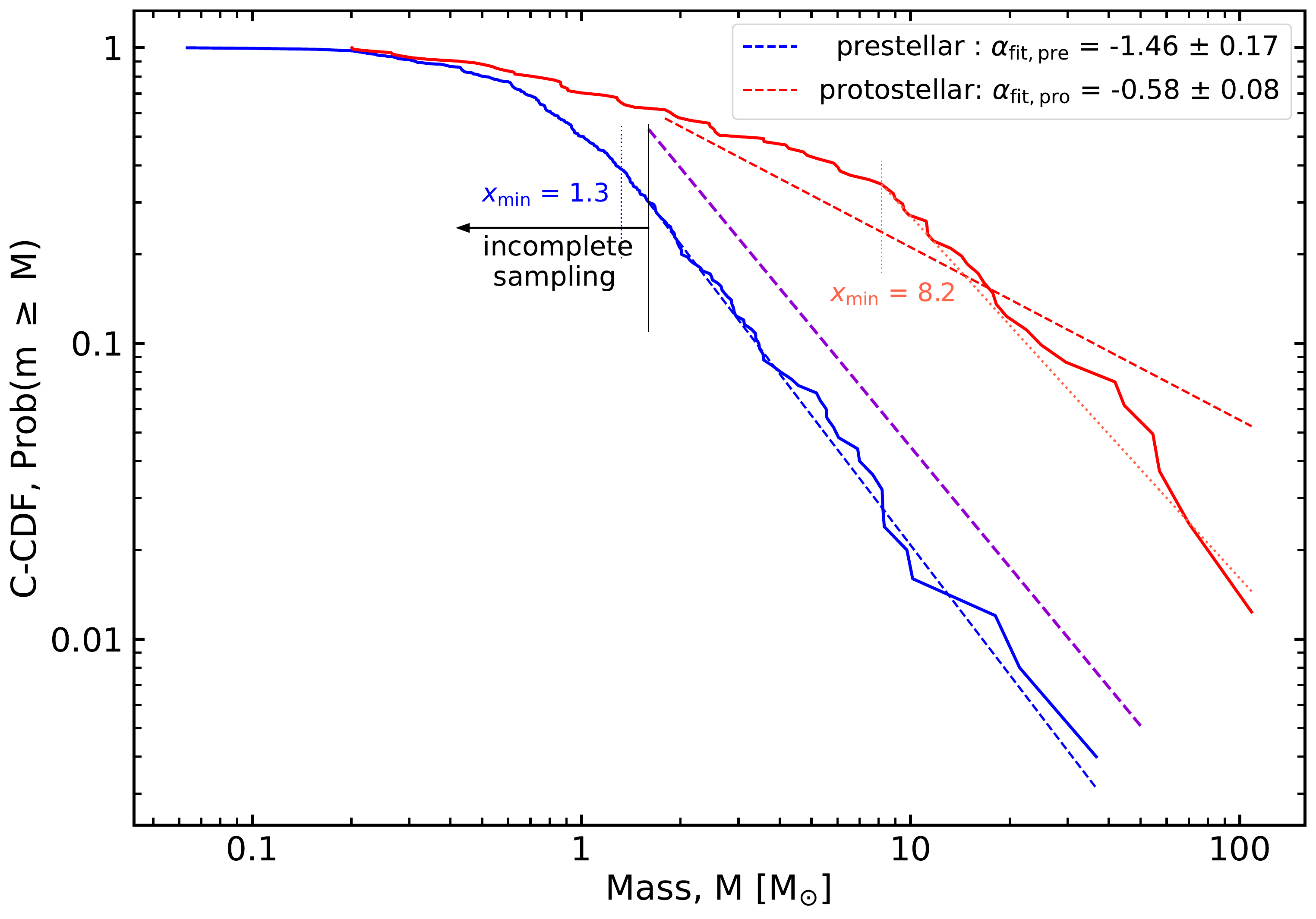}
    \caption{ CMFs for the prestellar (in blue) and protostellar (in red) cores in W43. CMFs are represented in the form of a complementary cumulative distribution function (C-CDF) fitted with the MLE method implemented in the \texttt{powerlaw} package \citep{Alstott14}.
    Dashed lines represent the fit made above the 1.6$\,\Msol$ completeness limit (black vertical line) with single power laws whose slopes are indicated. $x_{\rm min}$ values are indicated and the fit above $x_{\rm min}$ is shown in dotted line for the protostellar CMF (slope of $\alpha_{\rm fit,pro,2}=-1.22 \pm 0.23$). The canonical Salpeter slope of the IMF (-1.35 on this form) is represented with a dashed purple line.}
    \label{fig:CMFs_W43_MLE}
\end{minipage}\hfill
\begin{minipage}[t]{0.48\linewidth}
	\centering
    \includegraphics[width=\linewidth]{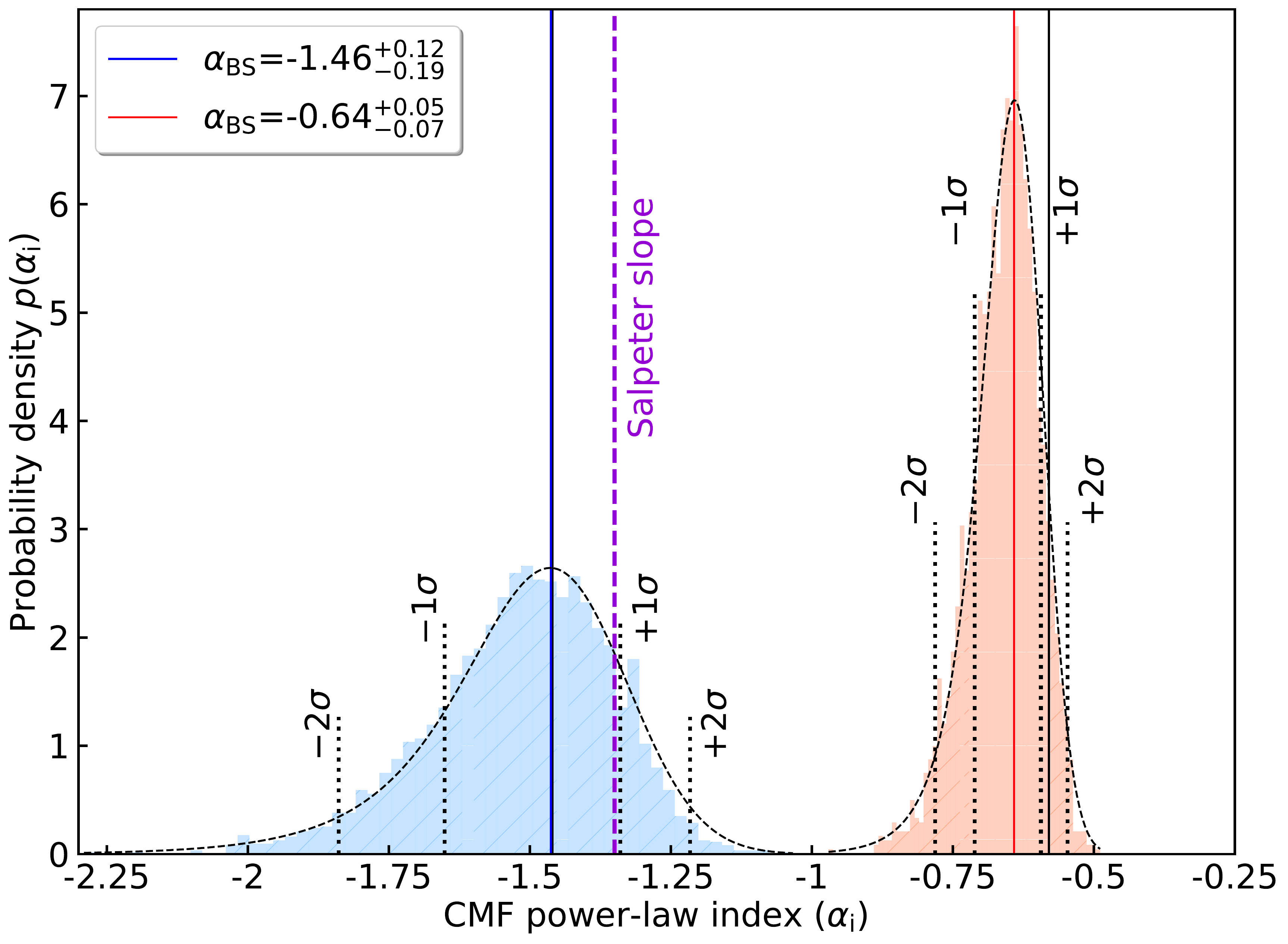}
    \caption{Determination of the index and uncertainty of the power-law fitted to the high-mass end of the cumulative CMF, measured for the populations of prestellar cores (in blue) and protostellar cores (in red).
    Each bootstrapping probability density function (coloured histograms) is built from 3000 slopes, $\alpha_{\rm i}$, fitted using the MLE method of the \texttt{powerlaw} package. 
    Power-law indices, $\alpha_{\rm BS}$, are taken to be the peak of the exponentially modified Gaussian (EMG) fits (blue and red solid lines) ; their asymmetric uncertainties are estimated from the -1$\sigma$ and +1$\sigma$ standard deviations of the EMG (dotted segments).
    The $\alpha_{\rm fit}$ values obtained by the MLE fit of the original data (see \cref{fig:CMFs_W43_MLE}) are shown by vertical black lines. The power-law index of canonical Salpeter IMF (-1.35, dashed magenta lines) is shown for comparison.}
    \label{fig:boot-KS}
\end{minipage}

\end{figure*}

The difference between prestellar and protostellar core populations also results in markedly distinct CMFs. Two cumulative representations of the CMF for the W43 core populations are shown in \cref{fig:CMFs_W43_Chi2,fig:CMFs_W43_MLE}. In addition, CMFs for the individual regions W43-MM1 and W43-MM2\&MM3 are shown in \cref{fig:app-CMFs}.
In agreement with the above analysis, the protostellar CMF extends up to much higher masses than the prestellar CMF, 109 versus 37$\,\Msol$, respectively. In \cref{fig:CMFs_W43_Chi2}, histograms comparing the counts of the prestellar, protostellar, and total core populations are represented and fitted with the polyfit method \citep[least square,][]{Sciutto89}. 
In \cref{fig:CMFs_W43_MLE}, normalised distributions (complementary cumulative distribution functions, C-CDFs) are represented. For the fit, we applied the Maximum Likelihood Estimator (MLE) method of \cite{Clauset09} implemented in the \texttt{powerlaw} python package \citep{Alstott14}. In this method, estimations of the power-law index $\alpha_{\rm fit}$ are performed for a given $x_{\rm min}$, the minimum value in the data to include in the fit, which can be provided by the user or set as a free parameter. 
When $x_{\rm min}$ is set at the completeness level, $x_{\rm min}=1.6\,\Msol$, the high-mass slopes of the prestellar and protostellar CMFs are $\alpha_{\rm fit,pre}=-1.46 \pm 0.17$ and $\alpha_{\rm fit,pro,1}=-0.58 \pm 0.08$, respectively.
When $x_{\rm min}$ is set as a free parameter, 
the couple of parameters ($x_{\rm min}$, $\alpha_{\rm fit}$) which minimises the KS distance between the data and the fit is provided.
For the prestellar CMF, the optimum ($x_{\rm min}$, $\alpha_{\rm fit}$) is ($1.3\,\Msol$, -1.43). The optimum $x_{\rm min}$ is slightly lower than the completeness and the associated slope is in very good agreement with the previous evaluation, $\alpha_{\rm fit,pre}=-1.46$.
In contrast, for the protostellar distribution, the optimum $x_{\rm min}$ is found to be 8.2$\,\Msol$, a mass much larger than the completeness level. The MLE slope on [8.2-109]$\,\Msol$, $\alpha_{\rm fit,pro,2}=-1.22 \pm 0.23$ is much steeper than the slope on [1.6-109]$\,\Msol$, $\alpha_{\rm fit,pro,1}=-0.58 \pm 0.08$, and therefore closer to the prestellar and Salpeter slopes.
This result indicates that the protostellar CMF of W43, in its cumulative form, is poorly represented by a single power-law from 1.6 to 109 $\,\Msol$.  
Interestingly, the protostellar CMF in its differential form is relatively flat, as suggested by the representation of \cref{fig:histomass}, which supports its top-heavy character. 

We also applied the bootstrap procedure, which provides a robust measurements of the most likely CMF power-law
index and its uncertainty. In detail, we built $N=3000$ synthetic set of mass generated from a random draw with discount of the measured core masses. 
The high-mass end of the $N$ associated CMF are then fitted using the MLE method of \cite{Alstott14}. The bootstrapping probability density function of the $N$ fitted slopes is shown in \cref{fig:boot-KS} for the prestellar and protostellar core populations. 
Two type of uncertainties are included in the procedure. The uncertainties associated with the estimation of the core mass are accounted by drawing according to a Gaussian law the mass of each core in a [$M_{\rm min}$ - $M_{\rm max}$] interval. The maximum and minimum masses of each core, $M_{\rm max}$ and $M_{\rm min}$ respectively, are computed from its measured flux, estimated temperature and dust opacity, plus or minus their associated $1\sigma$ uncertainties (see Sect. 4.2 of \citealt{Pouteau22a}).
The uncertainties associated with the sample incompleteness are accounted by allowing $x_{\rm min}$ to uniformly vary by $\pm 0.2\,\Msol$ from $1.6\,\Msol$, the 90\% completeness level.
The histograms of the bootstrapping probability density functions are fitted by exponentially modified Gaussians (EMG) with a negative skewness. Their peak $\alpha_{\rm BS}$ and the asymmetrical 1$\sigma$ uncertainties are also reported on \cref{fig:boot-KS}.
For the prestellar CMF, the bootstrap slope $\alpha_{\rm BS,pre}=-1.46_{-0.19}^{+0.12}$ is identical to the fitted slope and compatible at 1$\sigma$ with the Salpeter slope, -1.35. 
The bootstrap slope of the protostellar CMF, $\alpha_{\rm BS,pro}=-0.64_{-0.07}^{+0.09}$, is also consistent at 1$\sigma$ with the fitted slope $\alpha_{\rm fit,pro,1}=-0.58 \pm 0.08$. It deviates from the Salpeter and prestellar slopes by more than 3$\sigma$. 
The protostellar CMF fitted with a single power law above the completeness level is therefore significantly top-heavy. From the tests reported above, we nevertheless caution that the latter term should be understood as an over-abundance of cores at high masses rather than a statistical description of a power-law behaviour.

\section{Discussion}
\label{s:disc}

In \cref{su:disc-hm-prestellar}, we discuss the status of the high-mass prestellar core candidates in W43. In \cref{su:disc-cpl-proto}, we go through the possible observational biases in prestellar and protostellar core detection.  In \cref{su:disc-cores}, we discuss the implication of our results in the framework of star formation models.  

\subsection{High-mass prestellar core candidates}
\label{su:disc-hm-prestellar}

The scarcity of high-mass prestellar cores in various surveys of clumps suggests that their lifetime is, at most, very short \citep[see][]{motte18a}.
Prestellar cores on the verge of collapse, that is when their turbulent and/or magnetic supports have been dissipated, could live for as short as one free-fall time \citep{Galvan07,Bovino21}. 
The present study of W43 also reveals very few high-mass prestellar core candidates.
Out of the 18 cores more massive than 16$\,\Msol$, only four are not associated with any outflow or hot core signature (see \cref{su:cmf} and \cref{fig:histomass}).
Here we examine these four more closely. The first, MM1\#132 ($M=101\,\Msol$) is probably not a high-mass prestellar core. Its large size (FWHM$^{\rm dec}$=5000~au) and lower density (2.4 $\times 10^{7}$ cm$^{-3}$) compared to other cores of similar mass suggest that this source is not an individual, gravitationally bound cloud structure. Higher-resolution continuum images obtained using the longest baseline of the data (Brouillet, private com.) have confirmed this structure is not centrally concentrated. 
From its position at the immediate neighbourhood of the W43-MM1 centre (see \cref{fig:app-flow-MM1-zoom}), source MM1\#132 could be rather part of gas inflows. 

In contrast, core MM1\#6, with a mass $M=37 \pm 5 \,\Msol$ and a deconvolved size of FWHM$^{\rm dec}$=1800~au, was once considered to be an excellent high-mass prestellar core candidate.
Previously characterised\footnote{In \cite{Nony18}, it has a mass of $M=56 \pm 9 \,\Msol$ within a deconvolved size of FWHM$^{\rm dec}$=1300~au.} 
by \cite{Nony18}, it is located at the south-western tip of the main MM1 filament, in a region where outflow confusion is limited. 
\cite{Molet19}, however, concluded from a detailed chemical characterisation of core MM1\#6 that it could be at the beginning of the protostellar phase, casting some doubt on its true prestellar nature.  
In addition, two cores with lower masses are good prestellar candidates. Core MM1\#134 ($M=21\,\Msol$ and FWHM$^{\rm dec}$=2400~au) is located 
near the massive core MM1\#2 (see \cref{fig:app-flow-MM1-zoom}) and was not found earlier by \cite{Motte18b}, most likely because of the confusion between filaments and cores in this area. 
Core MM2\&MM3\#22 ($M=18\,\Msol$ and FWHM$^{\rm dec}$=3200~au) lies in the central part of the MM2 subregion (see \cref{fig:flow-MM2}).
In summary, we conclude that at most two to three massive cores in W43-MM1 and W43-MM2\&MM3 could be prestellar. This likely represents a statistic for the entire W43 molecular cloud complex, since the three studies regions - MM1, MM2 and MM3 - are its most massive fragments.

\subsection{Completeness of protostellar core detection}
\label{su:disc-cpl-proto}

Before discussing the physical interpretations of our results, we fist consider some possible biases 
and discuss the mass completeness of our outflow detections.
The comparison led in \cref{su:compar-regions} revealed a significant deficit of low-mass protostellar cores in W43-MM1 compared to W43-MM2\&MM3. In the [0.8-1.6]$\,\Msol$ interval, 11 $\pm$ 1 cores out of 56 (20 $\pm$ 2 \%) are protostellar in W43-MM2\&MM3 a fraction similar to that measured in the subsequent bins, between 1.6 and 3$\,\Msol$. In the same mass interval, a single
core out of 34 (3\%) has been characterised as protostellar in W43-MM1. This could result from a large number of misclassified protostellar cores, or from a large fraction of spurious low-mass sources which are currently accounted in the total prestellar population.
The first hypothesis is unlikely because the identification of outflows was carried out using the same methods in the two regions and with similarly complex CO lines (see below). 
The second hypothesis is more likely to play a role. The completeness of core extraction has been found to be worse in W43-MM1 than in W43-MM2\&MM3 (90\% level of 1.6$\,\Msol$ and 0.8$\,\Msol$, respectively), and contamination of the core sample by spurious sources is likely more problematic in the former region.
Indeed, as illustrated by the comparison in \cref{s:app-compar-cat}, low-mass core detection is affected by the quality of the interferometric map. 
The continuum emission is more centrally concentrated and brighter in W43-MM1 compared to W43-MM2\&MM3, which produces brighter sidelobes. It likely prevents for good detections of weakest sources and possibly produces more spurious low-mass prestellar core candidates.
Finally, the low number of low-mass protostellar cores in W43-MM1 compared to W43-MM2\&MM3 could also result from a different star-formation history (see \cref{su:disc-cores}). 

For the combined sample of cores in W43, we showed in \cref{su:cmf} that the fraction of protostellar cores increases with mass, from $f_{\rm pro} \simeq 15-20\%$ between 0.8 and 3 $\,\Msol$ to $f_{\rm pro} = 80 \pm 3 \%$ above 16$\,\Msol$. 
This result could be affected by the completeness of protostellar core detection, that is the limits of our ability to detect outflows from low-mass cores.
Outflow momentum and mass are known to be well correlated with the mass of the driving core \citep[e.g.][]{Bontemps96,Beuther02,Maud15}. 
To test whether the sensitivity limit of CO observations might prevent robust detection of the fainter outflows, we defined and measured a detection intensity for the outflow lobes associated with the lowest mass cores (see \cref{s:app-measure-outfl}). We found it to be significantly above 5$\sigma$ and conclude that sensitivity is not limiting the detection of low-mass outflows in W43.
Outflows detection is more likely limited by the complexity of CO observations. First, we are relatively insensitive to outflows emitting only at low velocity, where the CO spectra suffer from strong self-absorption or contamination (about 10$\,\kms$ from the central velocity, see \cref{fig:spec}). This effect, however, is not expected to affect preferentially low-mass outflows.
Then, \cite{Nony20} pointed out environmental effects as a possible bias for outflow detection. Outflows in the central part of the protocluster, with high gas densities and complex dynamics, are more difficult to identify and attribute because they are smaller and more prone to confusion with other lobes. 
In regions such as W43-MM1 and W43-MM2 with strong mass segregation 

\noindent \citep{Dib19,Pouteau22b}, this effect however affects preferentially high-mass cores, clustered in the central and densest areas.
We conclude that, although the number of detected protostellar cores is formally a lower limit, our outflow-based method does not introduce a significant bias as a function of core mass.

\subsection{CMF and core mass growth}
\label{su:disc-cores}

Most of the CMFs studies focusing on prestellar cores have been conducted in nearby ($d <$ 700 pc) regions. These have found good agreements between the CMFs high-mass ends and the Salpeter slope \citep[e.g.][]{Motte98,Polychroni13,Konyves15,Konyves20,Massi19}. Among the few studies that compared the prestellar and protostellar populations, \cite{Enoch08} used a combined sample of cores in three clouds of the Gould Belt and found a prestellar CMF with a high-mass end slope consistent with Salpeter ($\alpha=-1.3 \pm 0.4$) and a protostellar CMF that is flatter  ($\alpha= -0.8$ with constant $T=15$ K) and extends to higher masses. In Vela C, \cite{Massi19} found a starless CMF mostly consistent with the canonical IMF with some indications that the protostellar CMF could be flatter.
As for studies at larger distances with ALMA, few have separated protostellar from prestellar cores. In a combined survey of 12 IRDCs, \cite{Sanhueza19} found a prestellar CMF slope with a high-mass end slightly flatter than Salpeter (-1.17 $\pm$ 0.1). 


In this work, we study prestellar and protostellar core populations in a single, distant molecular complex with strong high-mass star formation activity. In \cref{su:cmf}, we showed that the prestellar CMF in W43 is significantly different from the protostellar CMF. The former does not cover the complete mass range of stars expected to form in W43 and its high-mass end is compatible with the canonical Salpeter slope. In contrast, the protostellar CMF is significantly top-heavy, up to core masses of 100$\,\Msol$. 
We also showed that the protostellar core population has a higher median mass compared to the prestellar core population (median mass of 8.9$\,\Msol$ and 2.7$\,\Msol$, resp.), and that the fraction of cores with protostellar activity increases with mass.

In the following, we first assume that the statistics performed in space reconcile with temporal statistics (i.e. the hypothesis of ergodicity). In that regard, the measured prestellar core population is assumed to be representative of the "parent" prestellar core population, which evolved into the measured protostellar core population. 
If the total mass reservoir available to form a star is mainly determined by the mass of its prestellar core, as assumed in "core fed" models, one would expect the mass of a  protostellar core to be lower than that of the prestellar core at the time of protostar formation, ($M_{\rm pre,end}$). 
Denoting $\dot{M}_{\rm out}$ as the mass loss rate of the gas core, through outflows or accretion onto the protostar, the mass of an isolated protostellar core would indeed evolve as $M_{\rm pro}(t)= M_{\rm pre,end} - \dot{M}_{\rm out}\,t$, with $t=0$ corresponding to the beginning of the first collapse. 
The existence of protostellar cores more massive than prestellar cores thus implies that cores are able to grow in mass during the protostellar phase through inflow and accretion from their environment, which can be included in the previous equation with an additional term $\dot{M}_{\rm core}$: 
\begin{equation}
    \label{eq:Mpro}
    M_{\rm pro}(t)= M_{\rm pre,end} + (\dot{M}_{\rm core} - \dot{M}_{\rm out})\,t.
\end{equation}
To observe a core at a given time $t_1$ such that $M_{\rm pro}(t_1) > M_{\rm pre,end}$, the core mass growth must be on average larger than the core mass loss ($\dot{M}_{\rm core} > \dot{M}_{\rm out}$) during the core lifetime. 
Continuous core mass growth during the protostellar phase has also been proposed by \cite{Kong21} to account for their observations, as well as in the analytical core evolution models developed by \cite{Hatchell08}
to explain the lack of high-mass prestellar cores in Perseus.
Such processes align with clump-fed models \citep[see e.g. ][]{Wang10,smith09,Vazquez19}.

We note that \cref{eq:Mpro} can easily be generalised to describe the mass evolution of prestellar cores, introducing $M_{\rm pre,ini}$ as the initial mass and assuming that the core mass growth processes represented in $\dot{M}_{\rm core}$ are similar in the prestellar and protostellar phases:
\begin{equation}
    \label{eq:Mpre}
    M_{\rm pre}(t)= M_{\rm pre,ini} + \dot{M}_{\rm core}\,t.
\end{equation}

In addition, the flattening of slope between the prestellar and protostellar CMFs implies that $\dot{M}_{\rm core}$ depends on the core mass, and more precisely that high-mass cores are able to grow in mass more strongly 
than low-mass cores. 
This is in line with observations showing a correlation between mass and infall rate 
\citep[see e.g. on clumps scale,][]{Yue21}.  

Beyond the discussion of possible mechanisms explaining the mass evolution of a single core from the prestellar to the protostellar stage, our results have also implications for the time evolution of core populations. 
From their study of the spatial variations of star formation in W43-MM2\&MM3, \cite{Pouteau22b} proposed that subregions undergoing a burst of star-formation, such as the centre of W43-MM2, could be associated with top-heavy CMFs, while others more quiescent would show Salpeter-like distributions. 
Following this approach, our results suggest that the population of protostellar cores which shows a top-heavy CMF is growing in mass during a burst of star formation. After this burst, the star-formation activity is expected to decrease and enter in a more quiescent state. In this context, the measured prestellar core population could evolve to the protostellar stage preserving a Salpeter-like CMF.
Similarly, the measured protostellar population could have inherited its top-heavy mass distribution from an alike prestellar CMF, that is, the parent prestellar core population could have been top-heavy compared to the current one. This would however contravene the hypothesis of ergodicity, as the observed prestellar core population would no longer be representative of the parent prestellar core population. 


Our results have implications for the issue of the origin of the stellar IMF.
After having constituted the main paradigm in star formation for decades , the universality of the IMF is now under serious debate, both from Galactic and extra-galactic studies. In the Milky Way, various claims of top-heavy IMFs have emerged in young massive clusters in the vicinity of the Galactic centre \citep{Hussmann12,Lu13,Hosek19}.
At the same time, several ALMA observations of top-heavy CMFs have been reported in distant, high-mass star-forming regions \citep[see, e.g.][]{Motte18b,Kong19,Pouteau22a}. 
Therefore, although simultaneous observations of CMF and IMF in the same region are challenging (see however \citealt{Takemura21}), we could reasonably expect that in some regions neither the CMF nor the IMF have a Salpeter slope. 
Interestingly, recent analytical models for the IMF obtain high-mass end slopes flatter than -1.35, such as -1 \citep{Ballesteros15} or -0.75 \citep{Lee18}, depending on the equation of state. 
Predicting the IMF shape from a CMF is a complicated question that involves various parameters such as the core-to-star mass efficiency, fragmentation, and time evolution. Some of these effects have been tested by \cite{Pouteau22a}, that suggested that the IMF resulting from the measured CMF in W43-MM2\&MM3 could remain top-heavy. 
Our results also support that the current protostellar core population will produce stars with a top-heavy IMF.
As discussed above, however, future populations of protostellar cores could evolve with a Salpeter-like CMF and produce stars with a canonical IMF. Therefore, the shape of the final IMF of W43, built from the accumulation of various episodes of star formation over time, will depend on the complex interplay between various processes in time and space. 


\section{Conclusion}
\label{s:conclu}

The study presented in this article is carried out in the context of the ALMA-IMF Large Program \citep{Motte22} and is motivated by previous observations of top-heavy CMFs \citep{Motte18b,Pouteau22a}.
We used ALMA observations of the CO (2--1) line to identify outflows and compare the properties of the prestellar and protostellar core populations in two regions of the W43 molecular complex, W43-MM1 and W43-MM2\&MM3. Our main results can be summarised as follows:

\begin{enumerate}
    \item We constructed a new continuum map of W43-MM1 using the pipeline developed by the ALMA-IMF consortium \citep{Ginsburg22} and obtained a significant reduction in the interferomotric sidelobes compared to that of \cite{Motte18b} (see \cref{s:app-compar-cat}). The new core catalogue of W43-MM1 consists of 127 cores with masses ranging from 0.2 to 109\,$\Msol$ (see \cref{sub:contMM1}). The core catalogue of W43-MM2\&MM3, presented by \cite{Pouteau22a}, is constituted of 205 cores with mass ranging from 0.1 to 70\,$\Msol$ (see \cref{sub:contMM23}).
    
    \item We found 51 cores in W43-MM2\&MM3 associated with outflows in CO (2--1), including 41 robust and 10 tentative detections. After re-examination of the work of \cite{Nony20}, we found 29 cores in W43-MM1 driving outflows, including 19 robust and 10 tentative detections. 
    
    \item We used the presence or absence of outflows and hot core signature as indicators of the protostellar and prestellar nature of cores. Once filtered with a common mass threshold of 1.6\,$\Msol$, 
    the two regions have similar fractions of protostellar cores among the total core populations, $f_{\rm pro} = 34 \pm 4$\% in W43-MM2\&MM3 and $f_{\rm pro} = 36 \pm 4$\% in W43-MM1. The core populations present similar properties regarding their mass between the two regions, although fewer low-mass protostellar cores are detected in W43-MM1
    
    \item The core sample analysed in W43 consists of 45 $\pm$ 5 protostellar cores and 80 $\pm$ 5 prestellar cores, with masses above 1.6\,$\Msol$ and lying in W43-MM2\&MM3 or W43-MM1. In addition, 11 $\pm$ 2 protostellar cores and 45 $\pm$ 2 prestellar cores with masses between 0.8 and 1.6\,$\Msol$ within W43-MM2\&MM3 are also considered. We found that the fraction of prostostellar cores is roughly constant at $f_{\rm pro} \simeq 15-20 $\% for low-mass cores, between 0.8 and 3\,$\Msol$. This fraction increases for intermediate- and high-mass cores, up to $f_{\rm pro} \simeq 80 $\% above 16\,$\Msol$. We also report the detection of two new high-mass prestellar core candidates with masses of 21$\,\Msol$ and 18$\,\Msol$, core \#134 in W43-MM1 and core \#22 in W43-MM2, respectively. 
    
    \item We measured significant differences between the prestellar and protostellar CMFs in W43. The high-mass end of the prestellar CMF is consistent with the Salpeter slope ($\alpha=-1.46_{-0.19}^{+0.12}$ 
    in the range  [1.6-37]\,$\Msol$), which implies that protostellar cores are the main contributors to the top-heavy form of the global CMF of these regions.
    The CMF of protostellar cores is more irregular, and its high-mass end can be approximately represented by a power-law of index $\alpha \simeq -0.6$.
    
    \item These results could be explained by clump-fed models, in which cores grow in mass, especially during the protostellar phase, through inflow from their environment.
    Moreover, the difference between the prestellar and protostellar CMF slopes implies that high-mass cores grow more in mass that low-mass cores. It also suggests that the resulting IMF in W43 will remain top-heavy, as initially suggested by \cite{Pouteau22a}. 

\end{enumerate}

Our results also call for more investigations to establish if the observed difference between prestellar and protostellar CMFs is a general property of high-mass star-forming regions, or if it is only observed in  particular environments such as W43. 
This will be a goal of future studies using the ALMA-IMF data sets. 

\begin{acknowledgements}
RGM and TN acknowledge support from UNAM-PAPIIT project IN108822 and from CONACyT Ciencia de Frontera project ID 86372. TN also acknowledges support from the postdoctoral fellowship program of the UNAM. Part of this work was performed using the high-performance computers at IRyA, Mexico, funded by CONACyT and UNAM. The work from the IT staff at this institute is acknowledged. TN is grateful for helpful discussions with Enrique V{\'a}zquez-Semadeni.
This project has received funding from the European Research Council (ERC) via the ERC Synergy Grant ECOGAL (grant 855130), from the French Agence Nationale de la Recherche (ANR) through the project COSMHIC (ANR-20-CE31-0009), and the French Programme National de Physique Stellaire and Physique et Chimie du Milieu Interstellaire (PNPS and PCMI) of CNRS/INSU (with INC/INP/IN2P3). The project leading to this publication has received support from 
ORP, that is funded by the European Union's Horizon 2020 research and innovation programme under grant agreement No 101004719 [ORP].
YP, BL and GB acknowledge funding from the European Research Council (ERC) under the European Union’s Horizon 2020 research and innovation programme, for the Project “The Dawn of Organic Chemistry” (DOC), grant agreement No 741002.
AS gratefully acknowledges support by the Fondecyt Regular (project code 1220610), and ANID BASAL projects ACE210002 and FB210003.
FL acknowledges the support of the Marie Curie Action of the European Union (project \textsl{MagiKStar}, Grant agreement number 841276).
S.B. acknowledges support by the French Agence Nationale de la Recherche (ANR) through the project \textit{GENESIS} (ANR-16-CE92-0035-01).
AG acknowledges support from the National Science Foundation under grant No. 2008101.
PS was supported by a Grant-in-Aid for Scientific Research (KAKENHI Number 18H01259 and 22H01271) of the Japan Society for the Promotion of Science. P.S. gratefully acknowledge the support from the NAOJ Visiting Fellow Program to visit the National Astronomical Observatory of Japan in 2019, February.
M.B. has received financial support from the French State in the framework of the IdEx Universit\'e de Bordeaux Investments for the future Program. 
TB acknowledges the support from S. N. Bose National Centre for Basic Sciences under the Department of Science and Technology, Govt. of India. 
GB also acknowledges funding from the State Agency for Research (AEI) of the Spanish MCIU through the AYA2017-84390-C2-2-R grant.
LB gratefully acknowledges support by the ANID BASAL projects ACE210002 and FB210003

This paper makes use of the ALMA data ADS/JAO.ALMA\#2017.1.01355.L. ALMA is a partnership of ESO (representing its member states), NSF (USA) and NINS (Japan), together with NRC (Canada), MOST and ASIAA (Taiwan), and KASI (Republic of Korea), in cooperation with the Republic of Chile. The Joint ALMA Observatory is operated by ESO, AUI/NRAO and NAOJ.
\end{acknowledgements}

\bibliographystyle{aa}
\bibliography{biblio-article}	

\begin{appendix}

\renewcommand{\thefigure}{A\arabic{figure}}
\renewcommand{\thetable}{A\arabic{table}}

\section{Continuum selection for the bsens$^*$ 1.3\,mm continuum map of W43-MM1}
\label{s:app-freq-MM1}

The $\bsens^*$ continuum image of W43-MM1 presented in this work was constructed with the pipeline developed for ALMA-IMF using subsections of the observed bandpass (see \cref{sub:contMM1} and \citealt{Ginsburg22}). Frequency selection was done visually to exclude the brightest lines (such as CO isotopolgues, SiO or SO) while maximising the total bandwidth. This method is an alternative to the solution presented by \cite{Ginsburg22}, which excludes the whole CO spectral window ("bsens-nobright", see their Appendix G). The frequencies included in the $\bsens^*$ continuum are provided below in GHz, LSRK frame. They add up to 2.8 GHz, compared to 3.8 GHz for the full bandwidth used for the \bsens continuum and 1.7 GHz for the \cleanest continuum selection. \\

SPW0:
216.016--216.032 ; 
216.052--216.070 ; 
216.090--216.130 ; 
216.155--216.245  

SPW1:
217.092--217.110 ; 
217.132--217.150 ; 
217.178--217.190  

SPW2:
219.813--219.830 ; 
219.858--219.867 ; 
219.889--219.906  

SPW3:
218.035--218.050 ; 
218.070--218.100 ; 
218.160--218.200 ; 
218.215--218.220 ; 
218.233--218.245  

SPW4:
219.445--219.466 ; 
219.500--219.540  

SPW5:
230.310--230.385 ; 
230.535--230.576 ; 
230.605--230.650  

SPW6:
231.000--231.130 ; 
231.176--231.187 ; 
231.215--231.228 ; 
231.260--231.365  

SPW7:
232.540--232.640 ; 
232.740--232.835 ; 
233.010--233.110 ; 
233.166--233.350 ; 
233.590--233.640 ; 
233.730--233.765 ; 
233.830--233.927 ; 
233.985--234.025 ; 
234.070--234.105 ; 
234.120--234.170 ; 
234.263--234.337

\renewcommand{\thefigure}{B\arabic{figure}}
\renewcommand{\thetable}{B\arabic{table}}

\section{Comparison between different core catalogues for W43-MM1}
\label{s:app-compar-cat}

\begin{figure*}
    \centering
    \includegraphics[width=0.92\hsize]{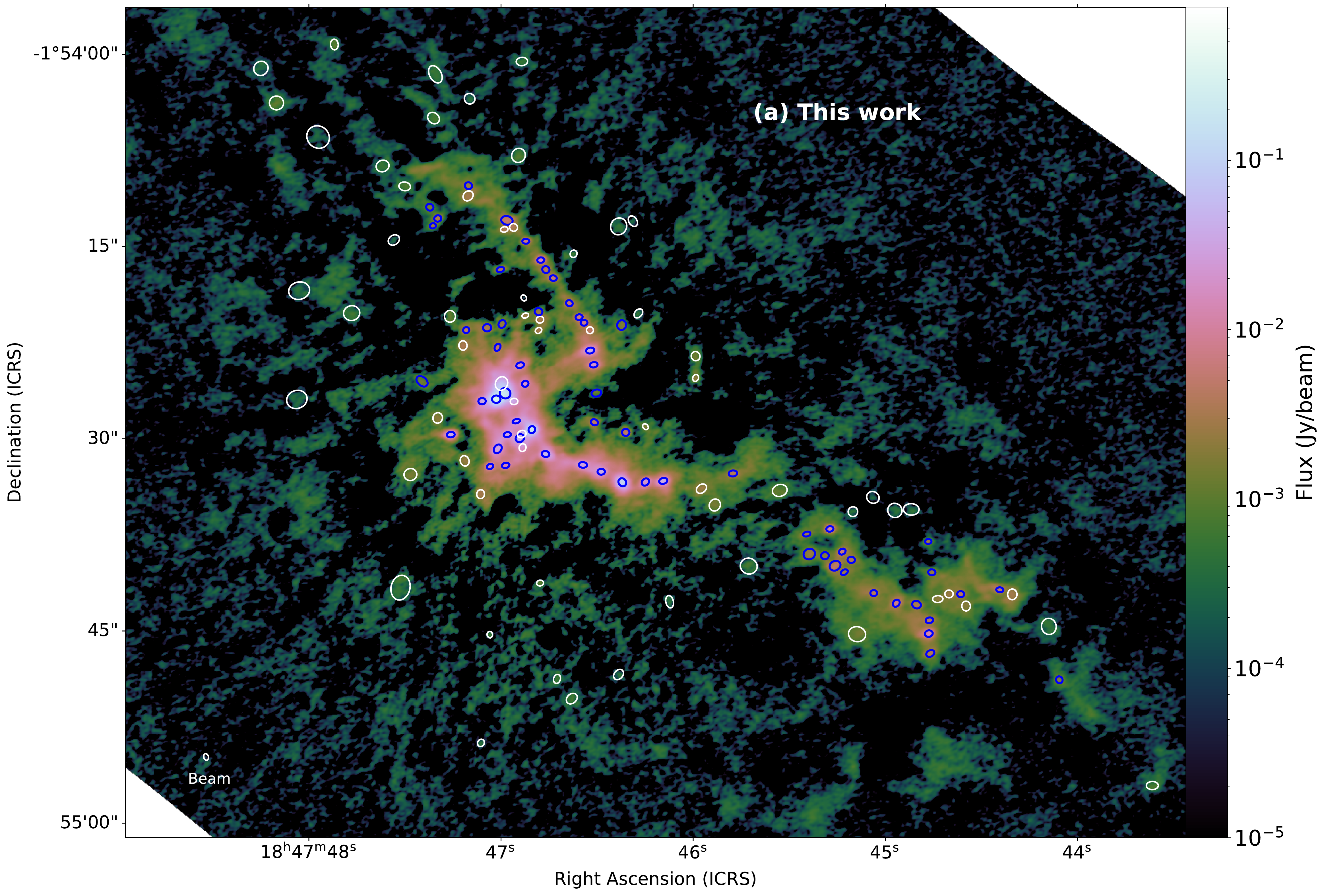}
    \includegraphics[width=0.92\hsize]{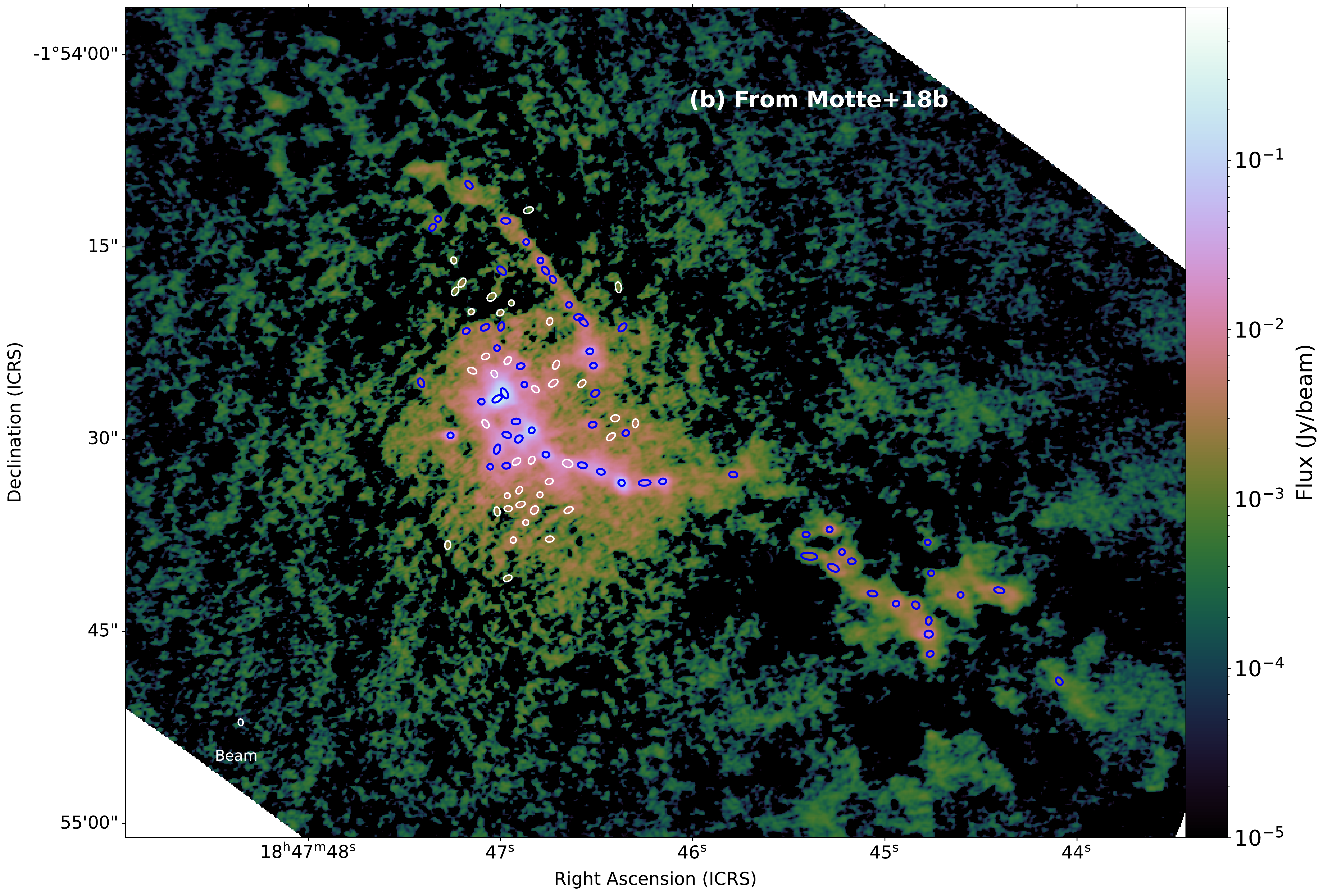}
    \caption{Continuum image of W43-MM1 obtained using the ALMA-IMF pipeline (top) compared with that shown in \cite{Motte18b} (bottom). 
    Cores common to the two catalogues are represented as blue ellipses, cores unique in each catalogue as white ellipses.
    Images are drawn with the same logarithmic colour scale and have almost equal beam, shown in the bottom left corner. A reduction of the noise level around the central part is visible in the top image, which should be attributed to the improvement in our data reduction methods. 
    }
    \label{fig:app-cont-MM1}
\end{figure*}

W43-MM1 Cycle 2 data were first reduced and analysed by \cite{Motte18b}, then reprocessed as part of the ALMA-IMF Large Program \citep[see also Appendix F of][]{Ginsburg22}. As can be seen from the comparison shown in \cref{fig:app-cont-MM1}, interferometric sidelobes are significantly reduced in the new continuum image, leading to a noise level about 30\% lower in the new image. 
In addition, the core-extraction method has been also improved. While \cite{Motte18b} applied the extraction software \textit{getsources} \citep{Mensh12} directly to the continuum map, for this work we apply \textit{getsf} \citep{Mensh21} to a continuum image that has first undergone a "denoising" procedure (see \cref{su:cont} and \citealt{Pouteau22a}). 

In the following, we compare the catalogue of \cite{Motte18b} (hereafter cat-o, 131 cores) with that presented here and resulting from the new data processing (hereafter cat-n, 127 cores). In total, 65 cores are common to the two catalogues, which represent half of their total contents. In detail, 58 cores out of the 94 cores considered to be more robust in cat-o have a counterpart in cat-n (62\%), but only 7 cores out of the 37 less robust cores (19\%). Cores from the two catalogues are shown on their associated continuum maps in \cref{fig:app-cont-MM1}. The most significant differences between the catalogues concern cores that are located in the outskirts of the central region, which can be attributed to differences in the quality of the maps. Interestingly, large cores are found in cat-n at larger distances from the filaments, compared to cat-o. 

\begin{figure}
    \centering
    \includegraphics[width=\hsize]{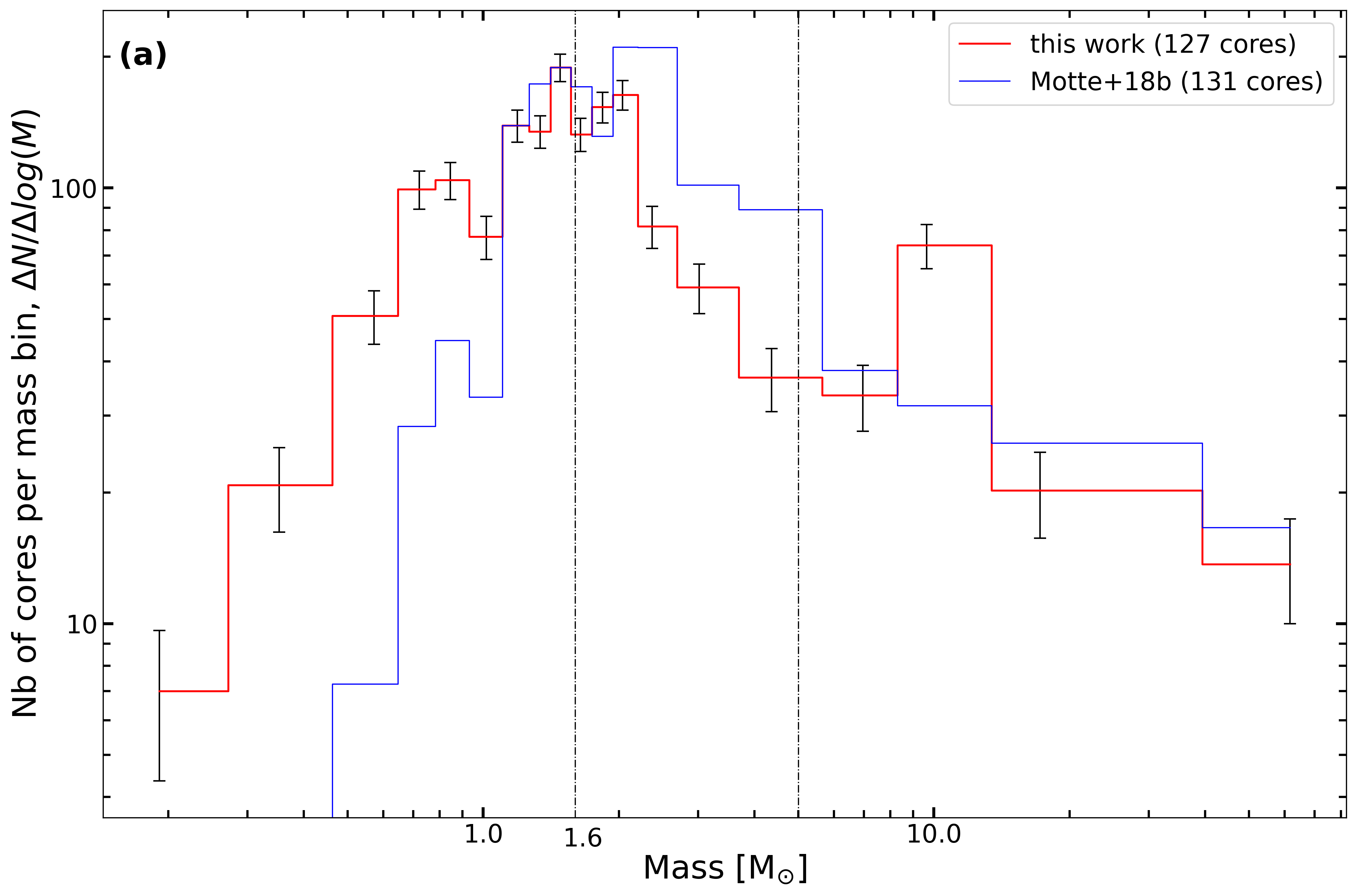}
    \includegraphics[width=\hsize]{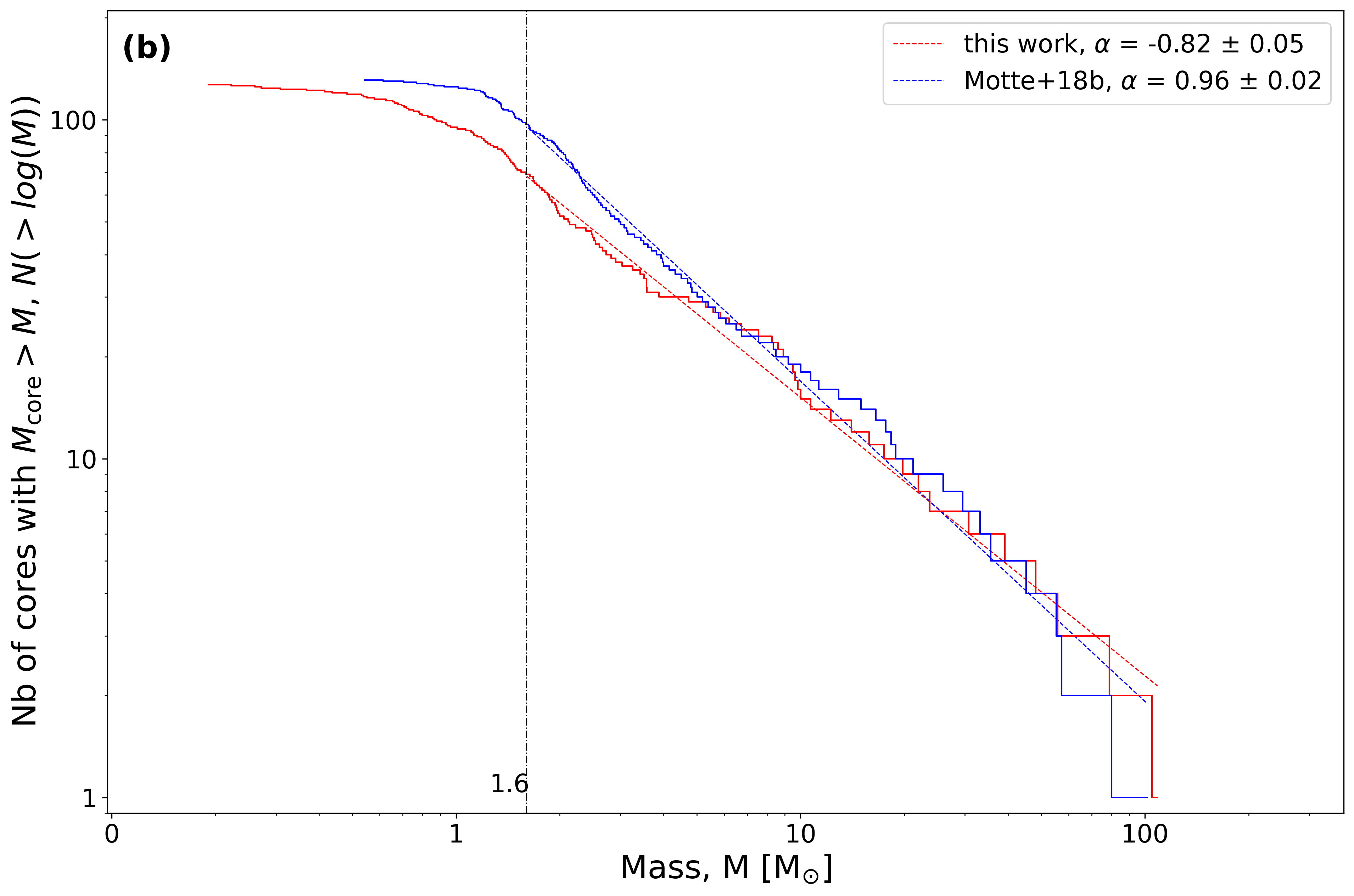}
    \caption{Comparison of the CMFs in W43-MM1 for this work (cat-n, 127 cores, in red) and for the catalogue of \cite{Motte18b} (cat-o, 131 cores, in blue).
    (a) The differential CMFs show that the peak of the distributions are located at similar masses and close to the adopted completeness (1.6$\,\Msol$, represented by a vertical dash line).
    (b) The cumulative CMFs are fitted by single power laws of the form N(>log(M) $\propto M^{\alpha}$ with $\alpha=-0.82 \pm 0.05$ for cat-n and $\alpha=-0.96 \pm 0.02$ for cat-o.}
    \label{fig:app-comp-CMF}
\end{figure}

\begin{figure}
    \centering
    \includegraphics[width=\hsize]{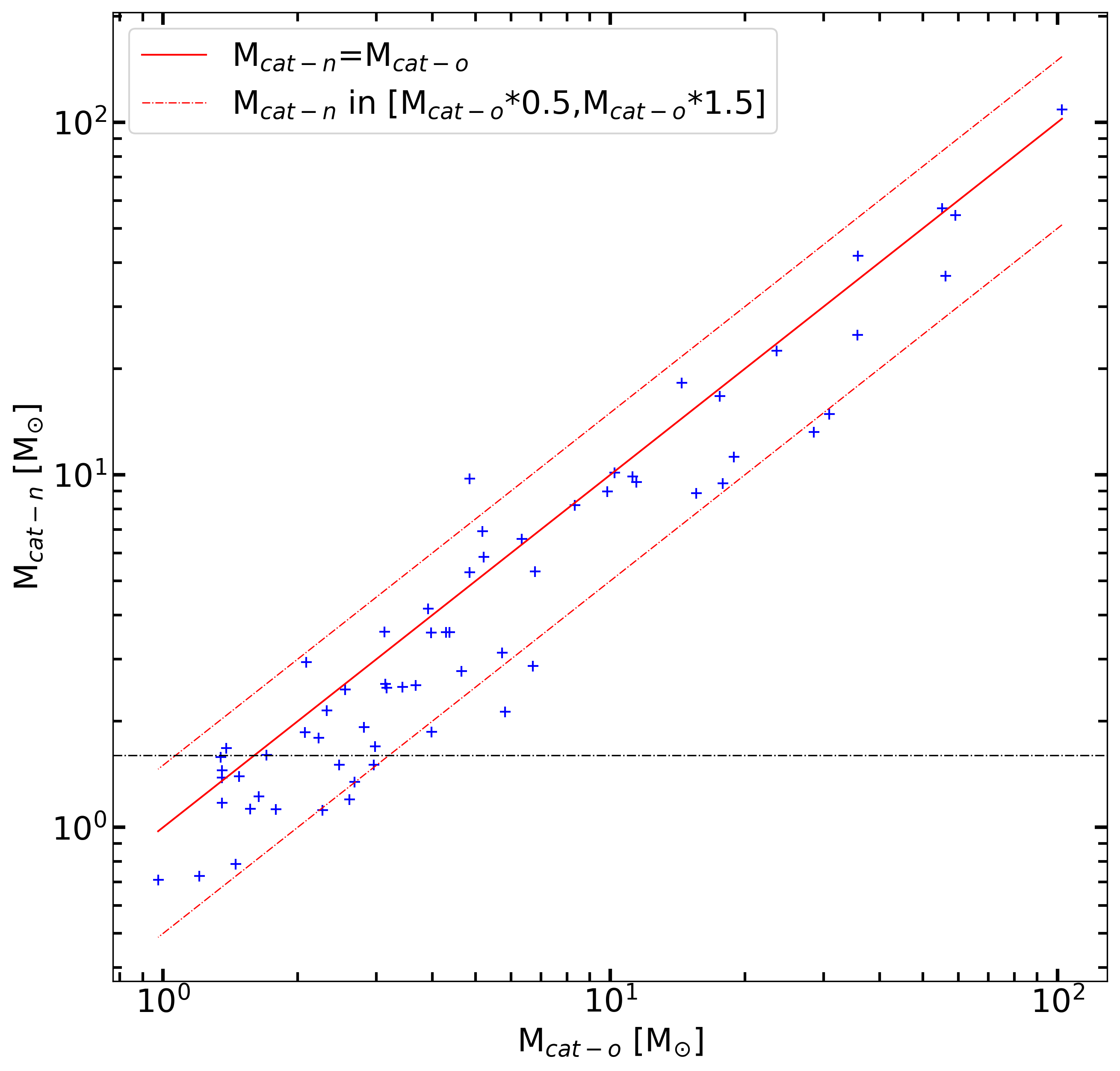}
    \caption{Comparison of the masses for the 65 cores detected in the two catalogues. Masses in the cat-n are typically within 50\% of their estimation in the cat-o.}
    \label{fig:app-comp-mass}
\end{figure}

The global (prestellar and protostellar cores) CMFs of W43-MM1 from the two catalogues are compared in \cref{fig:app-comp-CMF}. With a slope of $-0.82 \pm 0.15$ above 1.6$\,\Msol$, the CMF of cat-n is flatter than that of cat-o ($-0.96 \pm 0.12$), although the two values are compatible within 3$\sigma$ uncertainties. 
Two effects explain this difference. First, as mentioned above, the detection of cores changed from cat-o to cat-n. Among the 62 new cores in cat-n, 20 have masses above the 1.6$\,\Msol$ completeness limit and are included in the fit. 
Secondly, the mass of the cores detected in the two catalogues changed, a combined result of changes in the continuum map itself and changes in the way cores are defined and measured during the extraction.
\cref{fig:app-comp-mass} shows that masses in cat-n are typically within 50\% of their estimate in cat-o, with a median 17\% lower.

\FloatBarrier
\renewcommand{\thefigure}{C\arabic{figure}}
\renewcommand{\thetable}{C\arabic{table}}

\section{Completeness of the core catalogue}
\label{s:app-completeness}

\begin{figure}[ht]
    \centering
    \includegraphics[width=\hsize]{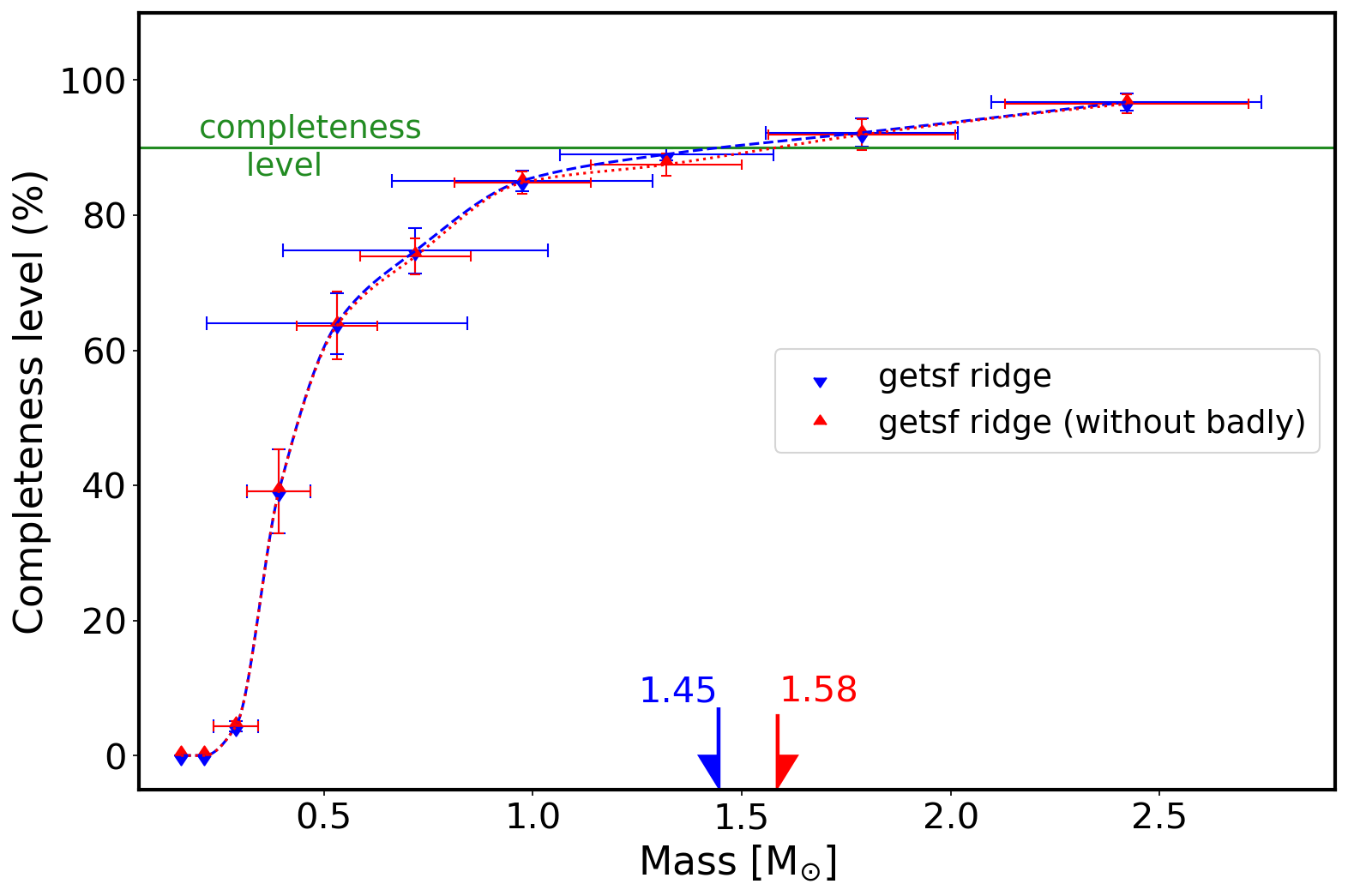}
    \caption{Completeness levels of the $\sim3200$ synthetic sources added on the centre of the background image of W43-MM1. The core catalogue is 90\% complete down to 1.6 $\pm$ 0.2 $\Msol$. Error bars represent the $\pm1\sigma$ uncertainties, measured from the dispersion of mass measurements across each bin (x-axis) and from the dispersion of detection rates between the four set of simulations (y-axis). Blue points represent the full sample of sources detected by \textit{getsf}, red points only measure the bin completeness for cores that have mass measurements within a factor of two of the reference.}
    \label{fig:app-completeness}
\end{figure}

We estimated the completeness level of W43-MM1 by injecting populations of about 1400 synthetic sources over the background image of W43-MM1. The background image is a product of \textit{getsf} created by subtracting from the detection image (denoised $\bsens^*$) the extracted sources. Synthetic sources were produced with a Gaussian profile whose FWHM size corresponds to the median size of observed cores, 0.6$\arcsec$. The synthetic sources of each population were split in ten bins of mass from 0.16 to 2.4~$\Msol$, with a constant number of about 140 sources per bin. We then created four independent synthetic images, in which both the location of the sources and their mass are randomly selected, in order to sample at best the variation of the background. In the following, we focus on sub-populations of about 800 synthetic sources located in the central part of the image (the ridge), where observed cores are detected. We ran the extraction algorithm \textit{getsf} on the four synthetic images with the same parameters as for the observations (see Sect.~\ref{su:cont}).
Figure \ref{fig:app-completeness} shows the detection rate of synthetic sources as a function of their mass. The 90\% completeness level, excluding sources with bad mass measurements, is found to be 1.6 $\pm$ 0.2 $\Msol$. The uncertainty is estimated from the error bars shown in \cref{fig:app-completeness} in the mass bins located near the point of intersection with the 90\% completeness level.

\FloatBarrier
\renewcommand{\thefigure}{D\arabic{figure}}
\renewcommand{\thetable}{D\arabic{table}}

\section{Measurements of low-mass outflows detection }
\label{s:app-measure-outfl}

In \cref{su:disc-cpl-proto} we mentioned the sensitivity of the CO (2--1) data as a possible limit for the detection of outflows from low-mass protostellar cores. To test this hypothesis, we propose in the following a quantitative measurement of outflow detection which is representative of our method, a visual inspection of the CO cube channel by channel.
We define the detection intensity of an outflow lobe, in Jy/beam, as the maximum intensity at which it is non-ambiguously detected. "Non-ambiguously" here signifies that the lobe can be clearly identified as a relatively narrow and elongated structure separated from large-scale CO emission. This criterion excludes the low-velocity channels (typically between 80 and 105 $\kms$, see spectra \cref{fig:spec}) in which CO outflows are brighter but also more contaminated by cloud emission. We also require the detection intensity to be reached in at least 2 consecutive channels. This detection intensity is then converted to a signal-to-noise by dividing it by the average RMS is a channel, 2 mJy/beam.

The 12 protostellar cores with masses of  [0.8-1.6]$\,\Msol$ in W43-MM2\&MM3 are associated with 18 outflow lobes. Their detection intensity, excluding a lobe with confusion, range from 5.7$\sigma$ to 49$\sigma$, with a median of 17$\sigma$. These detections are thus above the sensitivity limit, whether we put it at 3$\sigma$ or 5$\sigma$.
As a comparison, the eight outflow lobes for the six cores above 16$\,\Msol$ in W43-MM2\&MM3 have detection intensities ranging from 13$\sigma$ to 81$\sigma$ with a median of 40$\sigma$.
Therefore, we conclude that the protostellar cores of lowest mass included in the analysis of \cref{fig:histomass} are all well characterised. The result presented \cref{su:cmf} that the fraction of cores which are protostellar decreases with mass is not due to a sensitivity limitation.

\renewcommand{\thefigure}{E\arabic{figure}}
\renewcommand{\thetable}{E\arabic{table}}

\section{Complementary figures}
\label{s:app-complemnt-figures}

\begin{figure*}[htp]
    \centering
    \includegraphics[width=0.8\hsize]{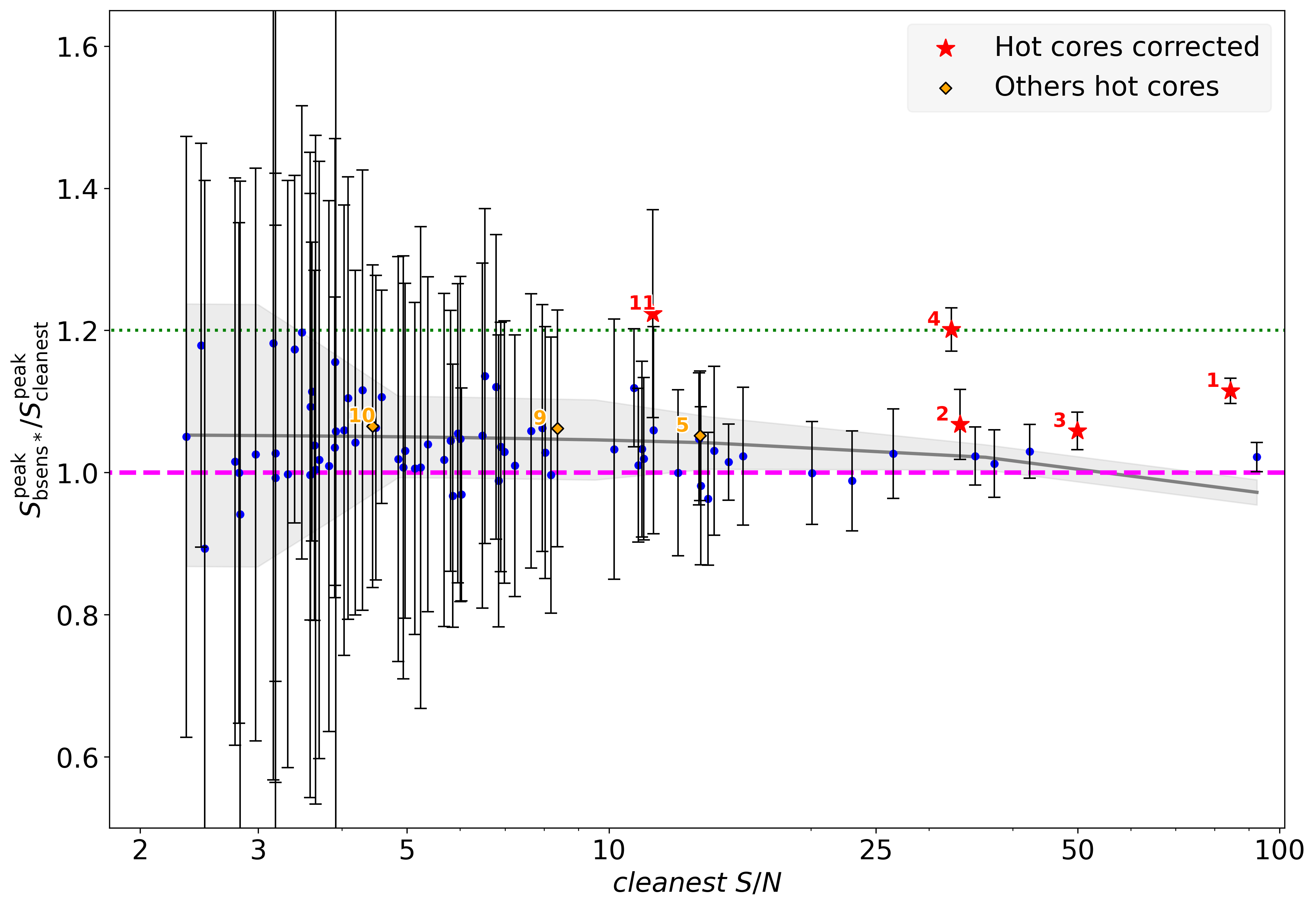}
    \caption{Line contamination of the 1.3~mm continuum fluxes of cores detected in W43-MM1, as estimated from the ratio of their $\bsens^*$ over \cleanest peak fluxes and shown as a function of the S/N in the \cleanest image. The grey curve indicates the median value of the core ratios, computed over bins of 20 adjacent cores as ranked by their S/N. The shaded grey area indicates the corresponding 3$\sigma$ dispersion in flux ratio values. Red and orange points locate cores identified as hot cores by \cite{Brouillet22}. \cleanest fluxes have been used instead of $\bsens^*$ for the five cores with significant contamination levels, shown in red. The horizontal lines indicate the contamination levels of 0\% (magenta dashed line) and 20\% (green dotted line).}
    \label{fig:app-hc-conta}
\end{figure*}

\begin{figure*}
    \centering
    \includegraphics[width=0.8\hsize]{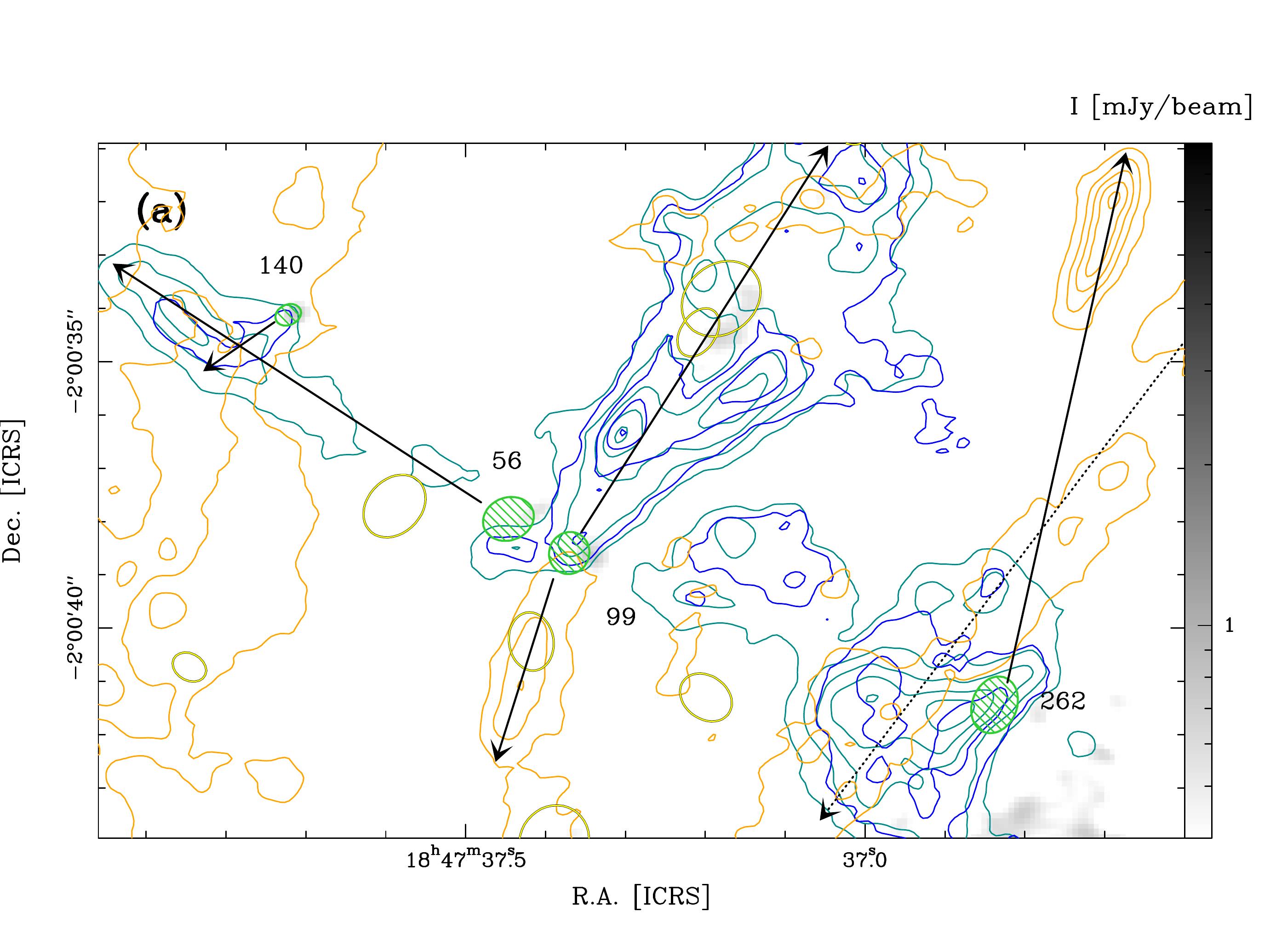}
    \subfloat{\includegraphics[width=0.53\hsize]{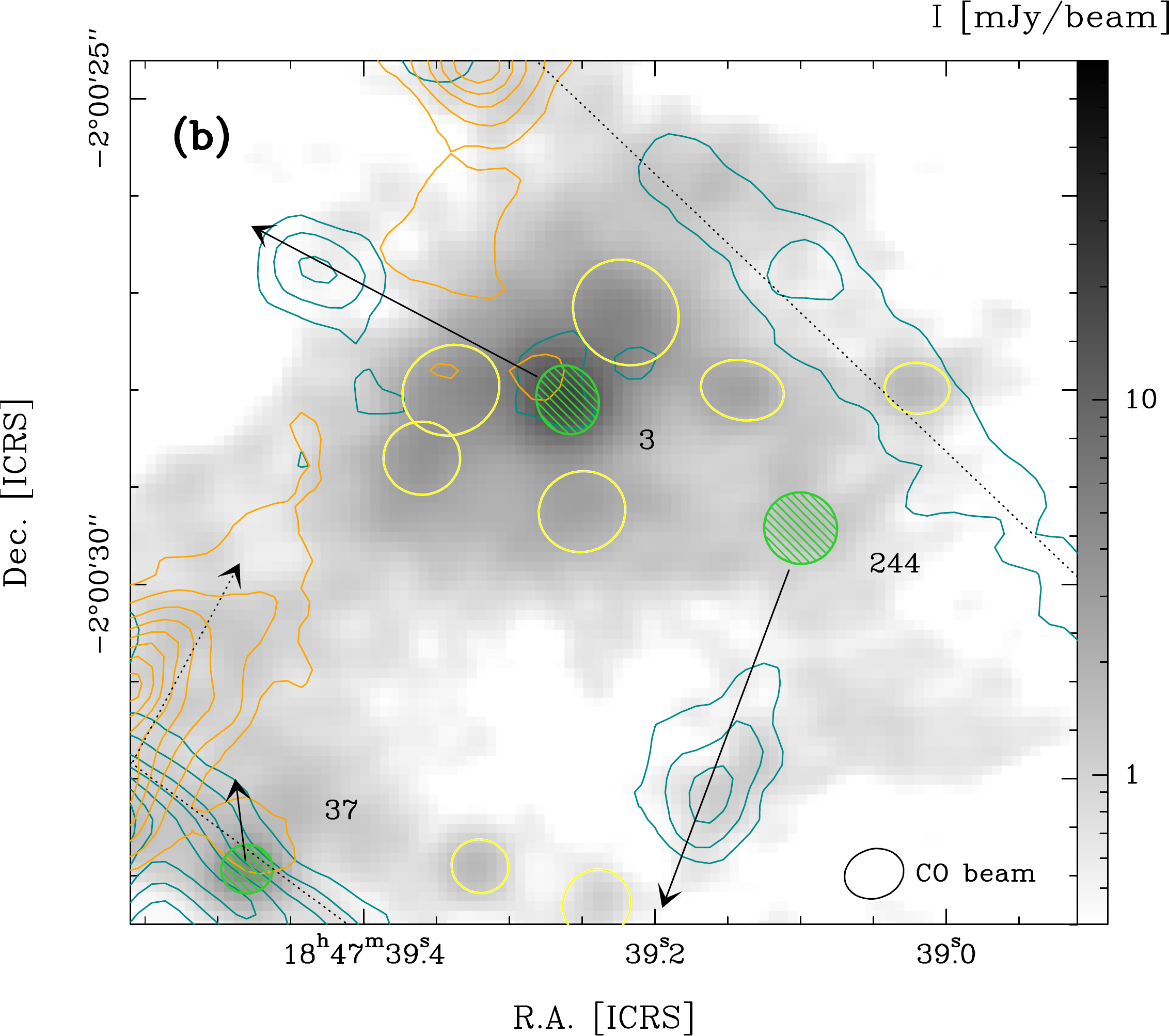}}
    \caption{Zooms from \cref{fig:flow-MM2} towards the north of W43-MM2 (in \textit{a}) and from \cref{fig:flow-MM3} toward the south of W43-MM3 (MM10) (in \textit{b}).
    In \textit{a}: The CO (2--1) blue-shifted line wing is integrated over 73.9-75.1 $\kms$ (blue contours) and 75.1-80.2 $\kms$ (cyan contours). The red-shifted line wing is integrated over 108.2-109.4 $\kms$ (orange contours). Contours are 4 to Max by steps of 10 in units of $\sigma=$ 4 mJy\,beam$^{-1}\,$km\,s$^{-1}$  (blue and orange) and 10 mJy\,beam$^{-1}\,$km\,s$^{-1}$  (cyan).
    In \textit{b}: The CO (2--1) blue-shifted line wing is integrated over 73.9-80.2 $\kms$ (cyan contours), the red-shifted line wing over 110.7-112.0 $\kms$ (orange contours). Contours are 4, 10 to Max by steps of 10 in units of $\sigma=$ 13 and 5 mJy\,beam$^{-1}\,$km\,s$^{-1}$, respectively.
    In \textit{a-b}: Cores driving outflows are represented by green ellipses showing their FWHM size and numbered. Arrows indicate the direction of their outflows, an ellipse representing the angular resolution of CO cube is shown in the lower right. Prestellar core candidates are represented by yellow ellipses.}
    \label{fig:app-flow-MM2-zooms}
\end{figure*}

\begin{figure*}
    \centering
    \includegraphics[width=0.6\hsize]{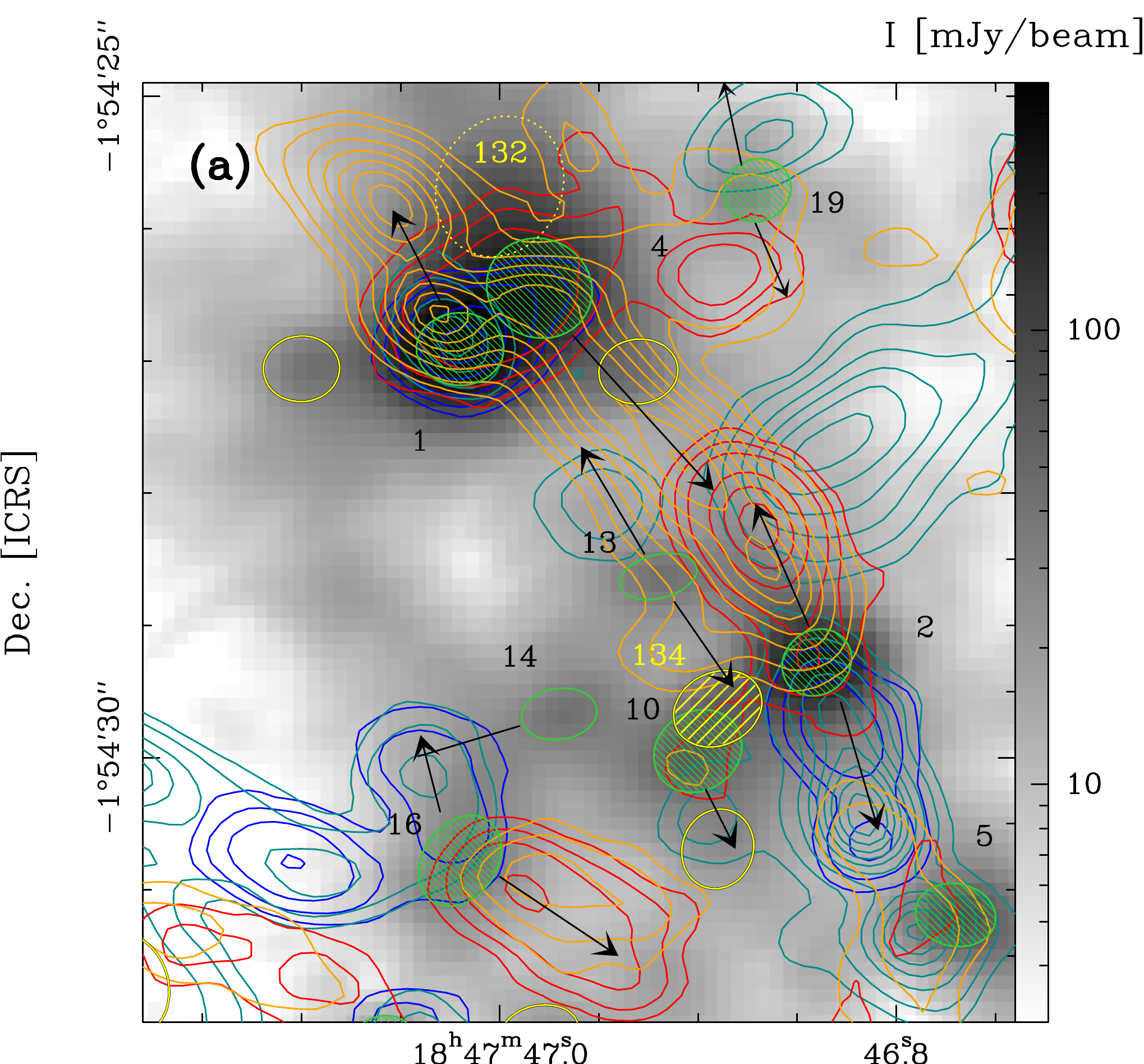}
    \includegraphics[width=0.75\hsize]{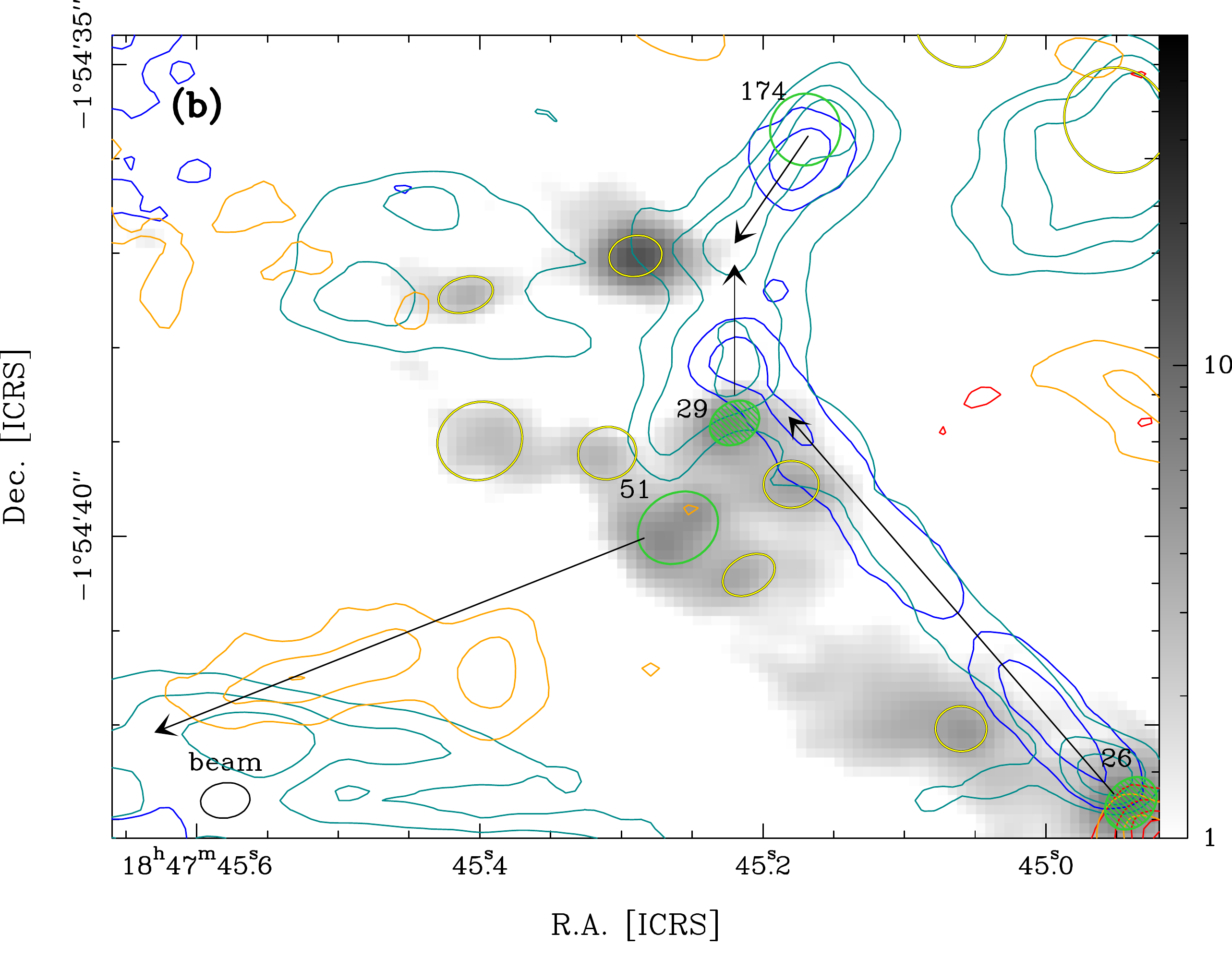}
    \caption{Zooms in towards areas of W43-MM1 with new cores and outflow detection. The CO (2--1) blue-shifted line wing is integrated over $42-64~\kms$ (HV, blue contours) and $82-88~\kms$ (LV, cyan contours), the red-shifted line wing over $108-119~\kms$ (LV, orange contours) and $128-158~\kms$ (HV, red contours). 
    In \textit{a}: For blue lobes, contours are 7, 15, 30 to 230 by steps of 40 (HV) and 7, 15 to 120 by steps of 15 (LV), in units of $\sigma=20\,\rm mJy\,beam^{-1}\,\kms$. For red lobes, contours are 10, 20 to 160 by steps of 20, in units of $\sigma_{\rm LV,\, R}=37\,\rm mJy\,beam^{-1}\,\kms$ (LV) and 7, 15, 30 to 280 by steps of 50, in units of $\sigma$ (HV). 
    In \textit{b}: For blue lobes, contours are 5, 15, 30 by steps of 15 (HV) and 10, 20, 30 (LV), in units of $\sigma= 20\,\rm mJy\,beam^{-1}\,\kms$. For red lobes, contours are 5, 15, 25, 35, in units of $\sigma_{\rm LV,\, R}= 37\,\rm mJy\,beam^{-1}\,\kms$ (LV) and 5, 15 to 75 by steps of 15, in units of $\sigma$ (HV). 
    In \textit{a} and \textit{b}: Green ellipses locate protostellar cores, arrows indicate the direction of their outflows. Prestellar core candidates are represented with yellow ellipses, hatched for the high-mass core \#134 in (a). The source \#132 discarded from the core sample is shown with a dotted yellow ellipse. Adapted from Figs. 3 and 4 of \cite{Nony20}.}
    \label{fig:app-flow-MM1-zoom}
\end{figure*}

\begin{figure*}
    \centering
    \includegraphics[width=0.5\hsize]{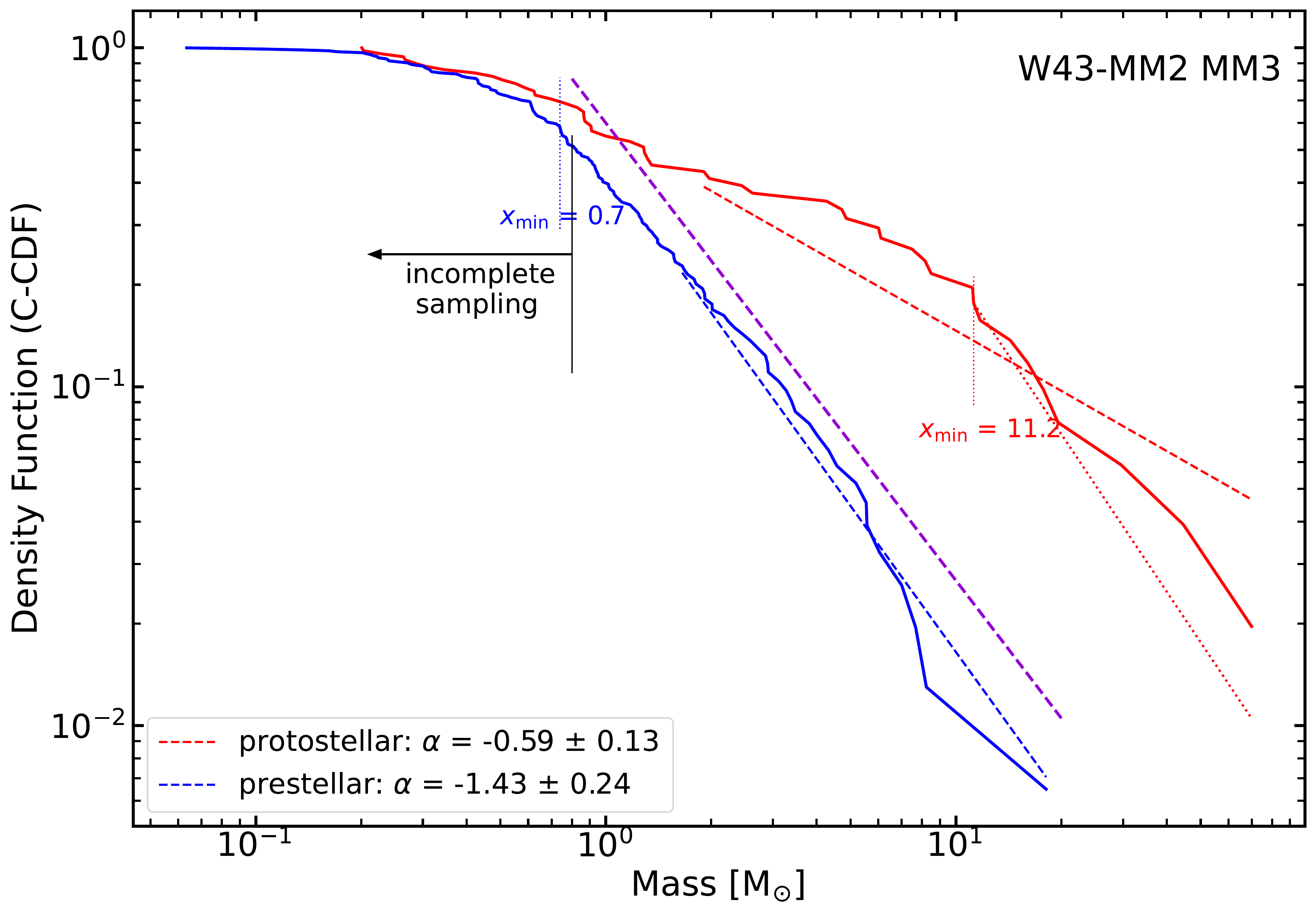}
    \includegraphics[width=0.49\hsize]{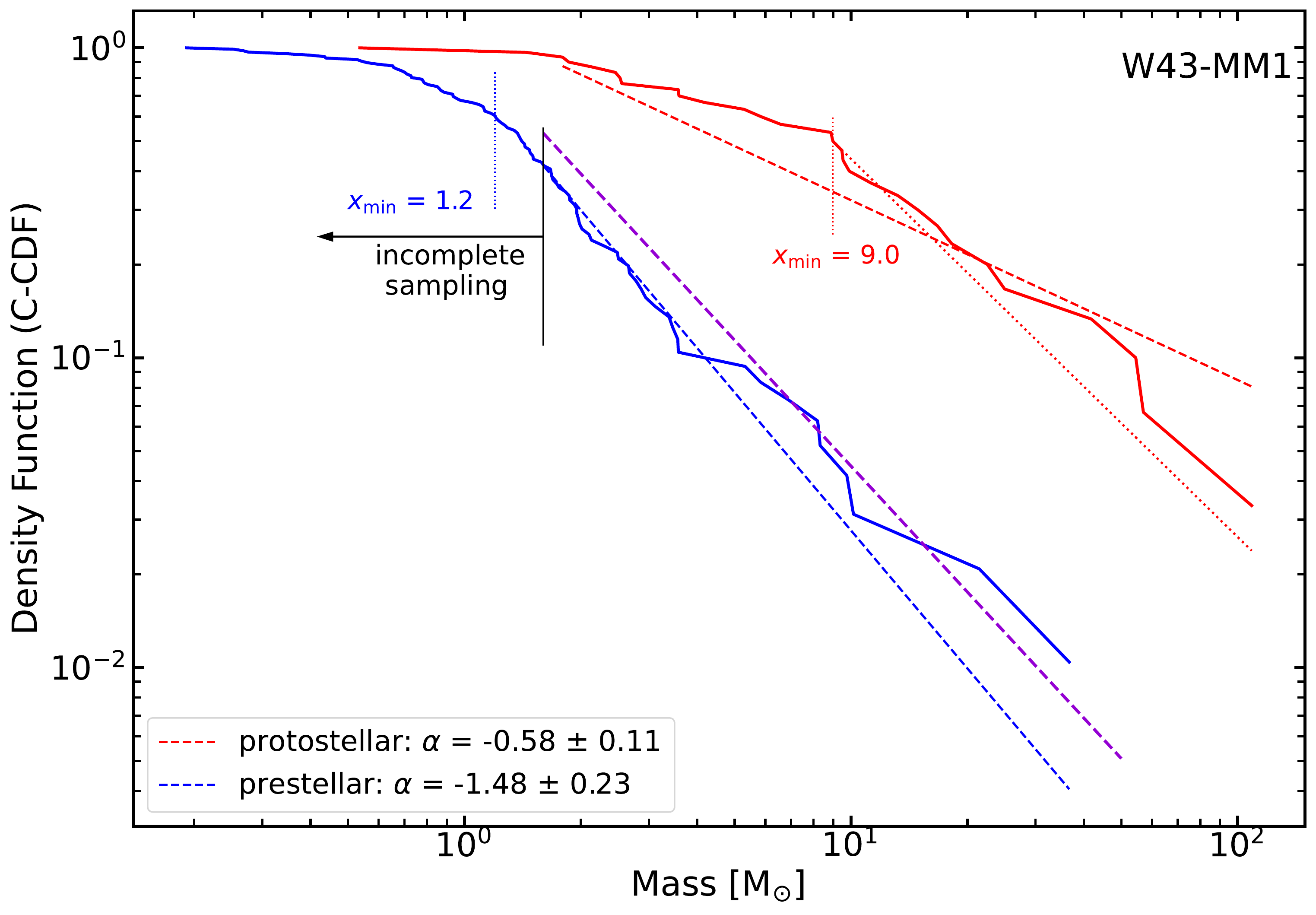}
    \caption{CMFs for the prestellar (in blue) and protostellar (in red) cores in W43-MM2\&MM3 (left) and in W43-MM1 (right). 
    CMFs are represented in the form of a complementary cumulative distribution function (C-CDF) fitted with the MLE method implemented in the \texttt{powerlaw} package \citep{Alstott14}. 
    Dashed lines represent the fit made above the completeness limit of W43-MM1, 1.6$\,\Msol$ (black vertical line) with single power law whose slopes are indicated. $x_{\rm min}$ values are indicated and the fit above $x_{\rm min}$ is shown in dotted line for the protostellar CMFs. The canonical Salpeter slope of the IMF (-1.35 on this form) is represented with dashed purple lines.}
    \label{fig:app-CMFs}
\end{figure*}

Figure~\ref{fig:app-hc-conta} shows the $\bsens^*$ over \cleanest flux ratios as a function of S/N and highlights the cores whose flux is contaminated by hot core emission. 
Figure~\ref{fig:app-flow-MM2-zooms} shows outflows in close-ups from \cref{fig:flow-MM2,fig:flow-MM3} towards W43-MM2\&MM3. Figure~\ref{fig:app-flow-MM1-zoom}
shows new outflows in W43-MM1.
Figure~\ref{fig:app-CMFs} presents CMFs similarly to \cref{fig:CMFs_W43_MLE} for the separate regions, W43-MM2\&MM3 and  W43-MM1.

\renewcommand{\thefigure}{F\arabic{figure}}
\renewcommand{\thetable}{F\arabic{table}}

\section{Core catalogues}
\label{s:app-tables}

Tables~\ref{tab:nature-coreMM23} and \ref{tab:nature-coreMM1} list the physical properties and outflow detection or non-detection of cores in W43-MM2\&MM3 and W43-MM1, respectively. All the protostellar cores are listed, while only prestellar cores above the completeness limit ($C_{90}=0.8\,\Msol$ and $1.6\,\Msol$, respectively) are listed. The full tables~\ref{tab:nature-coreMM23} and \ref{tab:nature-coreMM1} are available in electronic form through CDS.

\FloatBarrier

\begin{table*}[p]
\centering
\begin{threeparttable}
\caption{\label{tab:nature-coreMM23} Main characteristics of protostellar and  prestellar cores in W43-MM2\&MM3}
\begin{tabular}{cccc|cc|ccc}
\hline\hline
n\tnote{a} & Name\tnote{a} & Size\tnote{a} &  Mass\tnote{a} & Blue\tnote{b} & Red lobe\tnote{b} & Comments\tnote{c} \\
 &  W43-MM2\&3\_ALMAIMF* & [$\arcsec \times \arcsec$] & [$\Msol$] & & & \\
\hline
1 & 184736.80-20054.27  & 0.80$\,\times\,$0.65 & $  69.9\pm13.7$ &  N   &  Y  &              \\ 
  2 & 184741.71-20028.60  & 0.62$\,\times\,$0.51 & $  44.6\pm8.8 $ &  Y   & Y?  &              \\ 
  3 & 184739.26-20028.10  & 0.72$\,\times\,$0.63 & $  11.2\pm2.3 $ &  Y?  &  N  & tent. (detec) / HC \\ 
  5 & 184736.10-20115.98  & 0.90$\,\times\,$0.64 & $  17.8\pm3.3 $ & Y,Y  & Y,Y &  multiple   \\ 
  6 & 184736.03-20120.73  & 0.67$\,\times\,$0.48 & $   8.5\pm1.5 $ &  Y?  &  Y  &              \\ 
  7 & 184736.75-20053.75  & 0.81$\,\times\,$0.75 & $  14.3\pm1.9 $ & Y,Y? &  N  &              \\ 
  9 & 184741.73-20027.42  & 0.62$\,\times\,$0.50 & $  16.0\pm2.9 $ &  Y   &  Y  &              \\ 
 10 & 184736.28-20050.75  & 0.63$\,\times\,$0.51 & $   2.6\pm0.3 $ &  Y   &  Y  &              \\ 
 12 & 184736.70-20047.55  & 0.65$\,\times\,$0.49 & $  11.2\pm2.1 $ &  Y   & Y,Y? &              \\ 
 14 & 184735.10-20108.77  & 0.64$\,\times\,$0.49 & $   4.3\pm0.8 $ &  Y   &  Y  &              \\ 
 15 & 184736.84-20102.61  & 0.63$\,\times\,$0.44 & $   4.7\pm0.9 $ &  Y?  &  Y  &              \\ 
 20 & 184736.06-20127.82  & 0.67$\,\times\,$0.63 & $   4.9\pm0.8 $ &  Y?  &  N  &    tent.     \\ 
 24 & 184741.63-20025.37  & 0.63$\,\times\,$0.59 & $   7.5\pm1.3 $ &  Y   &  Y  &              \\ 
 25 & 184741.83-20029.32  & 0.98$\,\times\,$0.78 & $  19.6\pm3.5 $ &  N   & Y?  & tent. (detec) \\ 
 28 & 184736.68-20048.06  & 0.62$\,\times\,$0.52 & $   6.1\pm1.1 $ &  Y   &  N  &              \\ 
 33 & 184736.82-20052.88  & 1.22$\,\times\,$1.11 & $  29.6\pm5.6 $ &  Y   &  N  &              \\ 
 37 & 184739.48-20032.93  & 0.53$\,\times\,$0.50 & $   2.0\pm0.4 $ &  Y   & Y?  &              \\ 
 39 & 184736.14-20046.65  & 0.56$\,\times\,$0.44 & $   2.4\pm0.5 $ &  N   & Y?  &    tent.     \\ 
 41 & 184736.14-20129.16  & 0.64$\,\times\,$0.58 & $   1.9\pm0.3 $ &  Y   &  Y  &              \\ 
 44 & 184736.93-20054.79  & 1.10$\,\times\,$0.82 & $  11.7\pm2.1 $ &  Y?  &  N  &    tent.     \\ 
 50 & 184734.68-20103.92  & 0.64$\,\times\,$0.49 & $   1.3\pm0.2 $ &  Y   &  Y  &              \\ 
 51 & 184736.82-20050.45  & 0.66$\,\times\,$0.52 & $   6.0\pm1.2 $ &  Y   &  Y  &              \\ 
 56 & 184737.42-20037.95  & 1.02$\,\times\,$0.81 & $   0.5\pm0.1 $ &  Y   &  N  &   low vel.   \\ 
 62 & 184738.68-20044.44  & 0.80$\,\times\,$0.57 & $   1.2\pm0.2 $ &  Y   &  Y  &              \\ 
 71 & 184736.27-20029.67  & 0.94$\,\times\,$0.78 & $   1.0\pm0.2 $ &  Y   &  Y  &              \\ 
 75 & 184736.08-20113.11  & 0.69$\,\times\,$0.57 & $   1.3\pm0.2 $ &  Y   &  Y  &              \\ 
 78 & 184736.75-20055.73  & 0.95$\,\times\,$0.67 & $   8.2\pm1.5 $ &  Y?  &  Y  &              \\ 
 85 & 184737.22-20024.32  & 1.29$\,\times\,$1.08 & $   0.2\pm0.0 $ &  Y   &  Y  &              \\ 
 86 & 184741.73-15959.00  & 0.93$\,\times\,$0.83 & $   0.3\pm0.1 $ &  Y?  &  N  & tent (conf \#238) \\ 
 88 & 184734.96-20031.56  & 1.46$\,\times\,$1.34 & $   0.6\pm0.1 $ &  Y   &  Y  &              \\ 
 97 & 184735.07-20112.72  & 0.68$\,\times\,$0.41 & $   0.8\pm0.2 $ &  Y?  &  N  & tent. (detec) \\ 
 99 & 184737.34-20038.59  & 0.81$\,\times\,$0.78 & $   0.9\pm0.2 $ &  Y   &  Y  &              \\ 
112 & 184736.04-20049.55  & 0.56$\,\times\,$0.47 & $   1.4\pm0.3 $ &  Y?  &  Y  &              \\ 
113 & 184741.24-20022.60  & 0.90$\,\times\,$0.68 & $   0.6\pm0.1 $ &  Y   &  N  &              \\ 
133 & 184735.06-20110.72  & 0.50$\,\times\,$0.45 & $   0.3\pm0.1 $ &  Y?   &  Y  &              \\ 
140 & 184737.71-20034.12  & 0.52$\,\times\,$0.39 & $   0.2\pm0.0 $ &  Y   &  N  &              \\ 
147 & 184739.59-20032.42  & 0.86$\,\times\,$0.66 & $   0.9\pm0.2 $ &  N   &  Y  &              \\ 
157 & 184739.51-20021.78  & 1.04$\,\times\,$0.85 & $   0.9\pm0.2 $ &  Y?  &  N  &    tent.     \\ 
182 & 184738.68-20038.48  & 1.05$\,\times\,$1.01 & $   0.7\pm0.1 $ &  Y?  & Y?  &    tent.     \\ 
187 & 184739.66-20031.95  & 0.57$\,\times\,$0.45 & $   0.4\pm0.1 $ &  Y   &  N  &              \\ 
192 & 184736.60-20050.87  & 0.77$\,\times\,$0.66 & $   1.3\pm0.3 $ &  Y   &  N  &              \\ 
204 & 184736.00-20050.60  & 0.69$\,\times\,$0.53 & $   0.9\pm0.3 $ &  Y   &  N  &              \\ 
230 & 184738.23-20038.28  & 0.87$\,\times\,$0.76 & $   0.3\pm0.1 $ &  N   &  Y  &              \\ 
236 & 184737.00-20102.35  & 0.77$\,\times\,$0.62 & $   0.6\pm0.1 $ &  Y   &  N  &              \\ 
237 & 184739.75-20033.46  & 0.75$\,\times\,$0.60 & $   0.5\pm0.1 $ &  Y   &  N  &              \\ 
238 & 184742.58-15943.35  & 0.70$\,\times\,$0.48 & $   0.8\pm0.2 $ &  Y   &  N  &              \\ 
239 & 184736.07-20051.31  & 0.88$\,\times\,$0.78 & $   0.9\pm0.2 $ &  Y   &  Y  &              \\ 
244 & 184739.10-20029.42  & 0.75$\,\times\,$0.73 & $   0.6\pm0.1 $ &  Y   &  N  &              \\ 
262 & 184736.78-20041.44  & 1.10$\,\times\,$0.88 & $   0.3\pm0.1 $ &  N   &  Y  &   low vel.   \\ 
265 & 184735.36-20113.40  & 0.96$\,\times\,$0.93 & $   0.3\pm0.1 $ &  Y?  &  N  & tent (conf \#133, \#97) \\ 
287 & 184735.87-20051.97  & 0.55$\,\times\,$0.45 & $   0.2\pm0.1 $ &  Y   &  Y  &              \\
 \hline
\end{tabular} 
 \begin{tablenotes}
\item [a] Parameters taken from \cite{Pouteau22a} 
\item [b] Detection, tentative detection, and non detection of outflow lobes are denoted with "Y", "Y?" and "N", respectively.
\item [c] Cores whose outflow attribution, and therefore protostellar nature, is uncertain are denoted as "tent". Details are given in parenthesis when confusion arise from another outflow lobe ("conf") or when the detection of the lobe itself is tentative ("detec"). "low vel" indicate cores with a single lobe detected only at low velocity. The cores characterised by their molecular content are denoted as HC.
\end{tablenotes}
\end{threeparttable}
\end{table*}

\setcounter{table}{0}
\begin{table*}
\centering
\caption{continued}
\begin{tabular}{cccc|cc|ccc}
\hline\hline
n\tnote{a} & Name\tnote{a} & Size\tnote{a} &  Mass\tnote{a} & Blue\tnote{b} & Red lobe\tnote{b} & Comments\tnote{c} \\
 & W43-MM2\&3\_ALMAIMF* & [$\arcsec \times \arcsec$] & [$\Msol$] & & & \\
\hline
11 & 184740.97-20020.73  & 0.63$\,\times\,$0.60 & $   7.7\pm1.3 $ &  N   &  N  &              \\ 
 13 & 184736.15-20047.87  & 0.56$\,\times\,$0.47 & $   8.2\pm1.7 $ &  N   &  N  &              \\ 
 16 & 184735.69-20032.50  & 0.62$\,\times\,$0.46 & $   3.4\pm0.6 $ &  N   &  N  &              \\ 
 18 & 184740.23-20034.51  & 0.56$\,\times\,$0.52 & $   3.1\pm0.6 $ &  N   &  N  &              \\ 
 22 & 184736.65-20053.23  & 0.79$\,\times\,$0.68 & $  18.1\pm3.5 $ &  N   &  N  &              \\ 
 32 & 184738.30-20041.47  & 0.61$\,\times\,$0.44 & $   1.9\pm0.3 $ &  N   &  N  &              \\ 
 35 & 184733.73-20100.38  & 0.77$\,\times\,$0.62 & $   4.0\pm0.7 $ &  N   &  N  &              \\ 
 43 & 184739.22-20027.20  & 1.14$\,\times\,$1.04 & $   7.0\pm1.2 $ &  N   &  N  &              \\ 
 58 & 184741.24-20040.68  & 0.93$\,\times\,$0.85 & $   0.8\pm0.2 $ &  N   &  N  &              \\ 
 59 & 184741.58-20028.35  & 0.55$\,\times\,$0.46 & $   3.8\pm0.9 $ &  N   &  N  &              \\ 
 60 & 184742.00-20028.19  & 0.63$\,\times\,$0.49 & $   3.5\pm0.6 $ &  N   &  N  &              \\ 
 63 & 184741.88-20027.94  & 0.88$\,\times\,$0.80 & $   4.6\pm0.8 $ &  N   &  N  &              \\ 
 66 & 184737.36-20044.01  & 1.43$\,\times\,$1.32 & $   2.5\pm0.4 $ &  N   &  N  &              \\ 
 67 & 184734.66-20058.28  & 1.33$\,\times\,$1.24 & $   2.9\pm0.5 $ &  N   &  N  &              \\ 
 69 & 184742.56-20005.88  & 1.18$\,\times\,$1.05 & $   1.2\pm0.2 $ &  N   &  N  &              \\ 
 80 & 184739.36-20028.70  & 0.79$\,\times\,$0.75 & $   2.9\pm0.5 $ &  N   &  N  &              \\ 
 81 & 184734.60-20118.77  & 1.78$\,\times\,$1.60 & $   2.7\pm0.5 $ &  N   &  N  &              \\ 
 83 & 184736.50-20110.46  & 1.50$\,\times\,$1.18 & $   3.3\pm0.6 $ &  N   &  N  &              \\ 
 87 & 184736.45-20108.88  & 0.97$\,\times\,$0.83 & $   1.4\pm0.3 $ &  N   &  N  &              \\ 
 89 & 184734.78-20120.01  & 1.23$\,\times\,$0.95 & $   1.9\pm0.3 $ &  N   &  N  &              \\ 
 94 & 184741.70-20027.05  & 0.65$\,\times\,$0.52 & $   5.6\pm1.1 $ &  N   &  N  &              \\ 
 95 & 184737.57-20037.71  & 1.35$\,\times\,$1.03 & $   0.9\pm0.2 $ &  N   &  N  &              \\ 
 96 & 184740.59-20036.61  & 1.14$\,\times\,$0.96 & $   2.2\pm0.4 $ &  N   &  N  &              \\ 
 98 & 184741.73-20000.54  & 1.77$\,\times\,$1.39 & $   0.9\pm0.1 $ &  N   &  N  &              \\ 
100 & 184737.36-20022.68  & 2.09$\,\times\,$1.49 & $   1.1\pm0.2 $ &  N   &  N  &              \\ 
102 & 184735.26-20101.65  & 0.69$\,\times\,$0.48 & $   0.9\pm0.2 $ &  N   &  N  &              \\ 
103 & 184739.34-20028.00  & 1.02$\,\times\,$0.91 & $   5.5\pm1.0 $ &  N   &  N  &              \\ 
104 & 184741.40-20042.37  & 1.48$\,\times\,$1.21 & $   1.1\pm0.2 $ &  N   &  N  &              \\ 
105 & 184740.80-20036.84  & 0.96$\,\times\,$0.85 & $   1.4\pm0.3 $ &  N   &  N  &              \\ 
106 & 184738.65-20042.05  & 0.64$\,\times\,$0.56 & $   0.8\pm0.2 $ &  N   &  N  &              \\ 
107 & 184735.96-20102.36  & 0.92$\,\times\,$0.77 & $   0.8\pm0.1 $ &  N   &  N  &              \\ 
115 & 184736.05-20114.65  & 0.68$\,\times\,$0.63 & $   1.8\pm0.5 $ &  N   &  N  &              \\ 
117 & 184741.21-20026.17  & 0.68$\,\times\,$0.65 & $   1.0\pm0.2 $ &  N   &  N  &              \\ 
118 & 184736.05-20125.76  & 0.81$\,\times\,$0.64 & $   1.1\pm0.2 $ &  N   &  N  &              \\ 
119 & 184743.44-20018.06  & 1.83$\,\times\,$1.55 & $   1.7\pm0.3 $ &  N   &  N  &              \\ 
121 & 184739.14-20028.00  & 0.86$\,\times\,$0.61 & $   1.8\pm0.3 $ &  N   &  N  &              \\ 
122 & 184741.18-20039.81  & 1.18$\,\times\,$1.13 & $   1.6\pm0.3 $ &  N   &  N  &              \\ 
123 & 184736.38-20124.70  & 1.28$\,\times\,$1.09 & $   1.3\pm0.2 $ &  N   &  N  &              \\ 
125 & 184736.58-20047.51  & 0.61$\,\times\,$0.52 & $   1.7\pm0.3 $ &  N   &  N  &              \\ 
126 & 184736.67-20100.15  & 0.64$\,\times\,$0.47 & $   1.4\pm0.3 $ &  N   &  N  &              \\ 
127 & 184735.71-20101.45  & 1.09$\,\times\,$0.96 & $   1.0\pm0.2 $ &  N   &  N  &              \\ 
130 & 184734.86-20057.85  & 1.48$\,\times\,$1.14 & $   1.4\pm0.3 $ &  N   &  N  &              \\ 
131 & 184736.00-20049.59  & 0.67$\,\times\,$0.61 & $   2.3\pm0.5 $ &  N   &  N  &              \\ 
132 & 184738.86-20047.65  & 1.37$\,\times\,$1.26 & $   1.3\pm0.2 $ &  N   &  N  &              \\ 
135 & 184741.76-20028.25  & 0.74$\,\times\,$0.65 & $   6.0\pm1.1 $ &  N   &  N  &              \\ 
136 & 184737.09-20111.07  & 1.05$\,\times\,$0.93 & $   0.9\pm0.2 $ &  N   &  N  &              \\ 
137 & 184735.92-20056.56  & 1.20$\,\times\,$1.01 & $   1.0\pm0.2 $ &  N   &  N  &              \\ 
138 & 184737.56-20105.42  & 1.14$\,\times\,$1.06 & $   1.0\pm0.2 $ &  N   &  N  &              \\ 
141 & 184736.75-20057.40  & 0.74$\,\times\,$0.64 & $   2.2\pm0.6 $ &  N   &  N  &              \\ 
142 & 184741.77-20029.16  & 0.86$\,\times\,$0.79 & $   5.2\pm1.0 $ &  N   &  N  &              \\ 
146 & 184739.32-20032.90  & 0.59$\,\times\,$0.55 & $   0.9\pm0.2 $ &  N   &  N  &              \\ 
148 & 184743.03-20002.44  & 1.10$\,\times\,$0.64 & $   1.0\pm0.2 $ &  N   &  N  &              \\ 
\hline
\end{tabular} 
\end{table*}

\setcounter{table}{0}
\begin{table*}
\centering
\caption{continued}
\begin{tabular}{cccc|cc|ccc}
\hline\hline
n\tnote{a} & Name\tnote{a} & Size\tnote{a} &  Mass\tnote{a} & Blue\tnote{b} & Red lobe\tnote{b} & Comments\tnote{c} \\
 & W43-MM2\&3\_ALMAIMF* & [$\arcsec \times \arcsec$] & [$\Msol$] & & & \\
\hline
150 & 184740.61-20017.04  & 1.84$\,\times\,$1.79 & $   1.0\pm0.2 $ &  N   &  N  &              \\ 
152 & 184737.23-20028.79  & 1.49$\,\times\,$1.31 & $   1.5\pm0.3 $ &  N   &  N  &              \\ 
155 & 184741.50-20032.57  & 0.97$\,\times\,$0.86 & $   1.6\pm0.3 $ &  N   &  N  &              \\ 
156 & 184740.27-20016.13  & 1.46$\,\times\,$1.29 & $   0.9\pm0.2 $ &  N   &  N  &              \\ 
162 & 184740.55-20022.02  & 0.98$\,\times\,$0.61 & $   1.1\pm0.2 $ &  N   &  N  &              \\ 
165 & 184736.27-20051.50  & 0.86$\,\times\,$0.55 & $   2.9\pm0.6 $ &  N   &  N  &              \\ 
169 & 184736.76-20046.60  & 0.77$\,\times\,$0.51 & $   1.2\pm0.4 $ &  N   &  N  &              \\ 
170 & 184736.68-20102.51  & 1.04$\,\times\,$1.02 & $   1.3\pm0.3 $ &  N   &  N  &              \\ 
172 & 184741.64-20024.05  & 0.75$\,\times\,$0.50 & $   1.6\pm0.5 $ &  N   &  N  &              \\ 
177 & 184736.49-20050.45  & 0.79$\,\times\,$0.70 & $   1.3\pm0.3 $ &  N   &  N  &              \\ 
178 & 184742.58-20008.09  & 1.97$\,\times\,$1.74 & $   0.9\pm0.1 $ &  N   &  N  &              \\ 
179 & 184736.25-20050.03  & 0.70$\,\times\,$0.62 & $   2.0\pm0.4 $ &  N   &  N  &              \\ 
181 & 184739.02-20027.98  & 0.67$\,\times\,$0.53 & $   0.8\pm0.2 $ &  N   &  N  &              \\ 
186 & 184735.99-20039.75  & 2.14$\,\times\,$1.62 & $   1.1\pm0.2 $ &  N   &  N  &              \\ 
190 & 184742.45-15941.84  & 0.61$\,\times\,$0.33 & $   1.4\pm0.3 $ &  N   &  N  &              \\ 
196 & 184736.79-20105.05  & 1.05$\,\times\,$0.93 & $   0.9\pm0.2 $ &  N   &  N  &              \\ 
198 & 184736.26-20114.07  & 0.98$\,\times\,$0.78 & $   0.8\pm0.2 $ &  N   &  N  &              \\ 
200 & 184737.21-20058.46  & 1.35$\,\times\,$1.19 & $   1.2\pm0.2 $ &  N   &  N  &              \\ 
203 & 184736.83-20051.30  & 0.64$\,\times\,$0.46 & $   2.6\pm0.7 $ &  N   &  N  &              \\ 
208 & 184737.03-20100.97  & 0.80$\,\times\,$0.70 & $   0.9\pm0.2 $ &  N   &  N  &              \\ 
211 & 184739.25-20029.25  & 0.90$\,\times\,$0.83 & $   1.9\pm0.4 $ &  N   &  N  &              \\ 
212 & 184741.81-15956.33  & 1.83$\,\times\,$1.71 & $   1.0\pm0.2 $ &  N   &  N  &              \\ 
213 & 184736.14-20057.16  & 1.47$\,\times\,$1.14 & $   1.2\pm0.2 $ &  N   &  N  &              \\ 
217 & 184733.78-20100.96  & 0.95$\,\times\,$0.72 & $   1.7\pm0.3 $ &  N   &  N  &              \\ 
221 & 184741.81-20027.80  & 1.03$\,\times\,$0.73 & $   4.3\pm0.8 $ &  N   &  N  &              \\ 
223 & 184736.98-20100.25  & 0.80$\,\times\,$0.68 & $   1.2\pm0.3 $ &  N   &  N  &              \\ 
247 & 184736.79-20051.15  & 0.59$\,\times\,$0.56 & $   2.0\pm0.6 $ &  N   &  N  &              \\ 
\hline
\end{tabular} 
\end{table*}

\begin{table*}
\begin{threeparttable}
\caption{\label{tab:nature-coreMM1}  Main characteristics of prestellar and protostellar cores in W43-MM1}
\begin{tabular}{ccc|cccc|ccc} 
\hline\hline
n\tnote{a} & Name\tnote{b} & Size & $S^{\rm peak}$\,\tnote{c} & $S^{\rm int}$\,\tnote{c} & T$_{\rm dust}$\tnote{c} &  Mass\tnote{c} & Blue\tnote{d} & Red lobe\tnote{d} & Comments\tnote{e} \\
 & 1847*-154* & [$\arcsec \times \arcsec$] & [mJy.beam$^{-1}$] & [mJy.beam$^{-1}$] & [K] & [$\Msol$] & & & \\
 \hline
  1 & 47.02-26.91 & 0.66$\,\times\,$0.55 & $308.0\pm3.6$ & $599.1\pm4.7$ & $74\pm6$ & $108.7\pm14.2$ &  N  &  Y  &    HC*     \\ 
  2 & 46.84-29.28 & 0.55$\,\times\,$0.48 & $184.9\pm5.5$ & $272.8\pm6.3$ & $59\pm6$ & $ 57.0\pm8.5 $ &  Y  &  Y  &    HC*     \\ 
  3 & 46.37-33.40 & 0.67$\,\times\,$0.57 & $ 95.1\pm1.9$ & $210.7\pm2.7$ & $45\pm2$ & $ 54.5\pm3.3 $ &  Y  &  Y  &    HC*     \\ 
  4 & 46.98-26.45 & 0.81$\,\times\,$0.75 & $126.2\pm3.9$ & $358.8\pm4.5$ & $88\pm7$ & $ 41.8\pm4.0 $ &  N  & Y?  & tent. / HC* \\ 
  7 & 47.26-29.67 & 0.62$\,\times\,$0.43 & $ 40.2\pm0.6$ & $ 57.6\pm0.7$ & $30\pm2$ & $ 22.5\pm2.0 $ &  Y  &  Y  &            \\ 
  8 & 46.54-23.11 & 0.65$\,\times\,$0.46 & $ 44.9\pm1.1$ & $ 77.1\pm1.4$ & $45\pm5$ & $ 18.2\pm2.5 $ &  Y  &  Y  &            \\ 
  5 & 46.77-31.19 & 0.61$\,\times\,$0.47 & $ 48.4\pm3.2$ & $ 74.2\pm3.2$ & $47\pm2$ & $ 16.7\pm1.1 $ &  N  &  N  &     HC     \\ 
 15 & 44.77-45.20 & 0.62$\,\times\,$0.51 & $ 12.9\pm0.9$ & $ 21.0\pm0.9$ & $50\pm3$ & $  4.2\pm0.3 $ &  Y  &  Y  &            \\ 
 26 & 44.94-42.83 & 0.62$\,\times\,$0.48 & $  7.0\pm0.5$ & $ 11.8\pm0.5$ & $23\pm2$ & $  5.8\pm0.7 $ &  Y  &  Y  &            \\ 
 11 & 46.52-24.22 & 0.59$\,\times\,$0.41 & $ 14.7\pm1.3$ & $ 18.5\pm1.1$ & $93\pm11$ & $  1.9\pm0.3 $ & Y?  & Y?  & tent. / HC* \\ 
  9 & 46.48-32.57 & 0.58$\,\times\,$0.47 & $ 32.1\pm3.6$ & $ 46.4\pm3.3$ & $50\pm2$ & $  9.5\pm0.8 $ &  Y  &  Y  &     HC     \\ 
 18 & 46.25-33.38 & 0.64$\,\times\,$0.50 & $ 17.4\pm1.6$ & $ 25.1\pm1.3$ & $23\pm2$ & $ 13.2\pm1.7 $ &  Y  &  N  &            \\ 
 29 & 45.22-38.80 & 0.56$\,\times\,$0.44 & $  6.1\pm0.4$ & $  7.6\pm0.4$ & $24\pm2$ & $  3.6\pm0.4 $ &  Y  &  N  &            \\ 
 10 & 46.90-29.95 & 0.69$\,\times\,$0.60 & $ 23.7\pm4.8$ & $ 45.0\pm4.0$ & $51\pm2$ & $  8.9\pm0.9 $ & Y?  &  N  & tent. / HC \\ 
 51 & 45.26-39.91 & 0.89$\,\times\,$0.73 & $  3.8\pm0.5$ & $ 11.2\pm0.6$ & $24\pm2$ & $  5.3\pm0.6 $ &  N  & Y?  &   tent.    \\ 
 67 & 44.09-48.81 & 0.55$\,\times\,$0.51 & $  2.4\pm0.3$ & $  3.7\pm0.3$ & $23\pm2$ & $  1.8\pm0.2 $ &  Y  &  Y  &            \\ 
 39 & 44.61-42.13 & 0.55$\,\times\,$0.49 & $  3.3\pm0.4$ & $  4.4\pm0.3$ & $23\pm2$ & $  2.1\pm0.3 $ &  Y  &  Y  &            \\ 
 16 & 47.02-30.78 & 0.75$\,\times\,$0.56 & $ 21.0\pm4.9$ & $ 41.4\pm4.7$ & $21\pm2$ & $ 25.0\pm4.4 $ &  Y  &  Y  &            \\ 
 44 & 44.77-46.76 & 0.67$\,\times\,$0.49 & $  3.1\pm0.7$ & $  4.8\pm0.6$ & $22\pm2$ & $  2.5\pm0.4 $ &  Y  &  N  &            \\ 
 19 & 46.87-25.71 & 0.52$\,\times\,$0.47 & $ 14.8\pm4.2$ & $ 17.7\pm3.4$ & $23\pm2$ & $  9.0\pm2.0 $ &  Y  &  Y  &            \\ 
 12 & 46.57-32.04 & 0.65$\,\times\,$0.44 & $ 21.1\pm3.8$ & $ 28.7\pm3.2$ & $23\pm2$ & $ 14.9\pm2.4 $ &  N  &  Y  &            \\ 
 36 & 46.64-19.42 & 0.54$\,\times\,$0.47 & $  4.9\pm0.6$ & $  6.3\pm0.6$ & $27\pm2$ & $  2.6\pm0.3 $ &  Y  & Y?  &   tent.    \\ 
 31 & 46.73-17.46 & 0.56$\,\times\,$0.44 & $  5.7\pm0.6$ & $  7.1\pm0.4$ & $23\pm2$ & $  3.6\pm0.5 $ &  Y  & Y?  &   tent.    \\ 
 22 & 47.06-32.16 & 0.51$\,\times\,$0.40 & $ 11.5\pm2.6$ & $ 11.4\pm2.1$ & $21\pm2$ & $  6.6\pm1.5 $ &  Y  & Y,Y &            \\ 
 23 & 46.90-24.26 & 0.63$\,\times\,$0.43 & $ 14.2\pm2.9$ & $ 17.1\pm2.2$ & $22\pm2$ & $  9.5\pm1.7 $ & Y?  &  N  &   tent.    \\ 
 14 & 46.97-29.67 & 0.57$\,\times\,$0.39 & $ 19.0\pm5.0$ & $ 19.9\pm3.9$ & $22\pm2$ & $ 11.2\pm2.6 $ & Y?  &  N  &   tent.    \\ 
 49 & 46.59-20.50 & 0.56$\,\times\,$0.44 & $  5.6\pm0.6$ & $  6.5\pm0.5$ & $28\pm2$ & $  2.5\pm0.3 $ &  Y  &  Y  &            \\ 
 13 & 46.92-28.63 & 0.60$\,\times\,$0.33 & $ 18.6\pm4.7$ & $ 20.1\pm3.7$ & $24\pm2$ & $  9.9\pm2.1 $ & Y?  & Y?  &   tent.    \\ 
 59 & 44.77-44.16 & 0.60$\,\times\,$0.44 & $  2.4\pm0.9$ & $  3.0\pm0.7$ & $23\pm2$ & $  1.5\pm0.4 $ &  Y  &  Y  &            \\ 
174 & 45.17-35.69 & 0.76$\,\times\,$0.75 & $  0.5\pm0.2$ & $  1.1\pm0.2$ & $24\pm2$ & $  0.5\pm0.1 $ & Y?  &  N  &   tent.    \\ 
\hline
\end{tabular}
 \begin{tablenotes}
\item [a] Sources with a counterpart in the catalogue used by \cite{Nony20} are labelled with the original numbering of \cite{Motte18b}. New sources in this catalogue are labelled starting from the number 132. 
\item [b] Full source identification constructed from its right ascension and declination coordinates, W43-MM1\_ALMA-IMF1847*-154*.
\item [c] $S^{\rm peak}$ and $S^{\rm int}$: Peak and integrated fluxes measured by \textit{getsf} in the new 1.3~mm continuum map, T$_{\rm dust}$: dust temperature from \cite{Motte18b}.  
\item [d] Detection, tentative detection, and non detection of outflow lobes are denoted with "Y", "Y?" and "N", respectively. Information taken from \cite{Nony20}, except for cores \#51 and \#174 (this work).
\item [e] Cores whose outflow attribution is uncertain are denoted as "tent.", cores associated with hot core emission are denoted as HC \citep[see][]{Brouillet22}. The five cores corrected for hot core contamination are tagged as "HC*".
\end{tablenotes}
\end{threeparttable}
\end{table*} 

\setcounter{table}{1}
\begin{table*}
\caption{continued}
\centering
\begin{tabular}{ccc|cccc|ccc} 
\hline\hline
n\tnote{a} & Name\tnote{b} & Size & $S^{\rm peak}$\,\tnote{c} & $S^{\rm int}$\,\tnote{c} & T$_{\rm dust}$\tnote{c} &  Mass\tnote{c} & Blue\tnote{d} & Red lobe\tnote{d} & Comments\tnote{e} \\
 & 1847*-154* & [$\arcsec \times \arcsec$] & [mJy.beam$^{-1}$] & [mJy.beam$^{-1}$] & [K] & [$\Msol$] & & & \\
\hline
 6 & 46.16-33.29 & 0.66$\,\times\,$0.46 & $ 39.1\pm1.2$ & $ 62.6\pm1.3$ & $22\pm2$ & $ 36.7\pm4.9 $ &  N  &  N  &            \\ 
 20 & 45.29-37.03 & 0.56$\,\times\,$0.43 & $ 12.3\pm0.3$ & $ 16.2\pm0.3$ & $23\pm2$ & $  8.2\pm0.9 $ &  N  &  N  &            \\ 
 21 & 46.79-16.05 & 0.56$\,\times\,$0.43 & $ 15.5\pm0.7$ & $ 19.6\pm0.7$ & $23\pm2$ & $ 10.1\pm1.2 $ &  N  &  N  &            \\ 
132 & 47.00-25.68 & 1.09$\,\times\,$0.94 & $ 57.3\pm3.9$ & $262.1\pm6.0$ & $31\pm2$ & $100.8\pm9.2 $ &  N  &  N  & discarded  \\ 
 37 & 46.97-12.95 & 0.90$\,\times\,$0.67 & $  5.3\pm0.5$ & $ 14.2\pm0.6$ & $23\pm2$ & $  6.9\pm0.8 $ &  N  &  N  &            \\ 
 25 & 46.87-14.58 & 0.55$\,\times\,$0.39 & $  6.9\pm0.6$ & $  7.4\pm0.5$ & $24\pm2$ & $  3.6\pm0.5 $ &  N  &  N  &            \\ 
 40 & 46.35-29.50 & 0.58$\,\times\,$0.57 & $  5.0\pm0.5$ & $  7.7\pm0.4$ & $24\pm2$ & $  3.6\pm0.4 $ &  N  &  N  &            \\ 
 17 & 47.10-27.06 & 0.58$\,\times\,$0.50 & $ 24.7\pm3.7$ & $ 31.6\pm2.9$ & $35\pm2$ & $  9.7\pm1.1 $ &  N  &  N  &            \\ 
 63 & 47.33-12.80 & 0.57$\,\times\,$0.50 & $  3.7\pm0.3$ & $  5.3\pm0.3$ & $21\pm2$ & $  2.9\pm0.4 $ &  N  &  N  &            \\ 
133 & 48.17-03.79 & 1.11$\,\times\,$1.08 & $  1.0\pm0.1$ & $  6.3\pm0.2$ & $21\pm2$ & $  3.5\pm0.4 $ &  N  &  N  &            \\ 
134 & 46.89-29.63 & 0.69$\,\times\,$0.55 & $ 25.8\pm4.7$ & $ 39.4\pm3.6$ & $23\pm2$ & $ 21.4\pm3.3 $ &  N  &  N  &            \\ 
34 & 46.51-28.71 & 0.58$\,\times\,$0.47 & $  4.4\pm0.4$ & $  6.3\pm0.5$ & $25\pm2$ & $  2.9\pm0.4 $ &  N  &  N  &            \\ 
 54 & 45.06-42.04 & 0.54$\,\times\,$0.48 & $  2.8\pm0.4$ & $  3.4\pm0.3$ & $23\pm2$ & $  1.7\pm0.3 $ &  N  &  N  &            \\ 
 32 & 46.57-20.95 & 0.54$\,\times\,$0.46 & $  6.5\pm0.8$ & $  8.1\pm0.7$ & $28\pm2$ & $  3.1\pm0.4 $ &  N  &  N  &            \\ 
 74 & 45.18-39.45 & 0.59$\,\times\,$0.50 & $  2.5\pm0.5$ & $  3.5\pm0.4$ & $24\pm2$ & $  1.7\pm0.3 $ &  N  &  N  &            \\ 
 71 & 47.35-13.38 & 0.51$\,\times\,$0.41 & $  2.7\pm0.3$ & $  2.9\pm0.2$ & $21\pm2$ & $  1.6\pm0.2 $ &  N  &  N  &            \\ 
 73 & 44.84-42.95 & 0.70$\,\times\,$0.59 & $  2.3\pm0.7$ & $  4.0\pm0.6$ & $23\pm2$ & $  1.9\pm0.4 $ &  N  &  N  &            \\ 
 99 & 45.40-38.99 & 0.91$\,\times\,$0.82 & $  1.7\pm0.4$ & $  5.2\pm0.4$ & $23\pm2$ & $  2.5\pm0.3 $ &  N  &  N  &            \\ 
 28 & 44.40-41.78 & 0.57$\,\times\,$0.39 & $  4.1\pm0.9$ & $  4.3\pm0.7$ & $23\pm2$ & $  2.1\pm0.4 $ &  N  &  N  &            \\ 
136 & 46.93-27.08 & 0.60$\,\times\,$0.49 & $ 19.4\pm4.5$ & $ 23.3\pm3.5$ & $31\pm2$ & $  8.3\pm1.4 $ &  N  &  N  &            \\ 
 46 & 46.77-16.80 & 0.58$\,\times\,$0.54 & $  3.5\pm0.6$ & $  5.0\pm0.5$ & $21\pm2$ & $  2.8\pm0.4 $ &  N  &  N  &            \\ 
138 & 46.91-07.88 & 1.12$\,\times\,$1.05 & $  0.8\pm0.2$ & $  3.5\pm0.3$ & $21\pm2$ & $  2.0\pm0.3 $ &  N  &  N  &            \\ 
139 & 46.98-13.66 & 0.59$\,\times\,$0.42 & $  3.0\pm0.4$ & $  3.6\pm0.3$ & $23\pm2$ & $  1.8\pm0.2 $ &  N  &  N  &            \\ 
141 & 45.96-33.89 & 0.87$\,\times\,$0.64 & $  1.6\pm0.3$ & $  4.2\pm0.4$ & $24\pm2$ & $  2.0\pm0.3 $ &  N  &  N  &            \\ 
 62 & 45.79-32.71 & 0.67$\,\times\,$0.47 & $  2.8\pm0.5$ & $  4.0\pm0.4$ & $24\pm2$ & $  1.9\pm0.3 $ &  N  &  N  &            \\ 
 24 & 46.98-32.08 & 0.60$\,\times\,$0.41 & $  8.4\pm2.3$ & $  9.5\pm1.7$ & $21\pm2$ & $  5.3\pm1.2 $ &  N  &  N  &            \\ 
142 & 47.17-11.04 & 0.86$\,\times\,$0.71 & $  1.9\pm0.6$ & $  4.8\pm0.6$ & $21\pm2$ & $  2.7\pm0.5 $ &  N  &  N  &            \\ 
143 & 45.15-45.25 & 1.35$\,\times\,$1.17 & $  0.6\pm0.2$ & $  3.6\pm0.3$ & $22\pm2$ & $  1.9\pm0.3 $ &  N  &  N  &            \\ 
 64 & 47.00-16.79 & 0.61$\,\times\,$0.40 & $  3.9\pm0.6$ & $  4.8\pm0.5$ & $22\pm2$ & $  2.5\pm0.4 $ &  N  &  N  &            \\ 
144 & 46.93-13.51 & 0.63$\,\times\,$0.59 & $  2.4\pm0.5$ & $  4.4\pm0.5$ & $24\pm2$ & $  2.1\pm0.3 $ &  N  &  N  &            \\ 
146 & 44.34-42.14 & 0.83$\,\times\,$0.73 & $  1.8\pm0.6$ & $  4.4\pm0.6$ & $22\pm2$ & $  2.3\pm0.4 $ &  N  &  N  &            \\ 
148 & 47.20-22.72 & 0.73$\,\times\,$0.64 & $  2.2\pm0.8$ & $  4.5\pm0.8$ & $26\pm2$ & $  2.0\pm0.4 $ &  N  &  N  &            \\ 
151 & 47.33-28.37 & 0.82$\,\times\,$0.74 & $  1.2\pm0.3$ & $  3.1\pm0.3$ & $21\pm2$ & $  1.7\pm0.3 $ &  N  &  N  &            \\ 
153 & 47.95-06.45 & 1.83$\,\times\,$1.68 & $  0.3\pm0.1$ & $  3.6\pm0.1$ & $21\pm2$ & $  2.0\pm0.3 $ &  N  &  N  &            \\ 
154 & 47.19-31.72 & 0.82$\,\times\,$0.68 & $  1.5\pm0.4$ & $  3.6\pm0.4$ & $21\pm2$ & $  2.0\pm0.3 $ &  N  &  N  &            \\ 
155 & 45.89-35.17 & 0.95$\,\times\,$0.86 & $  1.0\pm0.4$ & $  3.5\pm0.5$ & $23\pm2$ & $  1.7\pm0.3 $ &  N  &  N  &            \\ 
156 & 48.06-26.93 & 1.62$\,\times\,$1.40 & $  0.4\pm0.1$ & $  3.4\pm0.1$ & $23\pm2$ & $  1.7\pm0.2 $ &  N  &  N  &            \\ 
158 & 46.39-13.42 & 1.34$\,\times\,$1.23 & $  1.0\pm0.2$ & $  6.7\pm0.3$ & $23\pm2$ & $  3.4\pm0.4 $ &  N  &  N  &            \\ 
162 & 47.11-34.32 & 0.69$\,\times\,$0.60 & $  1.9\pm0.8$ & $  3.4\pm0.7$ & $21\pm2$ & $  1.8\pm0.4 $ &  N  &  N  &            \\ 
173 & 47.52-41.62 & 1.95$\,\times\,$1.47 & $  0.4\pm0.2$ & $  4.9\pm0.4$ & $21\pm2$ & $  2.7\pm0.4 $ &  N  &  N  &            \\ 
177 & 46.89-30.69 & 0.61$\,\times\,$0.54 & $  7.9\pm3.4$ & $ 11.5\pm2.6$ & $23\pm2$ & $  5.8\pm1.5 $ &  N  &  N  &            \\ 
\hline
\end{tabular}
\end{table*}

\end{appendix}

\end{document}